\documentclass[twocolumn,superscriptaddress,showpacs,amsmath,amssymb,longbibliography,prb]{revtex4-1}
\usepackage[pdftex]{graphicx} % Include figure files
\usepackage{dcolumn}  % Align table columns on decimal point
\usepackage{bm}       % bold math
\usepackage{mhchem}
\usepackage[usenames,dvipsnames]{color}
\definecolor{URLCOL}{rgb}{0,0.52,0.83} %external link colort
\definecolor{LINKCOL}{rgb}{0.05,0.5,0} %internal link color
\definecolor{orange}{rgb}{0.6,0.3,0} %internal link color
\definecolor{CITECOL}{rgb}{0.25,0,0.48} %link to bibliography
\usepackage{epstopdf}
\usepackage{amsmath}
 \usepackage[table,xcdraw]{xcolor}
\usepackage{amsfonts}
\usepackage{booktabs}
\usepackage{siunitx}
\usepackage[symbol]{footmisc}

\usepackage{multirow}
\usepackage[pdftex,bookmarks,breaklinks,bookmarksopen,bookmarksnumbered,colorlinks,linkcolor=LINKCOL,linktocpage,citecolor=CITECOL,urlcolor=URLCOL,pdfpagemode=UseOutline,pdftex]{hyperref}
\usepackage{nicefrac}

\usepackage{fancyhdr}
\fancypagestyle{titlepage}
{\chead{file: \jobname}\rhead{Total pages: \pageref{page:end}}}

\definecolor{SEGreen}{rgb}{0.1,0.7,0.5}

\usepackage{booktabs}
\usepackage{subfigure}
\usepackage{natbib}
\definecolor{TITLECOL}{rgb}{0.1,0.2,0.7} %title color
\definecolor{SECOL}{rgb}{0.1,0.2,0.7} %sec color
\definecolor{CONTENTSCOL}{rgb}{0.1,0.2,0.7} %can choose the table of contents title to have same color as sec
\definecolor{SSECOL}{rgb}{0.25,0,0.48} %ssection color
\definecolor{SSSECOL}{rgb}{0.2,0.08,0.53} %subsubsection color  0.2,0.08,0.53
\definecolor{FINCOL}{rgb}{0.01,0.3,0.07} %subsubsection color  0.2,0.08,0.53

\def\coloredtitle#1{\title{\textcolor{TITLECOL}{#1}}} %title color
\def\coloredauthor#1{\author{\textcolor{CITECOL}{#1}}} %author color

\definecolor{URLCOL}{rgb}{0,0.17,0.43} %external link color
\definecolor{LINKCOL}{rgb}{0.05,0.4,0} %internal link color
\definecolor{CITECOL}{rgb}{0.35,0,0.48} %link to bibliography

\def\bea{\begin{eqnarray}}
\def\eea{\end{eqnarray}}
\def\ben{\begin{equation}}
\def\een{\end{equation}}
\def\benu{\begin{enumerate}}
\def\enu{\end{enumerate}}

\def\bei{\begin{itemize}}
\def\eei{\end{itemize}}
\def\beit{\begin{itemize}}
\def\eit{\end{itemize}}
\def\benu{\begin{enumerate}}
\def\enu{\end{enumerate}}

\def\sec#1{\section{\textcolor{SECOL}{#1}}}

\begin{document}
%Towards reliable hybrid functionals for fundamental and optical gaps of surfaces and bulk materials %OR Prediction of both bulk and surface semiconductor fundamental and optical gaps from hybrid functionals OR Reliable excitonic and electronic properties of surfaces with screened range separated hybrids
\coloredtitle{The reliability of hybrid functionals for accurate  fundamental and optical gap prediction of bulk solids and surfaces}

\coloredauthor{Francisca Sagredo}
%\email[email:]{fsagredo@lbl.gov}
\affiliation{Materials Sciences Division, Lawrence Berkeley National Laboratory, Berkeley, CA 94720, USA}
\affiliation{Department of Physics, University of California, Berkeley, CA 94720, USA}

\coloredauthor{Mar\'ia Camarasa-G\'omez}
\affiliation{Department of Molecular Chemistry and Materials Science, Weizmann Institute of Science, Rehovoth 76100, Israel}
\affiliation{Centro de Física de Materiales (CFM-MPC), CSIC-UPV/EHU, Donostia-San Sebastián 20018, Spain}

\coloredauthor{Francesco Ricci}
\affiliation{Materials Sciences Division, Lawrence Berkeley National Laboratory, Berkeley, CA 94720, USA}
\affiliation{Université catholique de Louvain (UCLouvain), Institute of Condensed Matter and Nanosciences (IMCN), B-1348 Louvain-la-Neuve, Belgium}
\affiliation{Matgenix SRL, A6K Advanced Engineering Centre, 6000 Charleroi, Belgium}

\coloredauthor{Aur\'elie Champagne}
\affiliation{Materials Sciences Division, Lawrence Berkeley National Laboratory, Berkeley, CA 94720, USA}
\affiliation{Institut de Chimie de la Matière Condensée de Bordeaux, CNRS, 33600 Pessac, France}

\coloredauthor{Leeor Kronik}
\affiliation{Department of Molecular Chemistry and Materials Science, Weizmann Institute of Science, Rehovoth 76100, Israel}

\coloredauthor{Jeffrey B. Neaton}
\affiliation{Materials Sciences Division, Lawrence Berkeley National Laboratory, Berkeley, CA 94720, USA}
\affiliation{Kavli Energy Nanosciences Institute at Berkeley, Berkeley, CA 94720, USA}
\affiliation{Department of Physics, University of California, Berkeley, CA 94720, USA}

\date{\today}
\begin{abstract}
\section*{Abstract} Hybrid functionals have been considered insufficiently reliable for the prediction of band gaps in solids and surfaces. We revisit this issue with a new generation of optimally-tuned range-separated hybrid functionals, focusing on the reconstructed Si(111)-(2$\times$1) and Ge(111)-(2$\times$1) surfaces. We show that certain hybrid functionals can accurately predict the surface-state and bulk fundamental and optical gaps, as well as projected band structures of these surfaces, by combining ground-state and time-dependent density functional theory.
\end{abstract}
\vspace{-2cm}
%%%%%%%%%%%%%%%%%%%%%%%%%%%%%%%%%%%%%%%%%%%%%%%%%%%%%%%%%%%%%%%%%%%%%%%%%
%%%%%%%%%%%%%%%%%%%%%%%%%%%%%%%%%%%%%%%%%%%%%%%
\maketitle
%\vspace{-5cm}
Density functional theory (DFT) has become a standard method to determine the electronic structure of both molecular and extended systems \cite{Parr1989, Dreizler1990, Martin2008,B12,Teale2003}. In practice, it is often used  within the Kohn-Sham (KS) framework \cite{KS1965}, where the original, many-electron system is mapped onto an equivalent system of non-interacting electrons. This KS mapping is exact in principle, in the sense of obtaining the same density as that of the original system. However, differences between the lowest unoccupied and highest occupied KS eigenvalues almost always underestimate the true fundamental band gap of the system, even if the exact functional is used \cite{PL83,SS83}. Most modern DFT calculations and codes use this KS scheme in practice with (semi-)local approximations to the exchange-correlation (XC) functional, which leads to severe failures in the prediction of optical properties of solids when used within time-dependent (TD) DFT \cite{M16}.

Due to the current limitations of standard TDDFT, \textit{ab initio} many-body perturbation theory approaches \cite{Hedin1965, Onida2002} have become a popular choice for the computation of electronic and optical excitations in solids. Specifically, calculation of charged excitations within the $GW$ approximation \cite{Hybertsen1986}, followed by calculations of neutral excitations using the Bethe-Salpeter equation (BSE) \cite{Rohlfing1998, Onida1998} are considered state-of-the-art \cite{Martin2016}. But the computationally costly `one-shot' $G_{0}W_{0}$ and the subsequent $G_0W_0$-BSE (dubbed $GW$-BSE in this letter), which are often used in practice to obtain optical absorption spectra, rely on the underlying density functional used. This starting point dependence is a known deficiency and an ongoing topic of research  \cite{Rinke2005, Fuchs2007, Chen2014,Leppert2019,Gant2022,Golze2019}. 
One solution to the limitations of KS theory is to use the generalized KS (GKS) scheme, which maps the original system onto one of {\it partially} interacting particles that are still represented by a single Slater determinant, thereby allowing the introduction of non-multiplicative XC  potential operators \cite{Seidl1996}.
Within this scheme, more advanced approximations, notably meta-generalized gradient approximations (MGGA)~\cite{CSDM19,SCAN15,PJCS19,TBLR19,YPSP16,LAK24,LASLK23,AK19} and hybrid functionals ~\cite{Becke1993, PBE0,Perdew2017,Shimazaki2009,Skone2014,Skone2016,Sun2020,JPCS20,Miceli2018,Yang2022}, have emerged as serious contenders for the solution of the band gap problem in solids.

An area in which the accuracy of advanced GKS-based functionals remains an important open issue is that of a functional that can consistently and accurately predict not only the fundamental gap, but also the optoelectronic properties of bulk materials, and more complex cases, like surfaces  \cite{JCL11}. In particular, Jain et al. \cite{JCL11} examined the performance of popular hybrid functionals such as PBE0~\cite{PBE0}, HSE~\cite{HSE, HSEerratum}, and B3LYP~\cite{Stephens1994} for the Si(111)-(2$\times$1) surfaces. They concluded that none can be used in a ground-state DFT calculation to consistently and accurately predict the fundamental gap, $E_g$, and the optical gap, $E_{opt}$ (quantities that differ by the exciton binding energy), of both bulks and surfaces. This task remains challenging even today, given that many advanced per-orbital correction methods that significantly improve the prediction of $E_g$ have yet to to be applied to optical excitations \cite{MahlerYang2022,Yang2022,NCFM18,DFM14,Ma2016}. 

Here, we reexamine the possibility of obtaining both bulk and surface $E_g$ and $E_{opt}$ using the above-mentioned hybrid functionals, as well as a new class of Wannier optimally-tuned screened range-separated hybrid functionals (WOT-SRSH) \cite{WOHF21} that are applicable to 3D solids, including layered materials \cite{Ohad2022,Ohad2023,Camarasa-Gomez2024,Sagredo2024}. We show that it is now possible to meet this challenge and predict accurate and reliable gaps for bulks and surfaces using slab calculations with certain hybrid functionals. In particular using WOT-SRSH, and for the standard semiconductors studied here also HSE, using ground-state DFT and TDDFT for charged and neutral excitations, respectively. 

The generalized Kohn-Sham (GKS) equation for a screened range-separated hybrid (SRSH) functional can be defined as \cite{Refaely2013}
\ben
\begin{split}
\bigg[-\frac{\bm{\nabla}^{2}}{2} + V(\bm{r})+ V_{H}([n];\bm{r}) + V_{X}^{SRSH}([n];\bm{r})
 +\\
 V_{SLc}([n],\bm{r})\Bigg] \phi_{i}(\bm{r})= \epsilon_{i} \phi_{i}(\bm{r})
 \label{eq1}
\end{split} 
\een
with
\ben
\begin{split}
V_{X}^{SRSH}=\alpha \hat{V}_{F}^{SR,\gamma}
+(1-\alpha) V_{SLx}^{SR} + \\ \frac{1}{\epsilon_{\infty}}\hat{V}_{F}^{LR,\gamma} + \bigg(1-\frac{1}{\epsilon_{\infty}}\bigg)V_{SLx}^{LR}.
\label{eq2}
\end{split} 
\een
Here $V, V_{H}, V_{X}^{SRSH}$, and $V_{c}$ are the external, Hartree, exchange, and correlation potentials, respectively. $\hat{V}_{F}$ is the Fock potential operator, SL  denotes semi-local exchange or correlation,  SR and LR  correspond to short or long range potentials, and $\epsilon_{i}$ and $\phi_{i}$ are the generalized KS eigen energies and associated orbitals. Finally, $\epsilon_{\infty}$ corresponds to the orientally averaged, clamped-ion dielectric constant, and $\gamma$ is the range separation parameter. 

A key aspect of range-separated hybrid functionals, which differentiates them from global hybrids (\textit{e.g.} PBE0 or B3LYP), is the different fraction of Fock exchange in the treatment of short- and long-range exchange associated with the $1/r$ Coulomb interaction, controlled by the  $\gamma$ parameter \cite{Leininger1997}. %In global hybrids, the Fock exchange $\alpha$ is typically taken to be fixed. 
We note that global hybrids are special cases of Eq.~(\ref{eq1}), \textit{e.g.}, PBE0~\cite{PBE0} is obtained when $\alpha= 1/\epsilon_{\infty}=1/4$, independent of $\gamma$. Similarly, HSE can be recovered for $\alpha=1/4$ and $\gamma= 0.2 {\rm \AA}^{-1}$, with $\epsilon_{\infty}\rightarrow \infty$. Different choices of $\alpha, \gamma, \epsilon_{\infty}$ lead to different hybrid functionals, and must be made judiciously. For example, it has been shown that using an `arbitrary amount' of Fock exchange $\alpha$ can lead to qualitatively incorrect defect physics in 2D materials \cite{CGRH22}.
%\vspace{-0.5cm}
\begin{table}[h]
\caption{Fundamental bulk and surface state gaps obtained from Si(111)-(2$\times$1) 
and Ge(111)-(2$\times$1), and compared to reference $GW$ and experiments.} 
\label{table5}
\centering
\begin{tabular}{ccc|ccc}
\hline \hline 
(eV)&{Si(111)2$\times$1}  &  &{Ge(111)2$\times$1}&  & \\ \hline 
    &   &  &     &  \\
 & Bulk  & Surface   & Bulk &  Surface   \\ \hline
PBE   & 0.61  & 0.37     &0.07 & 0.39 \\
AM05  &  0.50 & 0.40  & 0.08  & 0.43 \\
SCAN    &  0.84  & 0.45   &  0.18 & 0.50\\
$\alpha$-r2SCAN     & 1.72 & 0.77   &  1.37 & 1.80 \\
PBE0    & 1.78 & 1.00    & 1.47 & 1.20 \\
HSE    &   1.16 & 0.64 & 0.83 & 0.74 \\
SRSH(($\alpha,\gamma$)\cite{WOHF21})  & 1.15 & 0.64  &  0.86 & 0.75 \\
\hline
Reference       &   &  &     &  \\
\hline
$GW$  &  1.23~\cite{RL99}  & 0.69~\cite{CSCC86}    & 0.72~\cite{Aulbur2000}%1.18\footnote{Only $G_0W_0$ calculation computed here. Done with HSE as the starting point for bulk Ge. Agrees with computed spectra of ~\cite{Cunningham2018}.} 
& 0.67~\cite{zhu1991} \\
Exp   &   1.17~\cite{M04} & 0.75~\cite{CSCC86}    & 0.75~\cite{Aulbur2000}
%0.65~\cite{M04}   
& 0.65~\cite{NR89} \\
 \hline \hline
 \label{table1}
 \end{tabular}
\end{table}

%\vspace{-0.5cm}
For \textit{tuned} screened range-separated hybrid (SRSH) functionals, %with a more formal approach to GKS theory ~\cite{GNGK20},
the choice of parameters is flexible, but not arbitrary. Typically, $\alpha$ is  chosen \textit{along} with $\gamma$ to satisfy the ionization potential (IP) theorem (\textit{i.e.} the optimally-tuned [OT-]SRSH \cite{Refaely2013}), or an IP ansatz associated with removal of localized charge (\textit{i.e.} the Wannier-localized optimally-tuned [WOT-]SRSH \cite{WOHF21}). In other words, this ansatz serves as our physical constraint. By satisfying this known exact condition, using $\epsilon_{\infty}$ to constrain the long-range Coulomb interaction, and using an image-charge correction (when removing a charge from the Wannier function), the WOT-SRSH functional has shown improvements of the fundamental bands gaps, band structures and optical gaps of materials  \cite{WOHF21,Ohad2023,Ohad2022,Sagredo2024,Camarasa-Gomez2024}.

The WOT-SRSH functional is part of a recently emerging class of functionals~\cite{Kronik2012,Tal2020,Stein2010,Refaely2013,Autschbach2014,MCRP18}. A distinct aspect of this method is the ansatz used in the localization scheme for the wavefunction of extended systems, introduced in Ref. \cite{Ma2016}, specifically using maximally-localized Wannier functions. We note that HSE is also a range-separated hybrid (RSH) functional, albeit a special one because it lacks long-range exchange, and thus in general lacks the correct asymptotic potential. Therefore, while there is merit to exploring modifications in the HSE parameters \cite{MSC12},  it is unsuitable for the systematic tuning procedure presented above.%its lack of an explicit dielectric response makes

Even given the success of WOT-SRSH for bulk systems, surfaces pose additional challenges to the tuning procedure.  A major part of the added difficulty comes from the fact that the slab structures, used to simulate surfaces, are environments with a different dielectric screening than in the bulk. Therefore limits that can typically be taken for the dielectric medium in 2D systems, (particularly  $\epsilon_{\infty} \rightarrow 1 $ \cite{Qiu2016,Cudazzo2011,Andersen2015}), including the generalized tuning procedures for 2D systems based on them~\cite{Camarasa-Gomez2024}, cannot be used in surfaces. A natural question then arises: Is it possible to use tuned ($\alpha, \gamma$) parameters obtained for the bulk, even for surface-containing slabs? In other words, how transferable are the tuned ($\alpha, \gamma$) parameters outside of the ideal bulk scenario?  More generally, how reliable are various hybrid functionals in this context? 
%Here our first step to solving this problem is using the previously tuned bulk parameters on a homogoneous surface.
%First explain why it's a problem for a surface.Next - some recent with success with 2d, but not directly applicable to surfaces.Hence as a first step, we adopt the bulk parameters and apply to them a slab to assess how well that would work.
 %Semi-empirical tuning procedures and parameter transferibility have been proposed for 2D systems  ~\cite{Ramasubramaniam2019, Camarasa-Gomez2023}, and more recently a non-empirical tuning procedure for 2D materials has been formalized ~\cite{Camarasa-Gomez2024}.
%\vspace{-0.5cm}
\begin{figure*}[!htbp]
  \centering
  \begin{minipage}[b]{.45\linewidth}
{(a)\includegraphics[width=\linewidth]{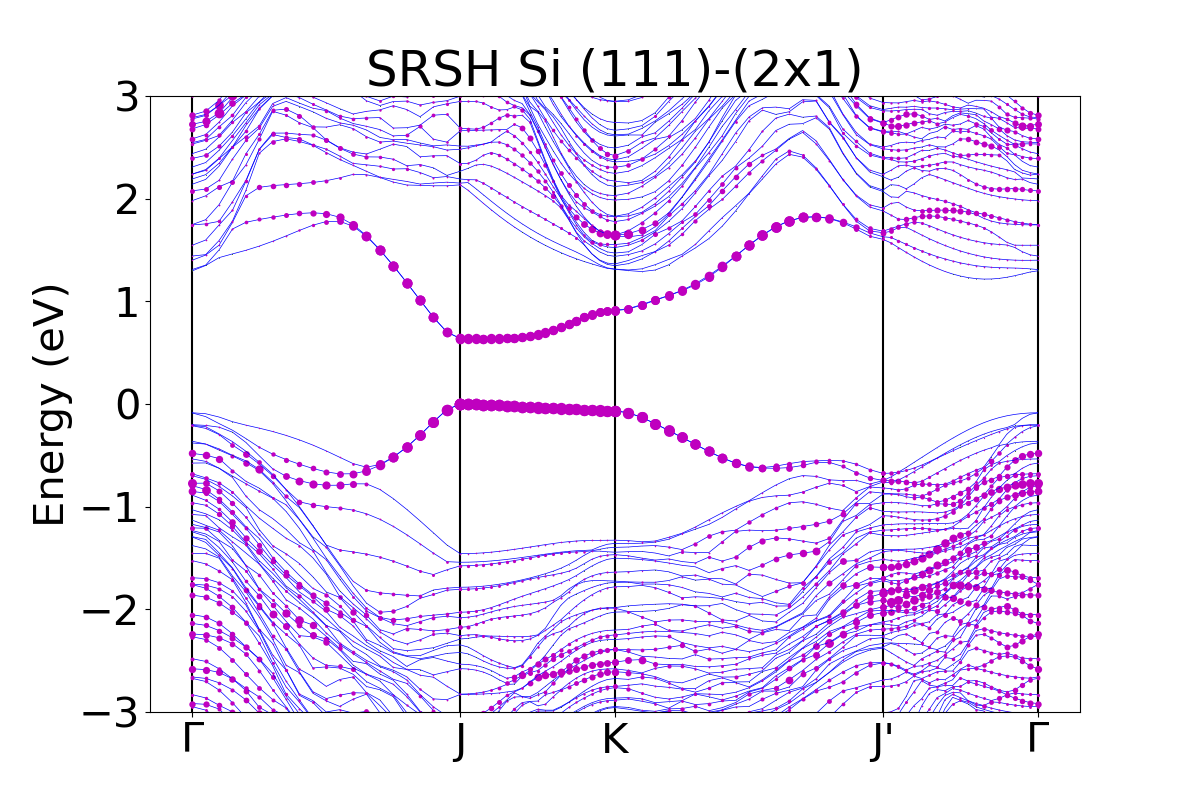}}
{(b)\includegraphics[width=\linewidth]{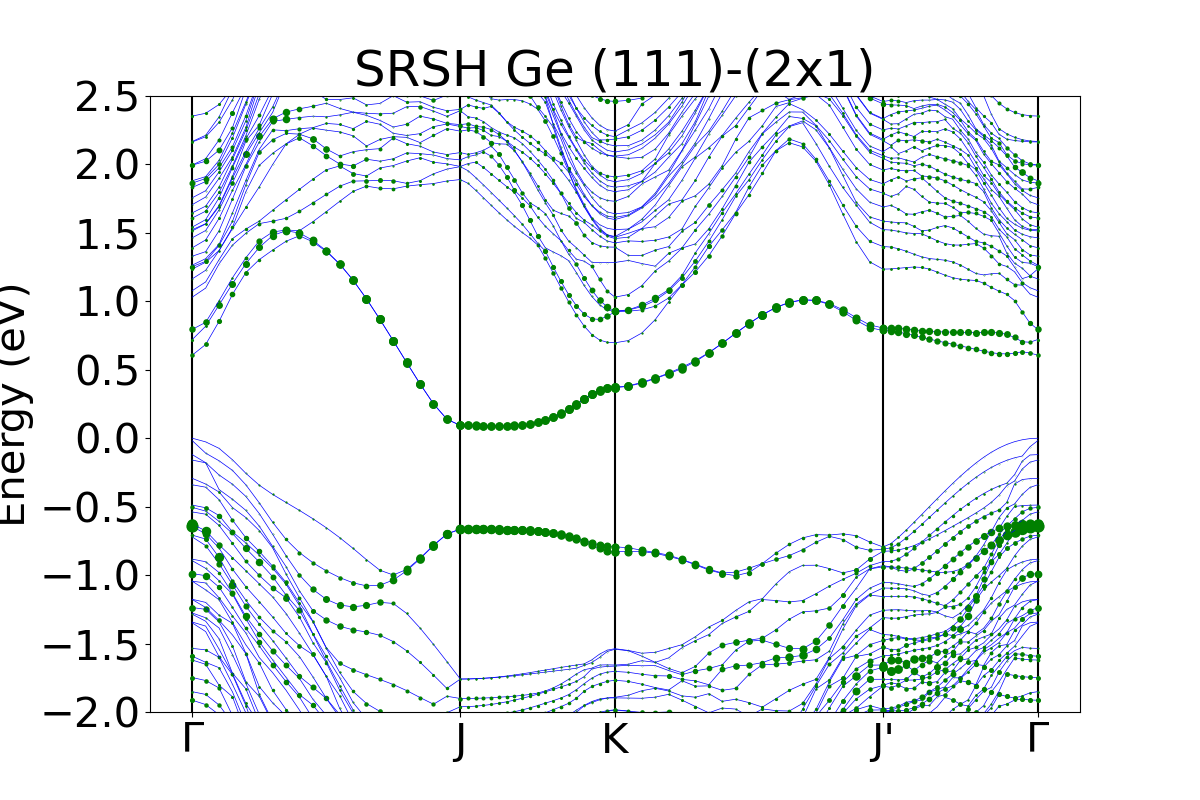}}%
  \end{minipage}%
 %\hfill
  \begin{minipage}[t]{.45\linewidth}
{(c,d)\includegraphics[width=\linewidth]{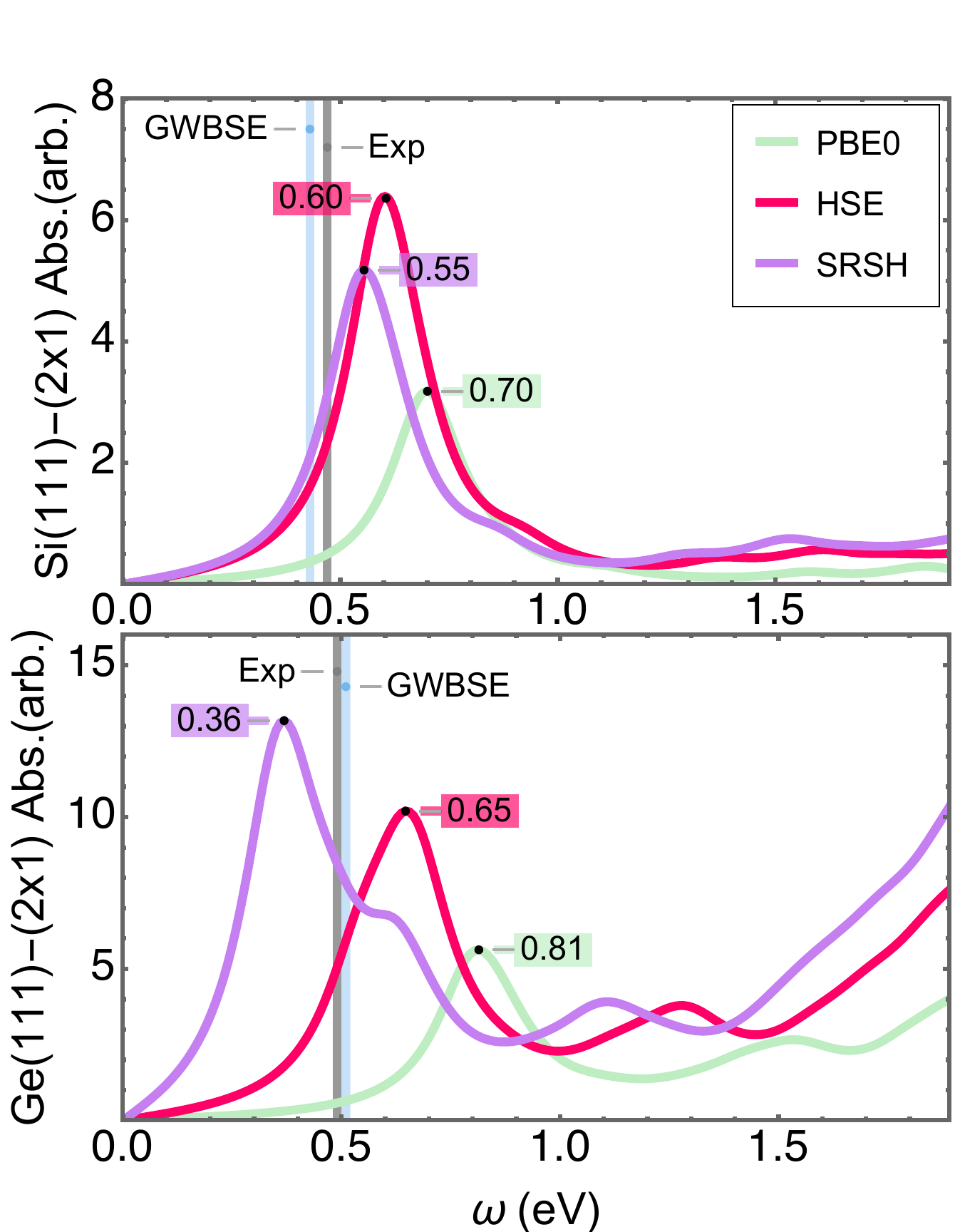}}%
  \end{minipage}%
  \caption
    {Projected band structure for slabs containing the reconstructed (a) Si(111) and (b) Ge(111)-(2$\times$1) surfaces (see SI for details on structures used). We show the band structure of the whole system with blue solid line, and highlight the contribution of the surface states with pink (Si) and green (Ge) dots that are proportional to their weight in the wavefunction, obtained using the SRSH functional with tuned bulk parameters. (c) Linear absorption spectrum, Im$[(\epsilon_{XX}+\epsilon_{YY}+\epsilon_{ZZ})/3]$, for Si(111)-(2$\times$1) and (d) Ge(111)-(2$\times$1), of the lowest-energy bright excitations associated with each surface, with a 0.1 eV broadening. Reported experimental values are denoted by the vertical gray bar at 0.47 eV \cite{MMQR57} (0.49 eV \cite{Rohlfing2000}), and previous $GW$-BSE \cite{JCL11,Rohlfing2000} results are given by the blue vertical line at 0.43 eV(0.51 eV) for Si (and Ge). The purple, magenta, and green lines are the results of the TDDFT calculations with SRSH, HSE, and PBE0, respectively, for the reconstructed Si (and Ge) (111)- (2$\times$1) surface represented by the slab. See SI for detailed discussion of Lorentzian broadening value used to average the plotted peaks, as well as analysis of asymmetry of the lowest peak of the HSE Ge spectrum.
      \label{fig proj bands}}
\end{figure*}%
%\vspace{-0.5cm}
We answer the above questions using two well-known reconstructed semiconductor surfaces \cite{S97}: Si(111)-(2$\times$1)~\cite{P81, RD91, NHL91}, and Ge(111)-(2$\times$1)~\cite{Rohlfing2000}. We model these surfaces with slab structures that contain properties of both bulk and surfaces. 
%--thus the ideal testing case for functional performance. 
Results for these systems can be readily compared to literature experimental \cite{UHNF82,PNR87,CSCC86,MMQR57,Nannarone1980, Aulbur2000} and theoretical \cite{JCL11,Rohlfing2000, Aulbur2000} values for bulk and surface fundamental and optical gaps.  

The (2$\times$1) reconstruction is produced by taking an ideal Si (or Ge) (111) surface and pushing two of the surface atoms toward each other. After surface relaxation, $p_z$ orbitals form $\pi$ bonded chains along the [001] direction of the surface. The dangling $p_z$ orbitals in turn, generate two surface-state bands inside the bulk-forbidden gap \cite{S99}, yielding distinctly different surface-state and bulk gaps of this material. For compatibility with past $GW$ results, we use the known positive buckling configuration for the Si case and the negative buckling for Ge ~\cite{NFLK04}. 

All calculations are performed using the \textit{Vienna Ab initio Simulation Package} VASP~\cite{vasp1, vasp2}, on a 48 atom unit cell (24 layers), with 20 {\AA} of vacuum. More computational details, including assessment of slab size, structural information, convergence tests, and comparison to results obtain with the \texttt{QUANTUM ESPRESSO} (QE) package \cite{Giannozzi2009}, are found in the supplementary information (SI)~\cite{SupplementalMaterial}.  We emphasize that for the optical spectra we are only concerned with the peak associated with the surface state, which is the lowest-energy one for both Si and Ge slabs. The peaks correspond to the lowest  excitations with non-zero oscillator strengths. Our reported gaps are an average of the lowest  bright excitons with a broadening of $0.1$ eV. See the SI \cite{SupplementalMaterial} for more details and analysis. Bulk optical peaks are known to appear at higher energies and are well-prediced by TDDFT with SRSH for both Si and Ge~\cite{Wing2019, Wing2020}. 
%We find agreement between QE and VASP results for all DFT calculations, within $0.05$ eV, except for the Ge slab with SRSH and HSE, in which the bulk gap differs by $0.18$ eV.

As a first step, we calculate the surface projected band structures, using the bulk-tuned \cite{WOHF21} WOT-SRSH functional. In Fig.~\ref{fig proj bands} (a) and (b), we plot the projected surface states of the Si and Ge surfaces in pink and green, respectively. Clearly, the bands of the surface states can be seen within the bulk bands of each material, as described above. More projected band structures calculated with other functionals can be found in the SI~\cite{SupplementalMaterial}. We find good agreement with previously computed projected band structures \cite{JCL11,zhu1991} for these surfaces. The fundamental gap for the surface states occurs at J. We observe a finite band gap at $\Gamma$, for Ge, whereas a negligible gap is found with standard semilocal or meta-GGAs tested for comparison (see SI).

In Table \ref{table1}, we present calculations for the bulk and surface state fundamental gaps for Si(111)-(2$\times$1) and Ge(111)-(2$\times$1) using PBE~\cite{PBE96}, PBE0~\cite{PBE0}, HSE06~\cite{HSE06, HSEerratum}, AM05~\cite{AM05}, SCAN~\cite{SCAN15}, $\alpha$-r\textsuperscript{2}SCAN (\textit{i.e.} r\textsuperscript{2}SCAN with Fock exchange)~\cite{DKPK23}. We further compare with previous $GW$ and $GW$-BSE calculations \cite{JCL11,Rohlfing2000}. We only investigate some currently popular meta-GGAs, but we make note of the recent success of newer functionals in similar bulk systems  \cite{AK19,LASLK23,LAK24}. We find the expected underestimations of the bulk and surface state gaps of the Si structure using the semi-local PBE functional, by $0.56$ and $0.38$ eV, respectively, when compared to experiment (as does AM05 with an absolute error (AE) of 0.67 and 0.35 eV, respectively). The global hybrid PBE0 overestimates both the bulk and surface state gap by $0.61$ and $0.25$  eV, respectively. The HSE06 hybrid functional performs with good accuracy, with an AE of $0.11$ and $0.01$ eV for the surface state and bulk gap, respectively. The meta-GGA SCAN underestimates the bulk gap by $0.33$ eV, and the surface state gap by $0.30$ eV. The SRSH functional, using previously tuned \cite{WOHF21} Si bulk ($\alpha$, $\gamma$) pairs from the WOT-SRSH functional, performs on par with HSE06, with an AE of $0.02$ and $0.13$ eV, for bulk and surface state gap respectively. Finally, using the recommended $\alpha=0.15$ Fock exchange \cite{DKPK23} the r\textsuperscript{2}SCAN \cite{r2SCAN} functional does get good agreement for the surface state fundamental gap, within 0.02 eV, but it overestimates the bulk gap by 0.55 eV with respect to the experimental value. Summarizing, the SRSH and HSE functionals outperform all other density approximations and are within 0.08 and 0.07 eV of previous $GW$ results \cite{RL99,CSCC86,Aulbur2000,zhu1991}, respectively.

We repeat this set of calculations with a similar slab representing the reconstructed Ge(111)-(2$\times$1) surface (see right side of table~\ref{table1}). We find the same general trends as in the Si case, but with a few caveats: First, as shown above, the fundamental gap of bulk Ge at $\Gamma$ is known to be almost zero for most semilocal and meta-GGA functionals. Second, even when using a hybrid, specifically HSE, as a starting point for the $GW$ calculations, an overestimation of the bulk Ge gap is seen (yields a gap of $1.18$ eV). This shows the sometimes inconsistent nature of `one-shot' $GW$, even with a hybrid functional starting point used in previous studies~\cite{Rinke2005,Gant2022,Golze2019}.
%\textcolor{red}{Maria what were the bulk Ge gap values when using WOT as a starting point to $GW$?} \textcolor{blue}{The bandgap at the $\Gamma$ point with WOT with the parameters reported in the PNAS paper is 0.84 eV, and the subsequent GW is 1.18 eV.} 

Finally, using linear response TDDFT ~\cite{Casida1996,TC2003}, within the Tamm-Dancoff approximation (TDA), it is possible to obtain the linear absorption spectra of the reconstructed surfaces, here in arbitrary units for the oscillator strength. The TDA has been found to be accurate for systems of similar excitonic effect,  compared to the full Casida equation~\cite{TD-approx,Byun2016}. The results of our calculations are seen in Fig.~\ref{fig proj bands} for the reconstructed Si (c) and Ge (d) slabs for PBE0, HSE, and SRSH functionals.  The attainment of the optical gap for these slabs, made clear by 
Fig.~\ref{fig proj bands}, goes beyond the expectations of Ref. \cite{JCL11} from hybrid functionals, owing to the use of TDDFT rather than ground-state DFT. %Importantly, for the case of Ge, the SRSH functional is the only one that predicts the ``shoulder'' in the Ge spectrum at $\sim$ 0.65 eV. 
We find that both the SRSH and HSE functionals perform with good accuracy, and get within 0.08 and and 0.13 eV of the experimental optical gap for Si respectively (for Ge the errors are 0.13 eV and 0.16 eV, respectively). When compared to previously reported $GW$-BSE optical gaps for the Si case, the SRSH and HSE are within 0.12 and 0.17 eV (for Ge those errors are 0.15 and 0.14 eV) respectively. PBE0 tends to overestimate the optical gap for the Si and Ge surface. 
%We take the oscillator strength of these TD-DFT calculations to be arbitrary, or a dimensionless quantity \cite{Ullrich2012}, 
%which is not the case in its $GW$-BSE counterpart. Furthermore, the optical spectra for our purposes is a demonstration of the accuracy and a comparison of functionals used, and they should not be expected to be replicated directly by experiment, unless experiments are able to reproduce an exact unit cell as the one used in these calculations. 
While the calculated spectra in Fig. 1 allow for a visual comparison of the TDDFT excited states associated with the surface states for different functionals, the magnitude of the peak depends on the supercell and slab size, and therefore are taken as arbitrary. The lowest-energy optical peaks for the Si(111)-(2$\times$1) and Ge(111)-(2$\times$1) slabs, at the WOT-SRSH level, are well within the $\pm$ 0.2 eV experimental uncertainty and prior GW-BSE calculations \cite{JCL11}. In the SI\cite{SupplementalMaterial}, we report the individual excitons that make up the absorbance peaks and analyze the effects of Lorentzian broadening. Other studies have shown that in-gap states induce novel low-lying excitons not present in the bulk, \textit{e.g.}, those originating with a mixture of bulk states and surface states \cite{Hernangomez2023-02}. A detailed study of surface excitons would be an interesting subject for future work.

To conclude, we find that it is possible to reliably predict both the bulk and surface state fundamental gaps, of Si(111)-(2$\times$1) and Ge(111)-(2$\times$1) with the SRSH (using previously tuned bulk $\alpha$, and $\gamma $ parameters) and HSE functionals, with good accuracy. More importantly, we find that it is possible to obtain accurate optical gaps, using these hybrid functionals, with TDDFT calculations. While both SRSH and HSE functionals perform equally well here, we expect to see deviations in the performance of larger gap materials, similar to the bulk case \textit{e.g.} Fig. 3 in SI of \cite{WOHF21}, and Ref.~\cite{Chen2018}. Such comparisons involving HSE and PBE0 of bulk materials have also been studied \textit{e.g.} \cite{Song2013}, and table SIII.1 of \cite{Gant2022}. Previous work more than a decade ago \cite{JCL11} concluded that hybrid functionals (at the time) did not give reliable band gaps for bulk systems or surfaces, and that fortuitously obtaining a good surface state and bulk gap with a hybrid functional did not guarantee the optical gap with ground state DFT calculation. Here, we overcome this by using the DFT and TDDFT frameworks with a new generation of SRSH, namely WOT-SRSH, where $\alpha$ and 
$\gamma$ are constrained by an IP ansatz, can lead to bulk and surface fundamental and optical gaps in excellent agreement with experiments and previous $GW$-BSE results. Finally, while a tuning procedure is not yet available for surfaces, the reliable performance seen with the transferability of bulk parameters for these reconstructed surfaces, provides an indication that the construction of this class of hybrids goes in the correct direction. 
\vspace{-.8cm}
%\begin{table}[h]
%\caption{Fundamental and optical gaps of Si(111) 2x1 surface, with entanglement (slight cheating)}
%\label{table1}
%\centering
%\begin{tabular}{ c c c c }
%\begin{tabular}{cS[table-format=2.2]S[table-format=1.2]}
%\hline
%\hline
%Si(111) 2x1 & {} & (eV) \textcolor{blue}{QE} & {}\\ \hline
%{} & {Bulk gap} & & {Surface gap}\\ \hline
%PBE & 0.61 \textcolor{blue}{($0.59$)} & & 0.37 \textcolor{blue}{($0.36$)}  \\
%PBE0 & \textcolor{blue}{1.80} & & 1.2 \textcolor{blue}{($1.16$)}\\
%HSE & \textcolor{blue}{1.17} & & 0.7 \textcolor{blue}{($0.62)$}\\
%RSH(Bulk Si($\alpha,\gamma$)\cite{WOHF21}) & 
% \textcolor{blue}{1.20} &
% & 0.76\textcolor{blue}{($0.69$)} \\
%AM05 & 0.50 & & {0.37 } \\ 
%SCAN & 0.84 & & 0.43 \\ \hline
%Reference & & & \\ \hline
%$GW$[20,31] & 1.23 & & 0.69  \\
%Exp[25,33] & 1.17 %& & 0.75\\ \hline
%Optical gaps & & \\ \hline
%$GW$-BSE[20,31] &1.23 & & 0.43 \\
%Exp[25,31] &1.16 & &0.47 \\ \hline %\hline
%\bottomrule
%\end{tabular}
%\end{table}
%\vspace{-0.5cm}
\sec{Supporting Information}
\vspace{-0.5cm}
Computational details, convergence tests, projected band structures for functionals not included in main text, and atomic structure details. 
\vspace{-0.5cm}
\sec{Acknowledgements}
\vspace{-0.5cm}
F.S. acknowledges discussions with Prof. M. Jain which clarified important aspects of past work. M.C.-G. and F.S. thank Prof. A. Ramasubramaniam for fruitful discussions. M.C.-G. is grateful to the Azrieli foundation for the award of an International Postdoctoral Fellowship and acknowledges support from the Diputación Foral de Gipuzkoa through Grant 2024-FELL-000007-01. F.R. acknowledges support from the BEWARE scheme of the Wallonia-Brussels Federation for funding under the European Commission’s Marie Curie-Skłodowska Action (COFUND 847587). L.K. acknowledges additional support from the Aryeh and Mintzi Katzman Professorial Chair and from the Helen and Martin Kimmel Award for Innovative Investigation. The formalism used in this work was funded through NSF–Binational Science Foundation Grant No. DMR-2015991 and the Israel Science Foundation; and our calculations were supported by the Theory of Materials Program at LBNL, funded by the U.S. Department of Energy, Office of Science, Basic Energy Sciences, Materials Sciences and Engineering Division, under Contract No. DE-AC02-05CH11231. Computational resources were provided by the NSF funded ACCESS program (frontera supercomputer),at the Texas Advanced Computing Center (TACC), the National Energy Research Scientific Computing Center (NERSC), a Department of Energy Office of Science User Facility at Lawrence Berkeley National Laboratory, and the Chemfarm facility at the Weizmann Institute of Science. 
\bibliography{wot-srsh-surfaces}

%merlin.mbs apsrev4-1.bst 2010-07-25 4.21a (PWD, AO, DPC) hacked
%Control: key (0)
%Control: author (0) dotless jnrlst
%Control: editor formatted (1) identically to author
%Control: production of article title (0) allowed
%Control: page (1) range
%Control: year (0) verbatim
%Control: production of eprint (0) enabled
\begin{thebibliography}{100}%
\makeatletter
\providecommand \@ifxundefined [1]{%
 \@ifx{#1\undefined}
}%
\providecommand \@ifnum [1]{%
 \ifnum #1\expandafter \@firstoftwo
 \else \expandafter \@secondoftwo
 \fi
}%
\providecommand \@ifx [1]{%
 \ifx #1\expandafter \@firstoftwo
 \else \expandafter \@secondoftwo
 \fi
}%
\providecommand \natexlab [1]{#1}%
\providecommand \enquote  [1]{``#1''}%
\providecommand \bibnamefont  [1]{#1}%
\providecommand \bibfnamefont [1]{#1}%
\providecommand \citenamefont [1]{#1}%
\providecommand \href@noop [0]{\@secondoftwo}%
\providecommand \href [0]{\begingroup \@sanitize@url \@href}%
\providecommand \@href[1]{\@@startlink{#1}\@@href}%
\providecommand \@@href[1]{\endgroup#1\@@endlink}%
\providecommand \@sanitize@url [0]{\catcode `\\12\catcode `\$12\catcode `\&12\catcode `\#12\catcode `\^12\catcode `\_12\catcode `\%12\relax}%
\providecommand \@@startlink[1]{}%
\providecommand \@@endlink[0]{}%
\providecommand \url  [0]{\begingroup\@sanitize@url \@url }%
\providecommand \@url [1]{\endgroup\@href {#1}{\urlprefix }}%
\providecommand \urlprefix  [0]{URL }%
\providecommand \Eprint [0]{\href }%
\providecommand \doibase [0]{http://dx.doi.org/}%
\providecommand \selectlanguage [0]{\@gobble}%
\providecommand \bibinfo  [0]{\@secondoftwo}%
\providecommand \bibfield  [0]{\@secondoftwo}%
\providecommand \translation [1]{[#1]}%
\providecommand \BibitemOpen [0]{}%
\providecommand \bibitemStop [0]{}%
\providecommand \bibitemNoStop [0]{.\EOS\space}%
\providecommand \EOS [0]{\spacefactor3000\relax}%
\providecommand \BibitemShut  [1]{\csname bibitem#1\endcsname}%
\let\auto@bib@innerbib\@empty
%</preamble>
\bibitem [{\citenamefont {Parr}\ and\ \citenamefont {Yang}(1989)}]{Parr1989}%
  \BibitemOpen
  \bibfield  {author} {\bibinfo {author} {\bibfnamefont {R.~G.}\ \bibnamefont {Parr}}\ and\ \bibinfo {author} {\bibfnamefont {W.}~\bibnamefont {Yang}},\ }\enquote {\bibinfo {title} {{Density Functional Theory of Atoms and Molecules}},}\ \ (\bibinfo  {publisher} {Oxford University Press, Oxford},\ \bibinfo {year} {1989})\BibitemShut {NoStop}%
\bibitem [{\citenamefont {Dreizler}\ and\ \citenamefont {Gross}(1990)}]{Dreizler1990}%
  \BibitemOpen
  \bibfield  {author} {\bibinfo {author} {\bibfnamefont {M.}~\bibnamefont {Dreizler}}\ and\ \bibinfo {author} {\bibfnamefont {E.~K.~U}\ \bibnamefont {Gross}},\ }\enquote {\bibinfo {title} {{Density Functional Theory: An Approach to the Quantum Many-Body Problem}},}\ \ (\bibinfo  {publisher} {Springer, Berlin},\ \bibinfo {year} {1990})\BibitemShut {NoStop}%
\bibitem [{\citenamefont {Martin}(2008)}]{Martin2008}%
  \BibitemOpen
  \bibfield  {author} {\bibinfo {author} {\bibfnamefont {R.~M.}\ \bibnamefont {Martin}},\ }\enquote {\bibinfo {title} {{Electronic Structure: Basic Theory and Practical Methods}},}\ \ (\bibinfo  {publisher} {Cambridge University Press},\ \bibinfo {year} {2008})\BibitemShut {NoStop}%
\bibitem [{\citenamefont {Burke}(2012)}]{B12}%
  \BibitemOpen
  \bibfield  {author} {\bibinfo {author} {\bibfnamefont {K.}~\bibnamefont {Burke}},\ }\bibfield  {title} {\enquote {\bibinfo {title} {Perspective on density functional theory},}\ }\href {\doibase 10.1063/1.4704546} {\bibfield  {journal} {\bibinfo  {journal} {J. Chem. Phys.}\ }\textbf {\bibinfo {volume} {136}},\ \bibinfo {pages} {150901} (\bibinfo {year} {2012})}\BibitemShut {NoStop}%
\bibitem [{\citenamefont {Teale}\ \emph {et~al.}(2022)\citenamefont {Teale}, \citenamefont {Helgaker}, \citenamefont {Savin}, \citenamefont {Adamo}, \citenamefont {Aradi}, \citenamefont {Arbuznikov}, \citenamefont {Ayers}, \citenamefont {Baerends}, \citenamefont {Barone}, \citenamefont {Calaminici}, \citenamefont {Cancès}, \citenamefont {Carter}, \citenamefont {Chattaraj}, \citenamefont {Chermette}, \citenamefont {Ciofini}, \citenamefont {Crawford}, \citenamefont {De~Proft}, \citenamefont {Dobson}, \citenamefont {Draxl}, \citenamefont {Frauenheim}, \citenamefont {Fromager}, \citenamefont {Fuentealba}, \citenamefont {Gagliardi}, \citenamefont {Galli}, \citenamefont {Gao}, \citenamefont {Geerlings}, \citenamefont {Gidopoulos}, \citenamefont {Gill}, \citenamefont {Gori-Giorgi}, \citenamefont {Görling}, \citenamefont {Gould}, \citenamefont {Grimme}, \citenamefont {Gritsenko}, \citenamefont {Jensen}, \citenamefont {Johnson}, \citenamefont {Jones}, \citenamefont {Kaupp}, \citenamefont {Köster}, \citenamefont
  {Kronik}, \citenamefont {Krylov}, \citenamefont {Kvaal}, \citenamefont {Laestadius}, \citenamefont {Levy}, \citenamefont {Lewin}, \citenamefont {Liu}, \citenamefont {Loos}, \citenamefont {Maitra}, \citenamefont {Neese}, \citenamefont {Perdew}, \citenamefont {Pernal}, \citenamefont {Pernot}, \citenamefont {Piecuch}, \citenamefont {Rebolini}, \citenamefont {Reining}, \citenamefont {Romaniello}, \citenamefont {Ruzsinszky}, \citenamefont {Salahub}, \citenamefont {Scheffler}, \citenamefont {Schwerdtfeger}, \citenamefont {Staroverov}, \citenamefont {Sun}, \citenamefont {Tellgren}, \citenamefont {Tozer}, \citenamefont {Trickey}, \citenamefont {Ullrich}, \citenamefont {Vela}, \citenamefont {Vignale}, \citenamefont {Wesolowski}, \citenamefont {Xu},\ and\ \citenamefont {Yang}}]{Teale2003}%
  \BibitemOpen
  \bibfield  {author} {\bibinfo {author} {\bibfnamefont {A.~M.}\ \bibnamefont {Teale}}, \bibinfo {author} {\bibfnamefont {T.}~\bibnamefont {Helgaker}}, \bibinfo {author} {\bibfnamefont {A.}~\bibnamefont {Savin}}, \bibinfo {author} {\bibfnamefont {C.}~\bibnamefont {Adamo}}, \bibinfo {author} {\bibfnamefont {B.}~\bibnamefont {Aradi}}, \bibinfo {author} {\bibfnamefont {A.~V.}\ \bibnamefont {Arbuznikov}}, \bibinfo {author} {\bibfnamefont {P.~W.}\ \bibnamefont {Ayers}}, \bibinfo {author} {\bibfnamefont {E.~J.}\ \bibnamefont {Baerends}}, \bibinfo {author} {\bibfnamefont {V.}~\bibnamefont {Barone}}, \bibinfo {author} {\bibfnamefont {P.}~\bibnamefont {Calaminici}}, \bibinfo {author} {\bibfnamefont {E.}~\bibnamefont {Cancès}}, \bibinfo {author} {\bibfnamefont {E.~A.}\ \bibnamefont {Carter}}, \bibinfo {author} {\bibfnamefont {P.~K.}\ \bibnamefont {Chattaraj}}, \bibinfo {author} {\bibfnamefont {H.}~\bibnamefont {Chermette}}, \bibinfo {author} {\bibfnamefont {I.}~\bibnamefont {Ciofini}}, \bibinfo {author} {\bibfnamefont
  {T.~D.}\ \bibnamefont {Crawford}}, \bibinfo {author} {\bibfnamefont {F.}~\bibnamefont {De~Proft}}, \bibinfo {author} {\bibfnamefont {J.~F.}\ \bibnamefont {Dobson}}, \bibinfo {author} {\bibfnamefont {C.}~\bibnamefont {Draxl}}, \bibinfo {author} {\bibfnamefont {T.}~\bibnamefont {Frauenheim}}, \bibinfo {author} {\bibfnamefont {E.}~\bibnamefont {Fromager}}, \bibinfo {author} {\bibfnamefont {P.}~\bibnamefont {Fuentealba}}, \bibinfo {author} {\bibfnamefont {L.}~\bibnamefont {Gagliardi}}, \bibinfo {author} {\bibfnamefont {G.}~\bibnamefont {Galli}}, \bibinfo {author} {\bibfnamefont {J.}~\bibnamefont {Gao}}, \bibinfo {author} {\bibfnamefont {P.}~\bibnamefont {Geerlings}}, \bibinfo {author} {\bibfnamefont {N.}~\bibnamefont {Gidopoulos}}, \bibinfo {author} {\bibfnamefont {P.~M.~W.}\ \bibnamefont {Gill}}, \bibinfo {author} {\bibfnamefont {P.}~\bibnamefont {Gori-Giorgi}}, \bibinfo {author} {\bibfnamefont {A.}~\bibnamefont {Görling}}, \bibinfo {author} {\bibfnamefont {T.}~\bibnamefont {Gould}}, \bibinfo {author}
  {\bibfnamefont {S.}~\bibnamefont {Grimme}}, \bibinfo {author} {\bibfnamefont {O.}~\bibnamefont {Gritsenko}}, \bibinfo {author} {\bibfnamefont {H.~J.~A.}\ \bibnamefont {Jensen}}, \bibinfo {author} {\bibfnamefont {E.~R.}\ \bibnamefont {Johnson}}, \bibinfo {author} {\bibfnamefont {R.~O.}\ \bibnamefont {Jones}}, \bibinfo {author} {\bibfnamefont {M.}~\bibnamefont {Kaupp}}, \bibinfo {author} {\bibfnamefont {A.~M.}\ \bibnamefont {Köster}}, \bibinfo {author} {\bibfnamefont {L.}~\bibnamefont {Kronik}}, \bibinfo {author} {\bibfnamefont {A.~I.}\ \bibnamefont {Krylov}}, \bibinfo {author} {\bibfnamefont {S.}~\bibnamefont {Kvaal}}, \bibinfo {author} {\bibfnamefont {A.}~\bibnamefont {Laestadius}}, \bibinfo {author} {\bibfnamefont {M.}~\bibnamefont {Levy}}, \bibinfo {author} {\bibfnamefont {M.}~\bibnamefont {Lewin}}, \bibinfo {author} {\bibfnamefont {S.}~\bibnamefont {Liu}}, \bibinfo {author} {\bibfnamefont {P.~F.}\ \bibnamefont {Loos}}, \bibinfo {author} {\bibfnamefont {N.~T.}\ \bibnamefont {Maitra}}, \bibinfo {author}
  {\bibfnamefont {F.}~\bibnamefont {Neese}}, \bibinfo {author} {\bibfnamefont {J.~P.}\ \bibnamefont {Perdew}}, \bibinfo {author} {\bibfnamefont {K.}~\bibnamefont {Pernal}}, \bibinfo {author} {\bibfnamefont {P.}~\bibnamefont {Pernot}}, \bibinfo {author} {\bibfnamefont {P.}~\bibnamefont {Piecuch}}, \bibinfo {author} {\bibfnamefont {E.}~\bibnamefont {Rebolini}}, \bibinfo {author} {\bibfnamefont {L.}~\bibnamefont {Reining}}, \bibinfo {author} {\bibfnamefont {P.}~\bibnamefont {Romaniello}}, \bibinfo {author} {\bibfnamefont {A.}~\bibnamefont {Ruzsinszky}}, \bibinfo {author} {\bibfnamefont {D.~R.}\ \bibnamefont {Salahub}}, \bibinfo {author} {\bibfnamefont {M.}~\bibnamefont {Scheffler}}, \bibinfo {author} {\bibfnamefont {P.}~\bibnamefont {Schwerdtfeger}}, \bibinfo {author} {\bibfnamefont {V.~N.}\ \bibnamefont {Staroverov}}, \bibinfo {author} {\bibfnamefont {J.}~\bibnamefont {Sun}}, \bibinfo {author} {\bibfnamefont {E.}~\bibnamefont {Tellgren}}, \bibinfo {author} {\bibfnamefont {D.~J.}\ \bibnamefont {Tozer}}, \bibinfo
  {author} {\bibfnamefont {S.~B.}\ \bibnamefont {Trickey}}, \bibinfo {author} {\bibfnamefont {C.~A.}\ \bibnamefont {Ullrich}}, \bibinfo {author} {\bibfnamefont {A.}~\bibnamefont {Vela}}, \bibinfo {author} {\bibfnamefont {G.}~\bibnamefont {Vignale}}, \bibinfo {author} {\bibfnamefont {T.~A.}\ \bibnamefont {Wesolowski}}, \bibinfo {author} {\bibfnamefont {X.}~\bibnamefont {Xu}}, \ and\ \bibinfo {author} {\bibfnamefont {W.}~\bibnamefont {Yang}},\ }\bibfield  {title} {\enquote {\bibinfo {title} {Dft exchange: sharing perspectives on the workhorse of quantum chemistry and materials science},}\ }\href {\doibase 10.1039/D2CP02827A} {\bibfield  {journal} {\bibinfo  {journal} {Phys. Chem. Chem. Phys.}\ }\textbf {\bibinfo {volume} {24}},\ \bibinfo {pages} {28700--28781} (\bibinfo {year} {2022})}\BibitemShut {NoStop}%
\bibitem [{\citenamefont {Kohn}\ and\ \citenamefont {Sham}(1965)}]{KS1965}%
  \BibitemOpen
  \bibfield  {author} {\bibinfo {author} {\bibfnamefont {W.}~\bibnamefont {Kohn}}\ and\ \bibinfo {author} {\bibfnamefont {L.~J.}\ \bibnamefont {Sham}},\ }\bibfield  {title} {\enquote {\bibinfo {title} {Self-consistent equations including exchange and correlation effects},}\ }\href {\doibase 10.1103/PhysRev.140.A1133} {\bibfield  {journal} {\bibinfo  {journal} {Phys. Rev.}\ }\textbf {\bibinfo {volume} {140}},\ \bibinfo {pages} {A1133--A1138} (\bibinfo {year} {1965})}\BibitemShut {NoStop}%
\bibitem [{\citenamefont {Perdew}\ and\ \citenamefont {Levy}(1983)}]{PL83}%
  \BibitemOpen
  \bibfield  {author} {\bibinfo {author} {\bibfnamefont {J.~P.}\ \bibnamefont {Perdew}}\ and\ \bibinfo {author} {\bibfnamefont {M.}~\bibnamefont {Levy}},\ }\bibfield  {title} {\enquote {\bibinfo {title} {Physical content of the exact kohn-sham orbital energies: Band gaps and derivative discontinuities},}\ }\href {\doibase 10.1103/PhysRevLett.51.1884} {\bibfield  {journal} {\bibinfo  {journal} {Phys. Rev. Lett.}\ }\textbf {\bibinfo {volume} {51}},\ \bibinfo {pages} {1884--1887} (\bibinfo {year} {1983})}\BibitemShut {NoStop}%
\bibitem [{\citenamefont {Sham}\ and\ \citenamefont {Schl\"uter}(1983)}]{SS83}%
  \BibitemOpen
  \bibfield  {author} {\bibinfo {author} {\bibfnamefont {L.~J.}\ \bibnamefont {Sham}}\ and\ \bibinfo {author} {\bibfnamefont {M.}~\bibnamefont {Schl\"uter}},\ }\bibfield  {title} {\enquote {\bibinfo {title} {Density-functional theory of the energy gap},}\ }\href {\doibase 10.1103/PhysRevLett.51.1888} {\bibfield  {journal} {\bibinfo  {journal} {Phys. Rev. Lett.}\ }\textbf {\bibinfo {volume} {51}},\ \bibinfo {pages} {1888--1891} (\bibinfo {year} {1983})}\BibitemShut {NoStop}%
\bibitem [{\citenamefont {Maitra}(2016)}]{M16}%
  \BibitemOpen
  \bibfield  {author} {\bibinfo {author} {\bibfnamefont {N.~T.}\ \bibnamefont {Maitra}},\ }\bibfield  {title} {\enquote {\bibinfo {title} {Perspective: Fundamental aspects of time-dependent density functional theory},}\ }\href {\doibase 10.1063/1.4953039} {\bibfield  {journal} {\bibinfo  {journal} {J. Chem. Phys.}\ }\textbf {\bibinfo {volume} {144}},\ \bibinfo {pages} {220901} (\bibinfo {year} {2016})}\BibitemShut {NoStop}%
\bibitem [{\citenamefont {Hedin}(1965)}]{Hedin1965}%
  \BibitemOpen
  \bibfield  {author} {\bibinfo {author} {\bibfnamefont {L.}~\bibnamefont {Hedin}},\ }\bibfield  {title} {\enquote {\bibinfo {title} {New method for calculating the one-particle green's function with application to the electron-gas problem},}\ }\href {\doibase 10.1103/PhysRev.139.A796} {\bibfield  {journal} {\bibinfo  {journal} {Phys. Rev.}\ }\textbf {\bibinfo {volume} {139}},\ \bibinfo {pages} {A796--A823} (\bibinfo {year} {1965})}\BibitemShut {NoStop}%
\bibitem [{\citenamefont {Onida}\ \emph {et~al.}(2002)\citenamefont {Onida}, \citenamefont {Reining},\ and\ \citenamefont {Rubio}}]{Onida2002}%
  \BibitemOpen
  \bibfield  {author} {\bibinfo {author} {\bibfnamefont {G.}~\bibnamefont {Onida}}, \bibinfo {author} {\bibfnamefont {L.}~\bibnamefont {Reining}}, \ and\ \bibinfo {author} {\bibfnamefont {A.}~\bibnamefont {Rubio}},\ }\bibfield  {title} {\enquote {\bibinfo {title} {Electronic excitations: density-functional versus many-body green’s-function approaches},}\ }\href {\doibase 10.1103/RevModPhys.74.601} {\bibfield  {journal} {\bibinfo  {journal} {Rev. Mod. Phys.}\ }\textbf {\bibinfo {volume} {74}},\ \bibinfo {pages} {601--659} (\bibinfo {year} {2002})}\BibitemShut {NoStop}%
\bibitem [{\citenamefont {Hybertsen}\ and\ \citenamefont {Louie}(1986)}]{Hybertsen1986}%
  \BibitemOpen
  \bibfield  {author} {\bibinfo {author} {\bibfnamefont {M.~S.}\ \bibnamefont {Hybertsen}}\ and\ \bibinfo {author} {\bibfnamefont {S.~G.}\ \bibnamefont {Louie}},\ }\bibfield  {title} {\enquote {\bibinfo {title} {Electron correlation in semiconductors and insulators: Band gaps and quasiparticle energies},}\ }\href {\doibase 10.1103/PhysRevB.34.5390} {\bibfield  {journal} {\bibinfo  {journal} {Phys. Rev. B}\ }\textbf {\bibinfo {volume} {34}},\ \bibinfo {pages} {5390--5413} (\bibinfo {year} {1986})}\BibitemShut {NoStop}%
\bibitem [{\citenamefont {Rohlfing}\ and\ \citenamefont {Louie}(1998)}]{Rohlfing1998}%
  \BibitemOpen
  \bibfield  {author} {\bibinfo {author} {\bibfnamefont {M.}~\bibnamefont {Rohlfing}}\ and\ \bibinfo {author} {\bibfnamefont {S.~G.}\ \bibnamefont {Louie}},\ }\bibfield  {title} {\enquote {\bibinfo {title} {Electron-hole excitations in semiconductors and insulators},}\ }\href {\doibase 10.1103/PhysRevLett.81.2312} {\bibfield  {journal} {\bibinfo  {journal} {Phys. Rev. Lett.}\ }\textbf {\bibinfo {volume} {81}},\ \bibinfo {pages} {2312--2315} (\bibinfo {year} {1998})}\BibitemShut {NoStop}%
\bibitem [{\citenamefont {Albrecht}\ \emph {et~al.}(1998)\citenamefont {Albrecht}, \citenamefont {Reining}, \citenamefont {Del~Sole},\ and\ \citenamefont {Onida}}]{Onida1998}%
  \BibitemOpen
  \bibfield  {author} {\bibinfo {author} {\bibfnamefont {S.}~\bibnamefont {Albrecht}}, \bibinfo {author} {\bibfnamefont {L.}~\bibnamefont {Reining}}, \bibinfo {author} {\bibfnamefont {R.}~\bibnamefont {Del~Sole}}, \ and\ \bibinfo {author} {\bibfnamefont {G.}~\bibnamefont {Onida}},\ }\bibfield  {title} {\enquote {\bibinfo {title} {Ab initio calculation of excitonic effects in the optical spectra of semiconductors},}\ }\href {\doibase 10.1103/PhysRevLett.80.4510} {\bibfield  {journal} {\bibinfo  {journal} {Phys. Rev. Lett.}\ }\textbf {\bibinfo {volume} {80}},\ \bibinfo {pages} {4510--4513} (\bibinfo {year} {1998})}\BibitemShut {NoStop}%
\bibitem [{\citenamefont {Martin}\ \emph {et~al.}(2016)\citenamefont {Martin}, \citenamefont {Reining},\ and\ \citenamefont {Ceperley}}]{Martin2016}%
  \BibitemOpen
  \bibfield  {author} {\bibinfo {author} {\bibfnamefont {R.~M.}\ \bibnamefont {Martin}}, \bibinfo {author} {\bibfnamefont {L.}~\bibnamefont {Reining}}, \ and\ \bibinfo {author} {\bibfnamefont {D.~M.}\ \bibnamefont {Ceperley}},\ }\enquote {\bibinfo {title} {{Interacting Electrons: Theory and Computational Approaches}},}\ \ (\bibinfo  {publisher} {Cambridge University Press},\ \bibinfo {year} {2016})\BibitemShut {NoStop}%
\bibitem [{\citenamefont {Rinke}\ \emph {et~al.}(2005)\citenamefont {Rinke}, \citenamefont {Qteish}, \citenamefont {Neugebauer}, \citenamefont {Freysoldt},\ and\ \citenamefont {Scheffler}}]{Rinke2005}%
  \BibitemOpen
  \bibfield  {author} {\bibinfo {author} {\bibfnamefont {P.}~\bibnamefont {Rinke}}, \bibinfo {author} {\bibfnamefont {A.}~\bibnamefont {Qteish}}, \bibinfo {author} {\bibfnamefont {J.}~\bibnamefont {Neugebauer}}, \bibinfo {author} {\bibfnamefont {C.}~\bibnamefont {Freysoldt}}, \ and\ \bibinfo {author} {\bibfnamefont {M.}~\bibnamefont {Scheffler}},\ }\bibfield  {title} {\enquote {\bibinfo {title} {Combining gw calculations with exact-exchange density-functional theory: an analysis of valence-band photoemission for compound semiconductors},}\ }\href {\doibase 10.1088/1367-2630/7/1/126} {\bibfield  {journal} {\bibinfo  {journal} {New J. Phys.}\ }\textbf {\bibinfo {volume} {7}},\ \bibinfo {pages} {126--126} (\bibinfo {year} {2005})}\BibitemShut {NoStop}%
\bibitem [{\citenamefont {Fuchs}\ \emph {et~al.}(2007)\citenamefont {Fuchs}, \citenamefont {Furthm\"uller}, \citenamefont {Bechstedt}, \citenamefont {Shishkin},\ and\ \citenamefont {Kresse}}]{Fuchs2007}%
  \BibitemOpen
  \bibfield  {author} {\bibinfo {author} {\bibfnamefont {F.}~\bibnamefont {Fuchs}}, \bibinfo {author} {\bibfnamefont {J.}~\bibnamefont {Furthm\"uller}}, \bibinfo {author} {\bibfnamefont {F.}~\bibnamefont {Bechstedt}}, \bibinfo {author} {\bibfnamefont {M.}~\bibnamefont {Shishkin}}, \ and\ \bibinfo {author} {\bibfnamefont {G.}~\bibnamefont {Kresse}},\ }\bibfield  {title} {\enquote {\bibinfo {title} {Quasiparticle band structure based on a generalized kohn-sham scheme},}\ }\href {\doibase 10.1103/PhysRevB.76.115109} {\bibfield  {journal} {\bibinfo  {journal} {Phys. Rev. B}\ }\textbf {\bibinfo {volume} {76}},\ \bibinfo {pages} {115109} (\bibinfo {year} {2007})}\BibitemShut {NoStop}%
\bibitem [{\citenamefont {Chen}\ and\ \citenamefont {Pasquarello}(2014)}]{Chen2014}%
  \BibitemOpen
  \bibfield  {author} {\bibinfo {author} {\bibfnamefont {W.}~\bibnamefont {Chen}}\ and\ \bibinfo {author} {\bibfnamefont {A.}~\bibnamefont {Pasquarello}},\ }\href {\doibase 10.1103/PhysRevB.90.165133} {\bibfield  {journal} {\bibinfo  {journal} {Phys. Rev. B}\ }\textbf {\bibinfo {volume} {90}},\ \bibinfo {pages} {165133} (\bibinfo {year} {2014})}\BibitemShut {NoStop}%
\bibitem [{\citenamefont {Leppert}\ \emph {et~al.}(2019)\citenamefont {Leppert}, \citenamefont {Rangel},\ and\ \citenamefont {Neaton}}]{Leppert2019}%
  \BibitemOpen
  \bibfield  {author} {\bibinfo {author} {\bibfnamefont {L.}~\bibnamefont {Leppert}}, \bibinfo {author} {\bibfnamefont {T.}~\bibnamefont {Rangel}}, \ and\ \bibinfo {author} {\bibfnamefont {J.~B.}\ \bibnamefont {Neaton}},\ }\bibfield  {title} {\enquote {\bibinfo {title} {Towards predictive band gaps for halide perovskites: Lessons from one-shot and eigenvalue self-consistent gw},}\ }\href {\doibase 10.1103/PhysRevMaterials.3.103803} {\bibfield  {journal} {\bibinfo  {journal} {Phys. Rev. Mater.}\ }\textbf {\bibinfo {volume} {3}},\ \bibinfo {pages} {103803} (\bibinfo {year} {2019})}\BibitemShut {NoStop}%
\bibitem [{\citenamefont {Gant}\ \emph {et~al.}(2022)\citenamefont {Gant}, \citenamefont {Haber}, \citenamefont {Filip}, \citenamefont {Sagredo}, \citenamefont {Wing}, \citenamefont {Ohad}, \citenamefont {Kronik},\ and\ \citenamefont {Neaton}}]{Gant2022}%
  \BibitemOpen
  \bibfield  {author} {\bibinfo {author} {\bibfnamefont {S.~E.}\ \bibnamefont {Gant}}, \bibinfo {author} {\bibfnamefont {J.~B.}\ \bibnamefont {Haber}}, \bibinfo {author} {\bibfnamefont {M.~R.}\ \bibnamefont {Filip}}, \bibinfo {author} {\bibfnamefont {F.}~\bibnamefont {Sagredo}}, \bibinfo {author} {\bibfnamefont {D.}~\bibnamefont {Wing}}, \bibinfo {author} {\bibfnamefont {G.}~\bibnamefont {Ohad}}, \bibinfo {author} {\bibfnamefont {L.}~\bibnamefont {Kronik}}, \ and\ \bibinfo {author} {\bibfnamefont {J.~B.}\ \bibnamefont {Neaton}},\ }\bibfield  {title} {\enquote {\bibinfo {title} {Optimally tuned starting point for single-shot gw calculations of solids},}\ }\href {\doibase 10.1103/PhysRevMaterials.6.053802} {\bibfield  {journal} {\bibinfo  {journal} {Phys. Rev. Mater.}\ }\textbf {\bibinfo {volume} {6}},\ \bibinfo {pages} {053802} (\bibinfo {year} {2022})}\BibitemShut {NoStop}%
\bibitem [{\citenamefont {Golze}\ \emph {et~al.}(2019)\citenamefont {Golze}, \citenamefont {Dvorak},\ and\ \citenamefont {Rinke}}]{Golze2019}%
  \BibitemOpen
  \bibfield  {author} {\bibinfo {author} {\bibfnamefont {D.}~\bibnamefont {Golze}}, \bibinfo {author} {\bibfnamefont {M.}~\bibnamefont {Dvorak}}, \ and\ \bibinfo {author} {\bibfnamefont {P.}~\bibnamefont {Rinke}},\ }\bibfield  {title} {\enquote {\bibinfo {title} {The gw compendium: A practical guide to theoretical photoemission spectroscopy},}\ }\href {\doibase 10.3389/fchem.2019.00377} {\bibfield  {journal} {\bibinfo  {journal} {Front. Chem.}\ }\textbf {\bibinfo {volume} {7}},\ \bibinfo {pages} {377} (\bibinfo {year} {2019})}\BibitemShut {NoStop}%
\bibitem [{\citenamefont {Seidl}\ \emph {et~al.}(1996)\citenamefont {Seidl}, \citenamefont {G\"orling}, \citenamefont {Vogl}, \citenamefont {Majewski},\ and\ \citenamefont {Levy}}]{Seidl1996}%
  \BibitemOpen
  \bibfield  {author} {\bibinfo {author} {\bibfnamefont {A.}~\bibnamefont {Seidl}}, \bibinfo {author} {\bibfnamefont {A.}~\bibnamefont {G\"orling}}, \bibinfo {author} {\bibfnamefont {P.}~\bibnamefont {Vogl}}, \bibinfo {author} {\bibfnamefont {J.~A.}\ \bibnamefont {Majewski}}, \ and\ \bibinfo {author} {\bibfnamefont {M.}~\bibnamefont {Levy}},\ }\bibfield  {title} {\enquote {\bibinfo {title} {Generalized kohn-sham schemes and the band-gap problem},}\ }\href {\doibase 10.1103/PhysRevB.53.3764} {\bibfield  {journal} {\bibinfo  {journal} {Phys. Rev. B}\ }\textbf {\bibinfo {volume} {53}},\ \bibinfo {pages} {3764--3774} (\bibinfo {year} {1996})}\BibitemShut {NoStop}%
\bibitem [{\citenamefont {Cangi}\ \emph {et~al.}(2019)\citenamefont {Cangi}, \citenamefont {Sagredo}, \citenamefont {Decolvenaere},\ and\ \citenamefont {Mattsson}}]{CSDM19}%
  \BibitemOpen
  \bibfield  {author} {\bibinfo {author} {\bibfnamefont {A.}~\bibnamefont {Cangi}}, \bibinfo {author} {\bibfnamefont {F.}~\bibnamefont {Sagredo}}, \bibinfo {author} {\bibfnamefont {E.}~\bibnamefont {Decolvenaere}}, \ and\ \bibinfo {author} {\bibfnamefont {A.~E.}\ \bibnamefont {Mattsson}},\ }\href {\doibase 10.2172/1569522} {\  (\bibinfo {year} {2019}),\ 10.2172/1569522}\BibitemShut {NoStop}%
\bibitem [{\citenamefont {Sun}\ \emph {et~al.}(2015)\citenamefont {Sun}, \citenamefont {Ruzsinszky},\ and\ \citenamefont {Perdew}}]{SCAN15}%
  \BibitemOpen
  \bibfield  {author} {\bibinfo {author} {\bibfnamefont {J.}~\bibnamefont {Sun}}, \bibinfo {author} {\bibfnamefont {A.}~\bibnamefont {Ruzsinszky}}, \ and\ \bibinfo {author} {\bibfnamefont {J.~P.}\ \bibnamefont {Perdew}},\ }\bibfield  {title} {\enquote {\bibinfo {title} {Strongly constrained and appropriately normed semilocal density functional},}\ }\href {\doibase 10.1103/PhysRevLett.115.036402} {\bibfield  {journal} {\bibinfo  {journal} {Phys. Rev. Lett.}\ }\textbf {\bibinfo {volume} {115}},\ \bibinfo {pages} {036402} (\bibinfo {year} {2015})}\BibitemShut {NoStop}%
\bibitem [{\citenamefont {Patra}\ \emph {et~al.}(2019)\citenamefont {Patra}, \citenamefont {Jana}, \citenamefont {Constantin},\ and\ \citenamefont {Samal}}]{PJCS19}%
  \BibitemOpen
  \bibfield  {author} {\bibinfo {author} {\bibfnamefont {B.}~\bibnamefont {Patra}}, \bibinfo {author} {\bibfnamefont {S.}~\bibnamefont {Jana}}, \bibinfo {author} {\bibfnamefont {L.~A.}\ \bibnamefont {Constantin}}, \ and\ \bibinfo {author} {\bibfnamefont {P.}~\bibnamefont {Samal}},\ }\bibfield  {title} {\enquote {\bibinfo {title} {Efficient band gap prediction of semiconductors and insulators from a semilocal exchange-correlation functional},}\ }\href {\doibase 10.1103/PhysRevB.100.155140} {\bibfield  {journal} {\bibinfo  {journal} {Phys. Rev. B}\ }\textbf {\bibinfo {volume} {100}},\ \bibinfo {pages} {155140} (\bibinfo {year} {2019})}\BibitemShut {NoStop}%
\bibitem [{\citenamefont {Traor\'e}\ \emph {et~al.}(2019)\citenamefont {Traor\'e}, \citenamefont {Bouder}, \citenamefont {Lafargue-Dit-Hauret}, \citenamefont {Rocquefelte}, \citenamefont {Katan}, \citenamefont {Tran},\ and\ \citenamefont {Kepenekian}}]{TBLR19}%
  \BibitemOpen
  \bibfield  {author} {\bibinfo {author} {\bibfnamefont {B.}~\bibnamefont {Traor\'e}}, \bibinfo {author} {\bibfnamefont {G.}~\bibnamefont {Bouder}}, \bibinfo {author} {\bibfnamefont {W.}~\bibnamefont {Lafargue-Dit-Hauret}}, \bibinfo {author} {\bibfnamefont {X.}~\bibnamefont {Rocquefelte}}, \bibinfo {author} {\bibfnamefont {C.}~\bibnamefont {Katan}}, \bibinfo {author} {\bibfnamefont {F.}~\bibnamefont {Tran}}, \ and\ \bibinfo {author} {\bibfnamefont {M.}~\bibnamefont {Kepenekian}},\ }\href {\doibase 10.1103/PhysRevB.99.035139} {\bibfield  {journal} {\bibinfo  {journal} {Phys. Rev. B}\ }\textbf {\bibinfo {volume} {99}},\ \bibinfo {pages} {035139} (\bibinfo {year} {2019})}\BibitemShut {NoStop}%
\bibitem [{\citenamefont {Yang}\ \emph {et~al.}(2016)\citenamefont {Yang}, \citenamefont {Peng}, \citenamefont {Sun},\ and\ \citenamefont {Perdew}}]{YPSP16}%
  \BibitemOpen
  \bibfield  {author} {\bibinfo {author} {\bibfnamefont {Z.}~\bibnamefont {Yang}}, \bibinfo {author} {\bibfnamefont {H.}~\bibnamefont {Peng}}, \bibinfo {author} {\bibfnamefont {J.}~\bibnamefont {Sun}}, \ and\ \bibinfo {author} {\bibfnamefont {J.~P.}\ \bibnamefont {Perdew}},\ }\bibfield  {title} {\enquote {\bibinfo {title} {More realistic band gaps from meta-generalized gradient approximations: Only in a generalized kohn-sham scheme},}\ }\href {\doibase 10.1103/PhysRevB.93.205205} {\bibfield  {journal} {\bibinfo  {journal} {Phys. Rev. B}\ }\textbf {\bibinfo {volume} {93}},\ \bibinfo {pages} {205205} (\bibinfo {year} {2016})}\BibitemShut {NoStop}%
\bibitem [{\citenamefont {Lebeda}\ \emph {et~al.}(2024)\citenamefont {Lebeda}, \citenamefont {Aschebrock},\ and\ \citenamefont {K\"ummel}}]{LAK24}%
  \BibitemOpen
  \bibfield  {author} {\bibinfo {author} {\bibfnamefont {T.}~\bibnamefont {Lebeda}}, \bibinfo {author} {\bibfnamefont {T.}~\bibnamefont {Aschebrock}}, \ and\ \bibinfo {author} {\bibfnamefont {S.}~\bibnamefont {K\"ummel}},\ }\bibfield  {title} {\enquote {\bibinfo {title} {Balancing the contributions to the gradient expansion: Accurate binding and band gaps with a nonempirical meta-gga},}\ }\href {\doibase 10.1103/PhysRevLett.133.136402} {\bibfield  {journal} {\bibinfo  {journal} {Phys. Rev. Lett.}\ }\textbf {\bibinfo {volume} {133}},\ \bibinfo {pages} {136402} (\bibinfo {year} {2024})}\BibitemShut {NoStop}%
\bibitem [{\citenamefont {Lebeda}\ \emph {et~al.}(2023)\citenamefont {Lebeda}, \citenamefont {Aschebrock}, \citenamefont {Sun}, \citenamefont {Leppert},\ and\ \citenamefont {K\"ummel}}]{LASLK23}%
  \BibitemOpen
  \bibfield  {author} {\bibinfo {author} {\bibfnamefont {T.}~\bibnamefont {Lebeda}}, \bibinfo {author} {\bibfnamefont {T.}~\bibnamefont {Aschebrock}}, \bibinfo {author} {\bibfnamefont {J.}~\bibnamefont {Sun}}, \bibinfo {author} {\bibfnamefont {L.}~\bibnamefont {Leppert}}, \ and\ \bibinfo {author} {\bibfnamefont {S.}~\bibnamefont {K\"ummel}},\ }\bibfield  {title} {\enquote {\bibinfo {title} {Right band gaps for the right reason at low computational cost with a meta-gga},}\ }\href {\doibase 10.1103/PhysRevMaterials.7.093803} {\bibfield  {journal} {\bibinfo  {journal} {Phys. Rev. Mater.}\ }\textbf {\bibinfo {volume} {7}},\ \bibinfo {pages} {093803} (\bibinfo {year} {2023})}\BibitemShut {NoStop}%
\bibitem [{\citenamefont {Aschebrock}\ and\ \citenamefont {K\"ummel}(2019)}]{AK19}%
  \BibitemOpen
  \bibfield  {author} {\bibinfo {author} {\bibfnamefont {T.}~\bibnamefont {Aschebrock}}\ and\ \bibinfo {author} {\bibfnamefont {S.}~\bibnamefont {K\"ummel}},\ }\href {\doibase 10.1103/PhysRevResearch.1.033082} {\bibfield  {journal} {\bibinfo  {journal} {Phys. Rev. Res.}\ }\textbf {\bibinfo {volume} {1}},\ \bibinfo {pages} {033082} (\bibinfo {year} {2019})}\BibitemShut {NoStop}%
\bibitem [{\citenamefont {Becke}(1993)}]{Becke1993}%
  \BibitemOpen
  \bibfield  {author} {\bibinfo {author} {\bibfnamefont {A.~D.}\ \bibnamefont {Becke}},\ }\bibfield  {title} {\enquote {\bibinfo {title} {A new mixing of hartree–fock and local density‐functional theories},}\ }\href {\doibase 10.1063/1.464304} {\bibfield  {journal} {\bibinfo  {journal} {J. Chem. Phys.}\ }\textbf {\bibinfo {volume} {98}},\ \bibinfo {pages} {1372--1377} (\bibinfo {year} {1993})}\BibitemShut {NoStop}%
\bibitem [{\citenamefont {Perdew}\ \emph {et~al.}(1996{\natexlab{a}})\citenamefont {Perdew}, \citenamefont {Ernzerhof},\ and\ \citenamefont {Burke}}]{PBE0}%
  \BibitemOpen
  \bibfield  {author} {\bibinfo {author} {\bibfnamefont {J.~P.}\ \bibnamefont {Perdew}}, \bibinfo {author} {\bibfnamefont {M.}~\bibnamefont {Ernzerhof}}, \ and\ \bibinfo {author} {\bibfnamefont {K.}~\bibnamefont {Burke}},\ }\bibfield  {title} {\enquote {\bibinfo {title} {Rationale for mixing exact exchange with density functional approximations},}\ }\href {\doibase 10.1063/1.472933} {\bibfield  {journal} {\bibinfo  {journal} {J. Chem. Phys.}\ }\textbf {\bibinfo {volume} {105}},\ \bibinfo {pages} {9982--9985} (\bibinfo {year} {1996}{\natexlab{a}})}\BibitemShut {NoStop}%
\bibitem [{\citenamefont {Perdew}\ \emph {et~al.}(2017)\citenamefont {Perdew}, \citenamefont {Yang}, \citenamefont {Burke}, \citenamefont {Yang}, \citenamefont {Gross}, \citenamefont {Scheffler}, \citenamefont {Scuseria}, \citenamefont {Henderson}, \citenamefont {Zhang}, \citenamefont {Ruzsinszky}, \citenamefont {Peng}, \citenamefont {Sun}, \citenamefont {Trushin},\ and\ \citenamefont {Görling}}]{Perdew2017}%
  \BibitemOpen
  \bibfield  {author} {\bibinfo {author} {\bibfnamefont {J.~P.}\ \bibnamefont {Perdew}}, \bibinfo {author} {\bibfnamefont {W.}~\bibnamefont {Yang}}, \bibinfo {author} {\bibfnamefont {K.}~\bibnamefont {Burke}}, \bibinfo {author} {\bibfnamefont {Z.}~\bibnamefont {Yang}}, \bibinfo {author} {\bibfnamefont {E.~K.~U.}\ \bibnamefont {Gross}}, \bibinfo {author} {\bibfnamefont {M.}~\bibnamefont {Scheffler}}, \bibinfo {author} {\bibfnamefont {G.~E.}\ \bibnamefont {Scuseria}}, \bibinfo {author} {\bibfnamefont {T.~M.}\ \bibnamefont {Henderson}}, \bibinfo {author} {\bibfnamefont {I.~Ying}\ \bibnamefont {Zhang}}, \bibinfo {author} {\bibfnamefont {A.}~\bibnamefont {Ruzsinszky}}, \bibinfo {author} {\bibfnamefont {H.}~\bibnamefont {Peng}}, \bibinfo {author} {\bibfnamefont {J.}~\bibnamefont {Sun}}, \bibinfo {author} {\bibfnamefont {E.}~\bibnamefont {Trushin}}, \ and\ \bibinfo {author} {\bibfnamefont {A.}~\bibnamefont {Görling}},\ }\bibfield  {title} {\enquote {\bibinfo {title} {Understanding band gaps of solids in generalized
  kohn–sham theory},}\ }\href {\doibase 10.1073/pnas.1621352114} {\bibfield  {journal} {\bibinfo  {journal} {PNAS}\ }\textbf {\bibinfo {volume} {114}},\ \bibinfo {pages} {2801--2806} (\bibinfo {year} {2017})}\BibitemShut {NoStop}%
\bibitem [{\citenamefont {Shimazaki}\ and\ \citenamefont {Asai}(2009)}]{Shimazaki2009}%
  \BibitemOpen
  \bibfield  {author} {\bibinfo {author} {\bibfnamefont {T.}~\bibnamefont {Shimazaki}}\ and\ \bibinfo {author} {\bibfnamefont {Y.}~\bibnamefont {Asai}},\ }\bibfield  {title} {\enquote {\bibinfo {title} {First principles band structure calculations based on self-consistent screened hartree–fock exchange potential},}\ }\href {\doibase 10.1063/1.3119259} {\bibfield  {journal} {\bibinfo  {journal} {J. Chem. Phys.}\ }\textbf {\bibinfo {volume} {130}},\ \bibinfo {pages} {164702} (\bibinfo {year} {2009})}\BibitemShut {NoStop}%
\bibitem [{\citenamefont {Skone}\ \emph {et~al.}(2014)\citenamefont {Skone}, \citenamefont {Govoni},\ and\ \citenamefont {Galli}}]{Skone2014}%
  \BibitemOpen
  \bibfield  {author} {\bibinfo {author} {\bibfnamefont {J.~H.}\ \bibnamefont {Skone}}, \bibinfo {author} {\bibfnamefont {M.}~\bibnamefont {Govoni}}, \ and\ \bibinfo {author} {\bibfnamefont {G.}~\bibnamefont {Galli}},\ }\bibfield  {title} {\enquote {\bibinfo {title} {Self-consistent hybrid functional for condensed systems},}\ }\href {\doibase 10.1103/PhysRevB.89.195112} {\bibfield  {journal} {\bibinfo  {journal} {Phys. Rev. B}\ }\textbf {\bibinfo {volume} {89}},\ \bibinfo {pages} {195112} (\bibinfo {year} {2014})}\BibitemShut {NoStop}%
\bibitem [{\citenamefont {Skone}\ \emph {et~al.}(2016)\citenamefont {Skone}, \citenamefont {Govoni},\ and\ \citenamefont {Galli}}]{Skone2016}%
  \BibitemOpen
  \bibfield  {author} {\bibinfo {author} {\bibfnamefont {J.~H.}\ \bibnamefont {Skone}}, \bibinfo {author} {\bibfnamefont {M.}~\bibnamefont {Govoni}}, \ and\ \bibinfo {author} {\bibfnamefont {G.}~\bibnamefont {Galli}},\ }\bibfield  {title} {\enquote {\bibinfo {title} {Nonempirical range-separated hybrid functionals for solids and molecules},}\ }\href {\doibase 10.1103/PhysRevB.93.235106} {\bibfield  {journal} {\bibinfo  {journal} {Phys. Rev. B}\ }\textbf {\bibinfo {volume} {93}},\ \bibinfo {pages} {235106} (\bibinfo {year} {2016})}\BibitemShut {NoStop}%
\bibitem [{\citenamefont {Sun}\ \emph {et~al.}(2020)\citenamefont {Sun}, \citenamefont {Yang},\ and\ \citenamefont {Ullrich}}]{Sun2020}%
  \BibitemOpen
  \bibfield  {author} {\bibinfo {author} {\bibfnamefont {J.}~\bibnamefont {Sun}}, \bibinfo {author} {\bibfnamefont {J.}~\bibnamefont {Yang}}, \ and\ \bibinfo {author} {\bibfnamefont {C.~A.}\ \bibnamefont {Ullrich}},\ }\bibfield  {title} {\enquote {\bibinfo {title} {Low-cost alternatives to the bethe-salpeter equation: Towards simple hybrid functionals for excitonic effects in solids},}\ }\href {\doibase 10.1103/PhysRevResearch.2.013091} {\bibfield  {journal} {\bibinfo  {journal} {Phys. Rev. Res.}\ }\textbf {\bibinfo {volume} {2}},\ \bibinfo {pages} {013091} (\bibinfo {year} {2020})}\BibitemShut {NoStop}%
\bibitem [{\citenamefont {Jana}\ \emph {et~al.}(2020)\citenamefont {Jana}, \citenamefont {Patra}, \citenamefont {Constantin},\ and\ \citenamefont {Samal}}]{JPCS20}%
  \BibitemOpen
  \bibfield  {author} {\bibinfo {author} {\bibfnamefont {S.}~\bibnamefont {Jana}}, \bibinfo {author} {\bibfnamefont {A.}~\bibnamefont {Patra}}, \bibinfo {author} {\bibfnamefont {L.~A.}\ \bibnamefont {Constantin}}, \ and\ \bibinfo {author} {\bibfnamefont {P.}~\bibnamefont {Samal}},\ }\bibfield  {title} {\enquote {\bibinfo {title} {Screened range-separated hybrid by balancing the compact and slowly varying density regimes: Satisfaction of local density linear response},}\ }\href {\doibase 10.1063/1.5131530} {\bibfield  {journal} {\bibinfo  {journal} {J. of Chem. Phy.}\ }\textbf {\bibinfo {volume} {152}},\ \bibinfo {pages} {044111} (\bibinfo {year} {2020})}\BibitemShut {NoStop}%
\bibitem [{\citenamefont {Miceli}\ \emph {et~al.}(2018{\natexlab{a}})\citenamefont {Miceli}, \citenamefont {Chen}, \citenamefont {Reshetnyak},\ and\ \citenamefont {Pasquarello}}]{Miceli2018}%
  \BibitemOpen
  \bibfield  {author} {\bibinfo {author} {\bibfnamefont {G.}~\bibnamefont {Miceli}}, \bibinfo {author} {\bibfnamefont {W.}~\bibnamefont {Chen}}, \bibinfo {author} {\bibfnamefont {I.}~\bibnamefont {Reshetnyak}}, \ and\ \bibinfo {author} {\bibfnamefont {A.}~\bibnamefont {Pasquarello}},\ }\bibfield  {title} {\enquote {\bibinfo {title} {Band-edge positions in gw: Effects of starting point and self-consistency},}\ }\href {\doibase 10.1103/PhysRevB.97.121112} {\bibfield  {journal} {\bibinfo  {journal} {Phys. Rev. B}\ }\textbf {\bibinfo {volume} {97}},\ \bibinfo {pages} {121112} (\bibinfo {year} {2018}{\natexlab{a}})}\BibitemShut {NoStop}%
\bibitem [{\citenamefont {Yang}\ \emph {et~al.}(2022)\citenamefont {Yang}, \citenamefont {Falletta},\ and\ \citenamefont {Pasquarello}}]{Yang2022}%
  \BibitemOpen
  \bibfield  {author} {\bibinfo {author} {\bibfnamefont {J.}~\bibnamefont {Yang}}, \bibinfo {author} {\bibfnamefont {S.}~\bibnamefont {Falletta}}, \ and\ \bibinfo {author} {\bibfnamefont {A.}~\bibnamefont {Pasquarello}},\ }\bibfield  {title} {\enquote {\bibinfo {title} {One-shot approach for enforcing piecewise linearity on hybrid functionals: Application to band gap predictions},}\ }\href@noop {} {\bibfield  {journal} {\bibinfo  {journal} {J. Phys. Chem. Lett.}\ }\textbf {\bibinfo {volume} {13}},\ \bibinfo {pages} {3066--3071} (\bibinfo {year} {2022})}\BibitemShut {NoStop}%
\bibitem [{\citenamefont {Jain}\ \emph {et~al.}(2011)\citenamefont {Jain}, \citenamefont {Chelikowsky},\ and\ \citenamefont {Louie}}]{JCL11}%
  \BibitemOpen
  \bibfield  {author} {\bibinfo {author} {\bibfnamefont {M.}~\bibnamefont {Jain}}, \bibinfo {author} {\bibfnamefont {J.~R.}\ \bibnamefont {Chelikowsky}}, \ and\ \bibinfo {author} {\bibfnamefont {S.~G.}\ \bibnamefont {Louie}},\ }\bibfield  {title} {\enquote {\bibinfo {title} {Reliability of hybrid functionals in predicting band gaps},}\ }\href {\doibase 10.1103/PhysRevLett.107.216806} {\bibfield  {journal} {\bibinfo  {journal} {Phys. Rev. Lett.}\ }\textbf {\bibinfo {volume} {107}},\ \bibinfo {pages} {216806} (\bibinfo {year} {2011})}\BibitemShut {NoStop}%
\bibitem [{\citenamefont {Heyd}\ \emph {et~al.}(2003)\citenamefont {Heyd}, \citenamefont {Scuseria},\ and\ \citenamefont {Ernzerhof}}]{HSE}%
  \BibitemOpen
  \bibfield  {author} {\bibinfo {author} {\bibfnamefont {J.}~\bibnamefont {Heyd}}, \bibinfo {author} {\bibfnamefont {G.~E.}\ \bibnamefont {Scuseria}}, \ and\ \bibinfo {author} {\bibfnamefont {M.}~\bibnamefont {Ernzerhof}},\ }\bibfield  {title} {\enquote {\bibinfo {title} {Hybrid functionals based on a screened coulomb potential},}\ }\href {\doibase 10.1063/1.1564060} {\bibfield  {journal} {\bibinfo  {journal} {J. Chem. Phys.}\ }\textbf {\bibinfo {volume} {118}},\ \bibinfo {pages} {8207--8215} (\bibinfo {year} {2003})}\BibitemShut {NoStop}%
\bibitem [{\citenamefont {Heyd}\ \emph {et~al.}(2006)\citenamefont {Heyd}, \citenamefont {Scuseria},\ and\ \citenamefont {Ernzerhof}}]{HSEerratum}%
  \BibitemOpen
  \bibfield  {author} {\bibinfo {author} {\bibfnamefont {J.}~\bibnamefont {Heyd}}, \bibinfo {author} {\bibfnamefont {G.~E.}\ \bibnamefont {Scuseria}}, \ and\ \bibinfo {author} {\bibfnamefont {M.}~\bibnamefont {Ernzerhof}},\ }\bibfield  {title} {\enquote {\bibinfo {title} {``erratum: “hybrid functionals based on a screened coulomb potentia''},}\ }\href {\doibase 10.1063/1.2204597} {\bibfield  {journal} {\bibinfo  {journal} {J. Chem. Phys.}\ }\textbf {\bibinfo {volume} {124}},\ \bibinfo {pages} {219906} (\bibinfo {year} {2006})}\BibitemShut {NoStop}%
\bibitem [{\citenamefont {Stephens}\ \emph {et~al.}(1994)\citenamefont {Stephens}, \citenamefont {Devlin}, \citenamefont {Chabalowski},\ and\ \citenamefont {Frisch}}]{Stephens1994}%
  \BibitemOpen
  \bibfield  {author} {\bibinfo {author} {\bibfnamefont {P.~J.}\ \bibnamefont {Stephens}}, \bibinfo {author} {\bibfnamefont {F.~J.}\ \bibnamefont {Devlin}}, \bibinfo {author} {\bibfnamefont {C.~F.}\ \bibnamefont {Chabalowski}}, \ and\ \bibinfo {author} {\bibfnamefont {M.~J.}\ \bibnamefont {Frisch}},\ }\bibfield  {title} {\enquote {\bibinfo {title} {Ab initio calculation of vibrational absorption and circular dichroism spectra using density functional force fields},}\ }\href {\doibase 10.1021/j100096a001} {\bibfield  {journal} {\bibinfo  {journal} {J. Phys. Chem.}\ }\textbf {\bibinfo {volume} {98}},\ \bibinfo {pages} {11623--11627} (\bibinfo {year} {1994})}\BibitemShut {NoStop}%
\bibitem [{\citenamefont {Mahler}\ \emph {et~al.}(2022)\citenamefont {Mahler}, \citenamefont {Williams}, \citenamefont {Su},\ and\ \citenamefont {Yang}}]{MahlerYang2022}%
  \BibitemOpen
  \bibfield  {author} {\bibinfo {author} {\bibfnamefont {A.}~\bibnamefont {Mahler}}, \bibinfo {author} {\bibfnamefont {J.}~\bibnamefont {Williams}}, \bibinfo {author} {\bibfnamefont {N.~Q.}\ \bibnamefont {Su}}, \ and\ \bibinfo {author} {\bibfnamefont {W.}~\bibnamefont {Yang}},\ }\bibfield  {title} {\enquote {\bibinfo {title} {Localized orbital scaling correction for periodic systems},}\ }\href {\doibase 10.1103/PhysRevB.106.035147} {\bibfield  {journal} {\bibinfo  {journal} {Phys. Rev. B}\ }\textbf {\bibinfo {volume} {106}},\ \bibinfo {pages} {035147} (\bibinfo {year} {2022})}\BibitemShut {NoStop}%
\bibitem [{\citenamefont {Nguyen}\ \emph {et~al.}(2018)\citenamefont {Nguyen}, \citenamefont {Colonna}, \citenamefont {Ferretti},\ and\ \citenamefont {Marzari}}]{NCFM18}%
  \BibitemOpen
  \bibfield  {author} {\bibinfo {author} {\bibfnamefont {N.~L.}\ \bibnamefont {Nguyen}}, \bibinfo {author} {\bibfnamefont {N.}~\bibnamefont {Colonna}}, \bibinfo {author} {\bibfnamefont {A.}~\bibnamefont {Ferretti}}, \ and\ \bibinfo {author} {\bibfnamefont {N.}~\bibnamefont {Marzari}},\ }\bibfield  {title} {\enquote {\bibinfo {title} {Koopmans-compliant spectral functionals for extended systems},}\ }\href {\doibase 10.1103/PhysRevX.8.021051} {\bibfield  {journal} {\bibinfo  {journal} {Phys. Rev. X}\ }\textbf {\bibinfo {volume} {8}},\ \bibinfo {pages} {021051} (\bibinfo {year} {2018})}\BibitemShut {NoStop}%
\bibitem [{\citenamefont {Dabo}\ \emph {et~al.}(2014)\citenamefont {Dabo}, \citenamefont {Ferretti},\ and\ \citenamefont {Marzari}}]{DFM14}%
  \BibitemOpen
  \bibfield  {author} {\bibinfo {author} {\bibfnamefont {I.}~\bibnamefont {Dabo}}, \bibinfo {author} {\bibfnamefont {A.}~\bibnamefont {Ferretti}}, \ and\ \bibinfo {author} {\bibfnamefont {N.}~\bibnamefont {Marzari}},\ }\enquote {\bibinfo {title} {Piecewise linearity and spectroscopic properties from koopmans-compliant functionals},}\ in\ \href {\doibase 10.1007/128_2013_504} {\emph {\bibinfo {booktitle} {First Principles Approaches to Spectroscopic Properties of Complex Materials}}}\ (\bibinfo  {publisher} {Springer Berlin Heidelberg},\ \bibinfo {address} {Berlin, Heidelberg},\ \bibinfo {year} {2014})\ pp.\ \bibinfo {pages} {193--233}\BibitemShut {NoStop}%
\bibitem [{\citenamefont {Ma}\ and\ \citenamefont {Wang}(2016)}]{Ma2016}%
  \BibitemOpen
  \bibfield  {author} {\bibinfo {author} {\bibfnamefont {J.}~\bibnamefont {Ma}}\ and\ \bibinfo {author} {\bibfnamefont {L.}~\bibnamefont {Wang}},\ }\bibfield  {title} {\enquote {\bibinfo {title} {Using wannier functions to improve solid band gap predictions in density functional theory},}\ }\href {\doibase 10.1038/srep24924} {\bibfield  {journal} {\bibinfo  {journal} {Sci. Rep.}\ }\textbf {\bibinfo {volume} {6}},\ \bibinfo {pages} {24924} (\bibinfo {year} {2016})}\BibitemShut {NoStop}%
\bibitem [{\citenamefont {Wing}\ \emph {et~al.}(2021)\citenamefont {Wing}, \citenamefont {Ohad}, \citenamefont {Haber}, \citenamefont {Filip}, \citenamefont {Gant}, \citenamefont {Neaton},\ and\ \citenamefont {Kronik}}]{WOHF21}%
  \BibitemOpen
  \bibfield  {author} {\bibinfo {author} {\bibfnamefont {D.}~\bibnamefont {Wing}}, \bibinfo {author} {\bibfnamefont {G.}~\bibnamefont {Ohad}}, \bibinfo {author} {\bibfnamefont {J.~B.}\ \bibnamefont {Haber}}, \bibinfo {author} {\bibfnamefont {M.~R.}\ \bibnamefont {Filip}}, \bibinfo {author} {\bibfnamefont {S.~E.}\ \bibnamefont {Gant}}, \bibinfo {author} {\bibfnamefont {J.~B.}\ \bibnamefont {Neaton}}, \ and\ \bibinfo {author} {\bibfnamefont {L.}~\bibnamefont {Kronik}},\ }\bibfield  {title} {\enquote {\bibinfo {title} {Band gaps of crystalline solids from wannier-localization–based optimal tuning of a screened range-separated hybrid functional},}\ }\href {\doibase 10.1073/pnas.2104556118} {\bibfield  {journal} {\bibinfo  {journal} {PNAS}\ }\textbf {\bibinfo {volume} {118}},\ \bibinfo {pages} {e2104556118} (\bibinfo {year} {2021})}\BibitemShut {NoStop}%
\bibitem [{\citenamefont {Ohad}\ \emph {et~al.}(2022)\citenamefont {Ohad}, \citenamefont {Wing}, \citenamefont {Gant}, \citenamefont {Cohen}, \citenamefont {Haber}, \citenamefont {Sagredo}, \citenamefont {Filip}, \citenamefont {Neaton},\ and\ \citenamefont {Kronik}}]{Ohad2022}%
  \BibitemOpen
  \bibfield  {author} {\bibinfo {author} {\bibfnamefont {G.}~\bibnamefont {Ohad}}, \bibinfo {author} {\bibfnamefont {D.}~\bibnamefont {Wing}}, \bibinfo {author} {\bibfnamefont {S.~E.}\ \bibnamefont {Gant}}, \bibinfo {author} {\bibfnamefont {A.~V.}\ \bibnamefont {Cohen}}, \bibinfo {author} {\bibfnamefont {J.~B.}\ \bibnamefont {Haber}}, \bibinfo {author} {\bibfnamefont {F.}~\bibnamefont {Sagredo}}, \bibinfo {author} {\bibfnamefont {M.~R.}\ \bibnamefont {Filip}}, \bibinfo {author} {\bibfnamefont {J.~B.}\ \bibnamefont {Neaton}}, \ and\ \bibinfo {author} {\bibfnamefont {L.}~\bibnamefont {Kronik}},\ }\bibfield  {title} {\enquote {\bibinfo {title} {Band gaps of halide perovskites from a wannier-localized optimally tuned screened range-separated hybrid functional},}\ }\href {\doibase 10.1103/PhysRevMaterials.6.104606} {\bibfield  {journal} {\bibinfo  {journal} {Phys. Rev. Materials}\ }\textbf {\bibinfo {volume} {6}},\ \bibinfo {pages} {104606} (\bibinfo {year} {2022})}\BibitemShut {NoStop}%
\bibitem [{\citenamefont {Ohad}\ \emph {et~al.}(2023)\citenamefont {Ohad}, \citenamefont {Gant}, \citenamefont {Wing}, \citenamefont {Haber}, \citenamefont {Camarasa-G\'omez}, \citenamefont {Sagredo}, \citenamefont {Filip}, \citenamefont {Neaton},\ and\ \citenamefont {Kronik}}]{Ohad2023}%
  \BibitemOpen
  \bibfield  {author} {\bibinfo {author} {\bibfnamefont {G.}~\bibnamefont {Ohad}}, \bibinfo {author} {\bibfnamefont {S.~E.}\ \bibnamefont {Gant}}, \bibinfo {author} {\bibfnamefont {D.}~\bibnamefont {Wing}}, \bibinfo {author} {\bibfnamefont {J.~B.}\ \bibnamefont {Haber}}, \bibinfo {author} {\bibfnamefont {M.}~\bibnamefont {Camarasa-G\'omez}}, \bibinfo {author} {\bibfnamefont {F.}~\bibnamefont {Sagredo}}, \bibinfo {author} {\bibfnamefont {M.~R.}\ \bibnamefont {Filip}}, \bibinfo {author} {\bibfnamefont {J.~B.}\ \bibnamefont {Neaton}}, \ and\ \bibinfo {author} {\bibfnamefont {L.}~\bibnamefont {Kronik}},\ }\bibfield  {title} {\enquote {\bibinfo {title} {Optical absorption spectra of metal oxides from time-dependent density functional theory and many-body perturbation theory based on optimally-tuned hybrid functionals},}\ }\href {\doibase 10.1103/PhysRevMaterials.7.123803} {\bibfield  {journal} {\bibinfo  {journal} {Phys. Rev. Mater.}\ }\textbf {\bibinfo {volume} {7}},\ \bibinfo {pages} {123803} (\bibinfo {year}
  {2023})}\BibitemShut {NoStop}%
\bibitem [{\citenamefont {Camarasa-G\'omez}\ \emph {et~al.}(2024)\citenamefont {Camarasa-G\'omez}, \citenamefont {Gant}, \citenamefont {Ohad}, \citenamefont {Neaton}, \citenamefont {Ramasubramaniam},\ and\ \citenamefont {Kronik}}]{Camarasa-Gomez2024}%
  \BibitemOpen
  \bibfield  {author} {\bibinfo {author} {\bibfnamefont {M.}~\bibnamefont {Camarasa-G\'omez}}, \bibinfo {author} {\bibfnamefont {S.~E.}\ \bibnamefont {Gant}}, \bibinfo {author} {\bibfnamefont {G.}~\bibnamefont {Ohad}}, \bibinfo {author} {\bibfnamefont {J.~B.}\ \bibnamefont {Neaton}}, \bibinfo {author} {\bibfnamefont {A.}~\bibnamefont {Ramasubramaniam}}, \ and\ \bibinfo {author} {\bibfnamefont {L.}~\bibnamefont {Kronik}},\ }\bibfield  {title} {\enquote {\bibinfo {title} {Excitations in layered materials from a non-empirical wannier-localized optimally-tuned screened range-separated hybrid functional},}\ }\href {\doibase 10.1038/s41524-024-01478-1} {\bibfield  {journal} {\bibinfo  {journal} {NPJ Comput Mater}\ } (\bibinfo {year} {2024}),\ 10.1038/s41524-024-01478-1}\BibitemShut {NoStop}%
\bibitem [{\citenamefont {Sagredo}\ \emph {et~al.}(2024)\citenamefont {Sagredo}, \citenamefont {Ohad}, \citenamefont {Gant}, \citenamefont {Haber}, \citenamefont {Filip}, \citenamefont {Kronik},\ and\ \citenamefont {Neaton}}]{Sagredo2024}%
  \BibitemOpen
  \bibfield  {author} {\bibinfo {author} {\bibfnamefont {F.}~\bibnamefont {Sagredo}}, \bibinfo {author} {\bibfnamefont {G.}~\bibnamefont {Ohad}}, \bibinfo {author} {\bibfnamefont {S.~E.}\ \bibnamefont {Gant}}, \bibinfo {author} {\bibfnamefont {J.~B.}\ \bibnamefont {Haber}}, \bibinfo {author} {\bibfnamefont {M.~R.}\ \bibnamefont {Filip}}, \bibinfo {author} {\bibfnamefont {L.}~\bibnamefont {Kronik}}, \ and\ \bibinfo {author} {\bibfnamefont {J.~B.}\ \bibnamefont {Neaton}},\ }\bibfield  {title} {\enquote {\bibinfo {title} {Electronic structure and optical properties of halide double perovskites from a wannier-localized optimally-tuned screened range-separated hybrid functional},}\ }\href {\doibase 10.1103/PhysRevMaterials.8.105401} {\bibfield  {journal} {\bibinfo  {journal} {Phys. Rev. Mater.}\ }\textbf {\bibinfo {volume} {8}},\ \bibinfo {pages} {105401} (\bibinfo {year} {2024})}\BibitemShut {NoStop}%
\bibitem [{\citenamefont {Refaely-Abramson}\ \emph {et~al.}(2013)\citenamefont {Refaely-Abramson}, \citenamefont {Sharifzadeh}, \citenamefont {Jain}, \citenamefont {Baer}, \citenamefont {Neaton},\ and\ \citenamefont {Kronik}}]{Refaely2013}%
  \BibitemOpen
  \bibfield  {author} {\bibinfo {author} {\bibfnamefont {S.}~\bibnamefont {Refaely-Abramson}}, \bibinfo {author} {\bibfnamefont {S.}~\bibnamefont {Sharifzadeh}}, \bibinfo {author} {\bibfnamefont {M.}~\bibnamefont {Jain}}, \bibinfo {author} {\bibfnamefont {R.}~\bibnamefont {Baer}}, \bibinfo {author} {\bibfnamefont {J.~B.}\ \bibnamefont {Neaton}}, \ and\ \bibinfo {author} {\bibfnamefont {L.}~\bibnamefont {Kronik}},\ }\bibfield  {title} {\enquote {\bibinfo {title} {Gap renormalization of molecular crystals from density-functional theory},}\ }\href {\doibase 10.1103/PhysRevB.88.081204} {\bibfield  {journal} {\bibinfo  {journal} {Phys. Rev. B}\ }\textbf {\bibinfo {volume} {88}},\ \bibinfo {pages} {081204} (\bibinfo {year} {2013})}\BibitemShut {NoStop}%
\bibitem [{\citenamefont {Leininger}\ \emph {et~al.}(1997)\citenamefont {Leininger}, \citenamefont {Stoll}, \citenamefont {Werner},\ and\ \citenamefont {Savin}}]{Leininger1997}%
  \BibitemOpen
  \bibfield  {author} {\bibinfo {author} {\bibfnamefont {T.}~\bibnamefont {Leininger}}, \bibinfo {author} {\bibfnamefont {H.}~\bibnamefont {Stoll}}, \bibinfo {author} {\bibfnamefont {H.-J.}\ \bibnamefont {Werner}}, \ and\ \bibinfo {author} {\bibfnamefont {A.}~\bibnamefont {Savin}},\ }\bibfield  {title} {\enquote {\bibinfo {title} {Combining long-range configuration interaction with short-range density functionals},}\ }\href {\doibase https://doi.org/10.1016/S0009-2614(97)00758-6} {\bibfield  {journal} {\bibinfo  {journal} {Chem. Phys. Lett.}\ }\textbf {\bibinfo {volume} {275}},\ \bibinfo {pages} {151--160} (\bibinfo {year} {1997})}\BibitemShut {NoStop}%
\bibitem [{\citenamefont {Chen}\ \emph {et~al.}(2022)\citenamefont {Chen}, \citenamefont {Griffin}, \citenamefont {Rignanese},\ and\ \citenamefont {Hautier}}]{CGRH22}%
  \BibitemOpen
  \bibfield  {author} {\bibinfo {author} {\bibfnamefont {W.}~\bibnamefont {Chen}}, \bibinfo {author} {\bibfnamefont {S.~M.}\ \bibnamefont {Griffin}}, \bibinfo {author} {\bibfnamefont {G.}~\bibnamefont {Rignanese}}, \ and\ \bibinfo {author} {\bibfnamefont {G.}~\bibnamefont {Hautier}},\ }\bibfield  {title} {\enquote {\bibinfo {title} {Nonunique fraction of fock exchange for defects in two-dimensional materials},}\ }\href {\doibase 10.1103/PhysRevB.106.L161107} {\bibfield  {journal} {\bibinfo  {journal} {Phys. Rev. B}\ }\textbf {\bibinfo {volume} {106}},\ \bibinfo {pages} {L161107} (\bibinfo {year} {2022})}\BibitemShut {NoStop}%
\bibitem [{\citenamefont {Rohlfing}\ and\ \citenamefont {Louie}(1999)}]{RL99}%
  \BibitemOpen
  \bibfield  {author} {\bibinfo {author} {\bibfnamefont {M.}~\bibnamefont {Rohlfing}}\ and\ \bibinfo {author} {\bibfnamefont {S.~G.}\ \bibnamefont {Louie}},\ }\bibfield  {title} {\enquote {\bibinfo {title} {Excitons and optical spectrum of the $\mathrm{Si}(111)\ensuremath{-}(2\ifmmode\times\else\texttimes\fi{}1)$ surface},}\ }\href {\doibase 10.1103/PhysRevLett.83.856} {\bibfield  {journal} {\bibinfo  {journal} {Phys. Rev. Lett.}\ }\textbf {\bibinfo {volume} {83}},\ \bibinfo {pages} {856--859} (\bibinfo {year} {1999})}\BibitemShut {NoStop}%
\bibitem [{\citenamefont {Ciccacci}\ \emph {et~al.}(1986)\citenamefont {Ciccacci}, \citenamefont {Selci}, \citenamefont {Chiarotti},\ and\ \citenamefont {Chiaradia}}]{CSCC86}%
  \BibitemOpen
  \bibfield  {author} {\bibinfo {author} {\bibfnamefont {F.}~\bibnamefont {Ciccacci}}, \bibinfo {author} {\bibfnamefont {S.}~\bibnamefont {Selci}}, \bibinfo {author} {\bibfnamefont {G.}~\bibnamefont {Chiarotti}}, \ and\ \bibinfo {author} {\bibfnamefont {P.}~\bibnamefont {Chiaradia}},\ }\bibfield  {title} {\enquote {\bibinfo {title} {Electron-phonon interaction in optical absorption at the \ce{Si}(111) 2 \ifmmode\times\else\texttimes\fi{} 1 surface},}\ }\href {\doibase 10.1103/PhysRevLett.56.2411} {\bibfield  {journal} {\bibinfo  {journal} {Phys. Rev. Lett.}\ }\textbf {\bibinfo {volume} {56}},\ \bibinfo {pages} {2411--2414} (\bibinfo {year} {1986})}\BibitemShut {NoStop}%
\bibitem [{\citenamefont {Aulbur}\ \emph {et~al.}(2000)\citenamefont {Aulbur}, \citenamefont {J{\"o}nsson},\ and\ \citenamefont {Wilkins}}]{Aulbur2000}%
  \BibitemOpen
  \bibfield  {author} {\bibinfo {author} {\bibfnamefont {W.~G.}\ \bibnamefont {Aulbur}}, \bibinfo {author} {\bibfnamefont {L.}~\bibnamefont {J{\"o}nsson}}, \ and\ \bibinfo {author} {\bibfnamefont {J.~W.}\ \bibnamefont {Wilkins}},\ }\enquote {\bibinfo {title} {Quasiparticle calculations in solids},}\ in\ \href {https://www.sciencedirect.com/science/article/pii/S0081194708602489} {\emph {\bibinfo {booktitle} {Solid State Physics}}},\ Vol.~\bibinfo {volume} {54}\ (\bibinfo  {publisher} {Academic Press},\ \bibinfo {year} {2000})\ pp.\ \bibinfo {pages} {1--218}\BibitemShut {NoStop}%
\bibitem [{\citenamefont {Zhu}\ and\ \citenamefont {Louie}(1991)}]{zhu1991}%
  \BibitemOpen
  \bibfield  {author} {\bibinfo {author} {\bibfnamefont {X.}~\bibnamefont {Zhu}}\ and\ \bibinfo {author} {\bibfnamefont {S.~G.}\ \bibnamefont {Louie}},\ }\bibfield  {title} {\enquote {\bibinfo {title} {Quasiparticle surface band structure and photoelectric threshold of \ce{Ge}(111)-2\ifmmode\times\else\texttimes\fi{}1},}\ }\href {\doibase 10.1103/PhysRevB.43.12146} {\bibfield  {journal} {\bibinfo  {journal} {Phys. Rev. B}\ }\textbf {\bibinfo {volume} {43}},\ \bibinfo {pages} {12146} (\bibinfo {year} {1991})}\BibitemShut {NoStop}%
\bibitem [{\citenamefont {Madelung}(2004)}]{M04}%
  \BibitemOpen
  \bibfield  {author} {\bibinfo {author} {\bibfnamefont {O.}~\bibnamefont {Madelung}},\ }\enquote {\bibinfo {title} {{Semiconductors:Data handbook}},}\ \ (\bibinfo  {publisher} {Springer Berlin, Heidelberg},\ \bibinfo {year} {2004})\BibitemShut {NoStop}%
\bibitem [{\citenamefont {Nicholls}\ and\ \citenamefont {Reihl}(1989)}]{NR89}%
  \BibitemOpen
  \bibfield  {author} {\bibinfo {author} {\bibfnamefont {J.M.}\ \bibnamefont {Nicholls}}\ and\ \bibinfo {author} {\bibfnamefont {B.}~\bibnamefont {Reihl}},\ }\bibfield  {title} {\enquote {\bibinfo {title} {Antibonding surface state band of the \ce{Ge}(111) 2 × 1 surface},}\ }\href {\doibase https://doi.org/10.1016/0039-6028(89)90630-4} {\bibfield  {journal} {\bibinfo  {journal} {Surface Science}\ }\textbf {\bibinfo {volume} {218}},\ \bibinfo {pages} {237--245} (\bibinfo {year} {1989})}\BibitemShut {NoStop}%
\bibitem [{\citenamefont {Kronik}\ \emph {et~al.}(2012)\citenamefont {Kronik}, \citenamefont {Stein}, \citenamefont {Refaely-Abramson},\ and\ \citenamefont {Baer}}]{Kronik2012}%
  \BibitemOpen
  \bibfield  {author} {\bibinfo {author} {\bibfnamefont {L.}~\bibnamefont {Kronik}}, \bibinfo {author} {\bibfnamefont {T.}~\bibnamefont {Stein}}, \bibinfo {author} {\bibfnamefont {S.}~\bibnamefont {Refaely-Abramson}}, \ and\ \bibinfo {author} {\bibfnamefont {R.}~\bibnamefont {Baer}},\ }\bibfield  {title} {\enquote {\bibinfo {title} {Excitation gaps of finite-sized systems from optimally tuned range-separated hybrid functionals},}\ }\href {\doibase 10.1021/ct2009363} {\bibfield  {journal} {\bibinfo  {journal} {J. Chem. Theory Comput.}\ }\textbf {\bibinfo {volume} {8}},\ \bibinfo {pages} {1515--1531} (\bibinfo {year} {2012})}\BibitemShut {NoStop}%
\bibitem [{\citenamefont {Tal}\ \emph {et~al.}(2020)\citenamefont {Tal}, \citenamefont {Liu}, \citenamefont {Kresse},\ and\ \citenamefont {Pasquarello}}]{Tal2020}%
  \BibitemOpen
  \bibfield  {author} {\bibinfo {author} {\bibfnamefont {A.}~\bibnamefont {Tal}}, \bibinfo {author} {\bibfnamefont {P.}~\bibnamefont {Liu}}, \bibinfo {author} {\bibfnamefont {G.}~\bibnamefont {Kresse}}, \ and\ \bibinfo {author} {\bibfnamefont {A.}~\bibnamefont {Pasquarello}},\ }\bibfield  {title} {\enquote {\bibinfo {title} {Accurate optical spectra through time-dependent density functional theory based on screening-dependent hybrid functionals},}\ }\href {\doibase 10.1103/PhysRevResearch.2.032019} {\bibfield  {journal} {\bibinfo  {journal} {Phys. Rev. Res.}\ }\textbf {\bibinfo {volume} {2}},\ \bibinfo {pages} {032019} (\bibinfo {year} {2020})}\BibitemShut {NoStop}%
\bibitem [{\citenamefont {Stein}\ \emph {et~al.}(2010)\citenamefont {Stein}, \citenamefont {Eisenberg}, \citenamefont {Kronik},\ and\ \citenamefont {Baer}}]{Stein2010}%
  \BibitemOpen
  \bibfield  {author} {\bibinfo {author} {\bibfnamefont {T.}~\bibnamefont {Stein}}, \bibinfo {author} {\bibfnamefont {H.}~\bibnamefont {Eisenberg}}, \bibinfo {author} {\bibfnamefont {L.}~\bibnamefont {Kronik}}, \ and\ \bibinfo {author} {\bibfnamefont {R.}~\bibnamefont {Baer}},\ }\bibfield  {title} {\enquote {\bibinfo {title} {Fundamental gaps in finite systems from eigenvalues of a generalized kohn-sham method},}\ }\href {\doibase 10.1103/PhysRevLett.105.266802} {\bibfield  {journal} {\bibinfo  {journal} {Phys. Rev. Lett.}\ }\textbf {\bibinfo {volume} {105}},\ \bibinfo {pages} {266802} (\bibinfo {year} {2010})}\BibitemShut {NoStop}%
\bibitem [{\citenamefont {Autschbach}\ and\ \citenamefont {Srebro}(2014)}]{Autschbach2014}%
  \BibitemOpen
  \bibfield  {author} {\bibinfo {author} {\bibfnamefont {J.}~\bibnamefont {Autschbach}}\ and\ \bibinfo {author} {\bibfnamefont {M.}~\bibnamefont {Srebro}},\ }\bibfield  {title} {\enquote {\bibinfo {title} {Delocalization error and “functional tuning” in kohn–sham calculations of molecular properties},}\ }\href {\doibase 10.1021/ar500171t} {\bibfield  {journal} {\bibinfo  {journal} {Accounts of Chemical Research}\ }\textbf {\bibinfo {volume} {47}},\ \bibinfo {pages} {2592--2602} (\bibinfo {year} {2014})}\BibitemShut {NoStop}%
\bibitem [{\citenamefont {Miceli}\ \emph {et~al.}(2018{\natexlab{b}})\citenamefont {Miceli}, \citenamefont {Chen}, \citenamefont {Reshetnyak},\ and\ \citenamefont {Pasquarello}}]{MCRP18}%
  \BibitemOpen
  \bibfield  {author} {\bibinfo {author} {\bibfnamefont {G.}~\bibnamefont {Miceli}}, \bibinfo {author} {\bibfnamefont {W.}~\bibnamefont {Chen}}, \bibinfo {author} {\bibfnamefont {I.}~\bibnamefont {Reshetnyak}}, \ and\ \bibinfo {author} {\bibfnamefont {A.}~\bibnamefont {Pasquarello}},\ }\bibfield  {title} {\enquote {\bibinfo {title} {Nonempirical hybrid functionals for band gaps and polaronic distortions in solids},}\ }\href {\doibase 10.1103/PhysRevB.97.121112} {\bibfield  {journal} {\bibinfo  {journal} {Phys. Rev. B}\ }\textbf {\bibinfo {volume} {97}},\ \bibinfo {pages} {121112} (\bibinfo {year} {2018}{\natexlab{b}})}\BibitemShut {NoStop}%
\bibitem [{\citenamefont {Moussa}\ \emph {et~al.}(2012)\citenamefont {Moussa}, \citenamefont {Schultz},\ and\ \citenamefont {Chelikowsky}}]{MSC12}%
  \BibitemOpen
  \bibfield  {author} {\bibinfo {author} {\bibfnamefont {J.~E.}\ \bibnamefont {Moussa}}, \bibinfo {author} {\bibfnamefont {P.~A.}\ \bibnamefont {Schultz}}, \ and\ \bibinfo {author} {\bibfnamefont {J.~R.}\ \bibnamefont {Chelikowsky}},\ }\bibfield  {title} {\enquote {\bibinfo {title} {Analysis of the heyd-scuseria-ernzerhof density functional parameter space},}\ }\href {\doibase 10.1063/1.4722993} {\bibfield  {journal} {\bibinfo  {journal} {J. Chem. Phys.}\ }\textbf {\bibinfo {volume} {136}},\ \bibinfo {pages} {204117} (\bibinfo {year} {2012})}\BibitemShut {NoStop}%
\bibitem [{\citenamefont {Qiu}\ \emph {et~al.}(2016)\citenamefont {Qiu}, \citenamefont {da~Jornada},\ and\ \citenamefont {Louie}}]{Qiu2016}%
  \BibitemOpen
  \bibfield  {author} {\bibinfo {author} {\bibfnamefont {D.~Y.}\ \bibnamefont {Qiu}}, \bibinfo {author} {\bibfnamefont {F.~H.}\ \bibnamefont {da~Jornada}}, \ and\ \bibinfo {author} {\bibfnamefont {S.~G.}\ \bibnamefont {Louie}},\ }\bibfield  {title} {\enquote {\bibinfo {title} {Screening and many-body effects in two-dimensional crystals: Monolayer ${\mathrm{mos}}_{2}$},}\ }\href {\doibase 10.1103/PhysRevB.93.235435} {\bibfield  {journal} {\bibinfo  {journal} {Phys. Rev. B}\ }\textbf {\bibinfo {volume} {93}},\ \bibinfo {pages} {235435} (\bibinfo {year} {2016})}\BibitemShut {NoStop}%
\bibitem [{\citenamefont {Cudazzo}\ \emph {et~al.}(2011)\citenamefont {Cudazzo}, \citenamefont {Tokatly},\ and\ \citenamefont {Rubio}}]{Cudazzo2011}%
  \BibitemOpen
  \bibfield  {author} {\bibinfo {author} {\bibfnamefont {P.}~\bibnamefont {Cudazzo}}, \bibinfo {author} {\bibfnamefont {I.~V.}\ \bibnamefont {Tokatly}}, \ and\ \bibinfo {author} {\bibfnamefont {A.}~\bibnamefont {Rubio}},\ }\bibfield  {title} {\enquote {\bibinfo {title} {Dielectric screening in two-dimensional insulators: Implications for excitonic and impurity states in graphane},}\ }\href {https://link.aps.org/doi/10.1103/PhysRevB.84.085406} {\bibfield  {journal} {\bibinfo  {journal} {Phys. Rev. B}\ }\textbf {\bibinfo {volume} {84}},\ \bibinfo {pages} {085406} (\bibinfo {year} {2011})}\BibitemShut {NoStop}%
\bibitem [{\citenamefont {Andersen}\ \emph {et~al.}(2015)\citenamefont {Andersen}, \citenamefont {Latini},\ and\ \citenamefont {Thygesen}}]{Andersen2015}%
  \BibitemOpen
  \bibfield  {author} {\bibinfo {author} {\bibfnamefont {K.}~\bibnamefont {Andersen}}, \bibinfo {author} {\bibfnamefont {S.}~\bibnamefont {Latini}}, \ and\ \bibinfo {author} {\bibfnamefont {K.~S.}\ \bibnamefont {Thygesen}},\ }\bibfield  {title} {\enquote {\bibinfo {title} {Dielectric genome of van der waals heterostructures},}\ }\href {\doibase 10.1021/acs.nanolett.5b01251} {\bibfield  {journal} {\bibinfo  {journal} {Nano Lett.}\ }\textbf {\bibinfo {volume} {15}},\ \bibinfo {pages} {4616--4621} (\bibinfo {year} {2015})}\BibitemShut {NoStop}%
\bibitem [{\citenamefont {Macfarlane}\ \emph {et~al.}(1957)\citenamefont {Macfarlane}, \citenamefont {McLean}, \citenamefont {Quarrington},\ and\ \citenamefont {Roberts}}]{MMQR57}%
  \BibitemOpen
  \bibfield  {author} {\bibinfo {author} {\bibfnamefont {G.~G.}\ \bibnamefont {Macfarlane}}, \bibinfo {author} {\bibfnamefont {T.~P.}\ \bibnamefont {McLean}}, \bibinfo {author} {\bibfnamefont {J.~E.}\ \bibnamefont {Quarrington}}, \ and\ \bibinfo {author} {\bibfnamefont {V.}~\bibnamefont {Roberts}},\ }\bibfield  {title} {\enquote {\bibinfo {title} {Fine structure in the absorption-edge spectrum of ge},}\ }\href {\doibase 10.1103/PhysRev.108.1377} {\bibfield  {journal} {\bibinfo  {journal} {Phys. Rev.}\ }\textbf {\bibinfo {volume} {108}},\ \bibinfo {pages} {1377--1383} (\bibinfo {year} {1957})}\BibitemShut {NoStop}%
\bibitem [{\citenamefont {Rohlfing}\ and\ \citenamefont {Louie}(2000)}]{Rohlfing2000}%
  \BibitemOpen
  \bibfield  {author} {\bibinfo {author} {\bibfnamefont {M.}~\bibnamefont {Rohlfing}}\ and\ \bibinfo {author} {\bibfnamefont {S.~G.}\ \bibnamefont {Louie}},\ }\bibfield  {title} {\enquote {\bibinfo {title} {Electron-hole excitations and optical spectra from first principles},}\ }\href {\doibase 10.1103/PhysRevB.62.4927} {\bibfield  {journal} {\bibinfo  {journal} {Phys. Rev. B}\ }\textbf {\bibinfo {volume} {62}},\ \bibinfo {pages} {4927--4944} (\bibinfo {year} {2000})}\BibitemShut {NoStop}%
\bibitem [{\citenamefont {Srivastava}(1997)}]{S97}%
  \BibitemOpen
  \bibfield  {author} {\bibinfo {author} {\bibfnamefont {G.~P.}\ \bibnamefont {Srivastava}},\ }\bibfield  {title} {\enquote {\bibinfo {title} {Theory of semiconductor surface reconstruction},}\ }\href {\doibase 10.1088/0034-4885/60/5/002} {\bibfield  {journal} {\bibinfo  {journal} {Rep. Prog. Phys.}\ }\textbf {\bibinfo {volume} {60}},\ \bibinfo {pages} {561} (\bibinfo {year} {1997})}\BibitemShut {NoStop}%
\bibitem [{\citenamefont {Pandey}(1981)}]{P81}%
  \BibitemOpen
  \bibfield  {author} {\bibinfo {author} {\bibfnamefont {K.~C.}\ \bibnamefont {Pandey}},\ }\bibfield  {title} {\enquote {\bibinfo {title} {New $\ensuremath{\pi}$-bonded chain model for si(111)-(2\ifmmode\times\else\texttimes\fi{}1) surface},}\ }\href {\doibase 10.1103/PhysRevLett.47.1913} {\bibfield  {journal} {\bibinfo  {journal} {Phys. Rev. Lett.}\ }\textbf {\bibinfo {volume} {47}},\ \bibinfo {pages} {1913--1917} (\bibinfo {year} {1981})}\BibitemShut {NoStop}%
\bibitem [{\citenamefont {Reining}\ and\ \citenamefont {Del~Sole}(1991)}]{RD91}%
  \BibitemOpen
  \bibfield  {author} {\bibinfo {author} {\bibfnamefont {L.}~\bibnamefont {Reining}}\ and\ \bibinfo {author} {\bibfnamefont {R.}~\bibnamefont {Del~Sole}},\ }\bibfield  {title} {\enquote {\bibinfo {title} {Quasi-one-dimensional excitons and the optical properties of si(111)2\ifmmode\times\else\texttimes\fi{}1},}\ }\href {\doibase 10.1103/PhysRevLett.67.3816} {\bibfield  {journal} {\bibinfo  {journal} {Phys. Rev. Lett.}\ }\textbf {\bibinfo {volume} {67}},\ \bibinfo {pages} {3816--3819} (\bibinfo {year} {1991})}\BibitemShut {NoStop}%
\bibitem [{\citenamefont {Northrup}\ \emph {et~al.}(1991)\citenamefont {Northrup}, \citenamefont {Hybertsen},\ and\ \citenamefont {Louie}}]{NHL91}%
  \BibitemOpen
  \bibfield  {author} {\bibinfo {author} {\bibfnamefont {J.~E.}\ \bibnamefont {Northrup}}, \bibinfo {author} {\bibfnamefont {M.~S.}\ \bibnamefont {Hybertsen}}, \ and\ \bibinfo {author} {\bibfnamefont {S.~G.}\ \bibnamefont {Louie}},\ }\bibfield  {title} {\enquote {\bibinfo {title} {Many-body calculation of the surface-state energies for \ce{Si}(111)2\ifmmode\times\else\texttimes\fi{}1},}\ }\href {\doibase 10.1103/PhysRevLett.66.500} {\bibfield  {journal} {\bibinfo  {journal} {Phys. Rev. Lett.}\ }\textbf {\bibinfo {volume} {66}},\ \bibinfo {pages} {500--503} (\bibinfo {year} {1991})}\BibitemShut {NoStop}%
\bibitem [{\citenamefont {Uhrberg}\ \emph {et~al.}(1982)\citenamefont {Uhrberg}, \citenamefont {Hansson}, \citenamefont {Nicholls},\ and\ \citenamefont {Flodstr\"om}}]{UHNF82}%
  \BibitemOpen
  \bibfield  {author} {\bibinfo {author} {\bibfnamefont {R.~I.~G.}\ \bibnamefont {Uhrberg}}, \bibinfo {author} {\bibfnamefont {G.~V.}\ \bibnamefont {Hansson}}, \bibinfo {author} {\bibfnamefont {J.~M.}\ \bibnamefont {Nicholls}}, \ and\ \bibinfo {author} {\bibfnamefont {S.~A.}\ \bibnamefont {Flodstr\"om}},\ }\bibfield  {title} {\enquote {\bibinfo {title} {Experimental evidence for one highly dispersive dangling-bond band on \ce{Si}(111) 2 \ifmmode\times\else\texttimes\fi{} 1},}\ }\href {\doibase 10.1103/PhysRevLett.48.1032} {\bibfield  {journal} {\bibinfo  {journal} {Phys. Rev. Lett.}\ }\textbf {\bibinfo {volume} {48}},\ \bibinfo {pages} {1032--1035} (\bibinfo {year} {1982})}\BibitemShut {NoStop}%
\bibitem [{\citenamefont {Perfetti}\ \emph {et~al.}(1987)\citenamefont {Perfetti}, \citenamefont {Nicholls},\ and\ \citenamefont {Reihl}}]{PNR87}%
  \BibitemOpen
  \bibfield  {author} {\bibinfo {author} {\bibfnamefont {P.}~\bibnamefont {Perfetti}}, \bibinfo {author} {\bibfnamefont {J.~M.}\ \bibnamefont {Nicholls}}, \ and\ \bibinfo {author} {\bibfnamefont {B.}~\bibnamefont {Reihl}},\ }\bibfield  {title} {\enquote {\bibinfo {title} {Unoccupied surface-state band on \ce{Si}(111) 2\ifmmode\times\else\texttimes\fi{}1},}\ }\href {\doibase 10.1103/PhysRevB.36.6160} {\bibfield  {journal} {\bibinfo  {journal} {Phys. Rev. B}\ }\textbf {\bibinfo {volume} {36}},\ \bibinfo {pages} {6160--6163} (\bibinfo {year} {1987})}\BibitemShut {NoStop}%
\bibitem [{\citenamefont {Nannarone}\ \emph {et~al.}(1980)\citenamefont {Nannarone}, \citenamefont {Chiaradia}, \citenamefont {Ciccacci}, \citenamefont {Memeo}, \citenamefont {Sassaroli}, \citenamefont {Selci},\ and\ \citenamefont {Chiarotti}}]{Nannarone1980}%
  \BibitemOpen
  \bibfield  {author} {\bibinfo {author} {\bibfnamefont {S.}~\bibnamefont {Nannarone}}, \bibinfo {author} {\bibfnamefont {P.}~\bibnamefont {Chiaradia}}, \bibinfo {author} {\bibfnamefont {F.}~\bibnamefont {Ciccacci}}, \bibinfo {author} {\bibfnamefont {R.}~\bibnamefont {Memeo}}, \bibinfo {author} {\bibfnamefont {P.}~\bibnamefont {Sassaroli}}, \bibinfo {author} {\bibfnamefont {S.}~\bibnamefont {Selci}}, \ and\ \bibinfo {author} {\bibfnamefont {G.}~\bibnamefont {Chiarotti}},\ }\bibfield  {title} {\enquote {\bibinfo {title} {Surface states in \ce{Si}(111)2×1 and \ce{Ge}(111)2×1 by optical reflectivity},}\ }\href {\doibase https://doi.org/10.1016/0038-1098(80)90731-0} {\bibfield  {journal} {\bibinfo  {journal} {Solid State Commun.}\ }\textbf {\bibinfo {volume} {33}},\ \bibinfo {pages} {593--595} (\bibinfo {year} {1980})}\BibitemShut {NoStop}%
\bibitem [{\citenamefont {Srivastava}(1999)}]{S99}%
  \BibitemOpen
  \bibfield  {author} {\bibinfo {author} {\bibfnamefont {G~P}\ \bibnamefont {Srivastava}},\ }\href {\doibase 10.1142/3635} {\emph {\bibinfo {title} {Theoretical Modelling of Semiconductor Surfaces}}}\ (\bibinfo  {publisher} {WORLD SCIENTIFIC},\ \bibinfo {year} {1999})\BibitemShut {NoStop}%
\bibitem [{\citenamefont {Nie}\ \emph {et~al.}(2004)\citenamefont {Nie}, \citenamefont {Feenstra}, \citenamefont {Lee},\ and\ \citenamefont {Kang}}]{NFLK04}%
  \BibitemOpen
  \bibfield  {author} {\bibinfo {author} {\bibfnamefont {S.}~\bibnamefont {Nie}}, \bibinfo {author} {\bibfnamefont {R.~M.}\ \bibnamefont {Feenstra}}, \bibinfo {author} {\bibfnamefont {J.}~\bibnamefont {Lee}}, \ and\ \bibinfo {author} {\bibfnamefont {M.}~\bibnamefont {Kang}},\ }\bibfield  {title} {\enquote {\bibinfo {title} {Buckling of \ce{Si} and \ce{Ge}(111)2×1 surfaces},}\ }\href {\doibase 10.1116/1.1705647} {\bibfield  {journal} {\bibinfo  {journal} {J. Vac. Sci. Technol. A}\ }\textbf {\bibinfo {volume} {22}},\ \bibinfo {pages} {1671--1674} (\bibinfo {year} {2004})}\BibitemShut {NoStop}%
\bibitem [{\citenamefont {Kresse}\ and\ \citenamefont {Furthmüller}(1996)}]{vasp1}%
  \BibitemOpen
  \bibfield  {author} {\bibinfo {author} {\bibfnamefont {G.}~\bibnamefont {Kresse}}\ and\ \bibinfo {author} {\bibfnamefont {J.}~\bibnamefont {Furthmüller}},\ }\bibfield  {title} {\enquote {\bibinfo {title} {Efficiency of ab-initio total energy calculations for metals and semiconductors using a plane-wave basis set},}\ }\href {\doibase https://doi.org/10.1016/0927-0256(96)00008-0} {\bibfield  {journal} {\bibinfo  {journal} {Comput. Mater. Sci.}\ }\textbf {\bibinfo {volume} {6}},\ \bibinfo {pages} {15--50} (\bibinfo {year} {1996})}\BibitemShut {NoStop}%
\bibitem [{\citenamefont {Kresse}\ and\ \citenamefont {Furthm\"uller}(1996)}]{vasp2}%
  \BibitemOpen
  \bibfield  {author} {\bibinfo {author} {\bibfnamefont {G.}~\bibnamefont {Kresse}}\ and\ \bibinfo {author} {\bibfnamefont {J.}~\bibnamefont {Furthm\"uller}},\ }\bibfield  {title} {\enquote {\bibinfo {title} {Efficient iterative schemes for ab initio total-energy calculations using a plane-wave basis set},}\ }\href {\doibase 10.1103/PhysRevB.54.11169} {\bibfield  {journal} {\bibinfo  {journal} {Phys. Rev. B}\ }\textbf {\bibinfo {volume} {54}},\ \bibinfo {pages} {11169--11186} (\bibinfo {year} {1996})}\BibitemShut {NoStop}%
\bibitem [{\citenamefont {Giannozzi}\ \emph {et~al.}(2009)\citenamefont {Giannozzi}, \citenamefont {Baroni}, \citenamefont {Bonini}, \citenamefont {Calandra}, \citenamefont {Car}, \citenamefont {Cavazzoni}, \citenamefont {{D. Ceresoli}}, \citenamefont {Chiarotti}, \citenamefont {Cococcioni}, \citenamefont {Dabo}, \citenamefont {Corso}, \citenamefont {de~Gironcoli}, \citenamefont {Fabris}, \citenamefont {Fratesi}, \citenamefont {Gebauer}, \citenamefont {Gerstmann}, \citenamefont {Gougoussis}, \citenamefont {{A. Kokalj}}, \citenamefont {Lazzeri}, \citenamefont {{Martin-Samos}}, \citenamefont {Marzari}, \citenamefont {Mauri}, \citenamefont {Mazzarello}, \citenamefont {{S. Paolini}}, \citenamefont {Pasquarello}, \citenamefont {Paulatto}, \citenamefont {Sbraccia}, \citenamefont {Scandolo}, \citenamefont {Sclauzero}, \citenamefont {Seitsonen}, \citenamefont {Smogunov}, \citenamefont {Umari},\ and\ \citenamefont {Wentzcovitch}}]{Giannozzi2009}%
  \BibitemOpen
  \bibfield  {author} {\bibinfo {author} {\bibfnamefont {P.}~\bibnamefont {Giannozzi}}, \bibinfo {author} {\bibfnamefont {S.}~\bibnamefont {Baroni}}, \bibinfo {author} {\bibfnamefont {N.}~\bibnamefont {Bonini}}, \bibinfo {author} {\bibfnamefont {M.}~\bibnamefont {Calandra}}, \bibinfo {author} {\bibfnamefont {R.}~\bibnamefont {Car}}, \bibinfo {author} {\bibfnamefont {C.}~\bibnamefont {Cavazzoni}}, \bibinfo {author} {\bibnamefont {{D. Ceresoli}}}, \bibinfo {author} {\bibfnamefont {G.~L.}\ \bibnamefont {Chiarotti}}, \bibinfo {author} {\bibfnamefont {M.}~\bibnamefont {Cococcioni}}, \bibinfo {author} {\bibfnamefont {I.}~\bibnamefont {Dabo}}, \bibinfo {author} {\bibfnamefont {A.~Dal}\ \bibnamefont {Corso}}, \bibinfo {author} {\bibfnamefont {S.}~\bibnamefont {de~Gironcoli}}, \bibinfo {author} {\bibfnamefont {S.}~\bibnamefont {Fabris}}, \bibinfo {author} {\bibfnamefont {G.}~\bibnamefont {Fratesi}}, \bibinfo {author} {\bibfnamefont {R.}~\bibnamefont {Gebauer}}, \bibinfo {author} {\bibfnamefont {U.}~\bibnamefont
  {Gerstmann}}, \bibinfo {author} {\bibfnamefont {C.}~\bibnamefont {Gougoussis}}, \bibinfo {author} {\bibnamefont {{A. Kokalj}}}, \bibinfo {author} {\bibfnamefont {M.}~\bibnamefont {Lazzeri}}, \bibinfo {author} {\bibfnamefont {L.}~\bibnamefont {{Martin-Samos}}}, \bibinfo {author} {\bibfnamefont {N.}~\bibnamefont {Marzari}}, \bibinfo {author} {\bibfnamefont {F.}~\bibnamefont {Mauri}}, \bibinfo {author} {\bibfnamefont {R.}~\bibnamefont {Mazzarello}}, \bibinfo {author} {\bibnamefont {{S. Paolini}}}, \bibinfo {author} {\bibfnamefont {A.}~\bibnamefont {Pasquarello}}, \bibinfo {author} {\bibfnamefont {L.}~\bibnamefont {Paulatto}}, \bibinfo {author} {\bibfnamefont {C.}~\bibnamefont {Sbraccia}}, \bibinfo {author} {\bibfnamefont {S.}~\bibnamefont {Scandolo}}, \bibinfo {author} {\bibfnamefont {G.}~\bibnamefont {Sclauzero}}, \bibinfo {author} {\bibfnamefont {A.~P.}\ \bibnamefont {Seitsonen}}, \bibinfo {author} {\bibfnamefont {A.}~\bibnamefont {Smogunov}}, \bibinfo {author} {\bibfnamefont {P.}~\bibnamefont {Umari}}, \
  and\ \bibinfo {author} {\bibfnamefont {R.~M.}\ \bibnamefont {Wentzcovitch}},\ }\bibfield  {title} {\enquote {\bibinfo {title} {Quantum espresso: a modular and open-source software project for quantum simulations of materials},}\ }\href {\doibase 10.1088/0953-8984/21/39/395502} {\bibfield  {journal} {\bibinfo  {journal} {J. Phys.: Condens. Matter}\ }\textbf {\bibinfo {volume} {21}},\ \bibinfo {pages} {395502} (\bibinfo {year} {2009})}\BibitemShut {NoStop}%
\bibitem [{Sup()}]{SupplementalMaterial}%
  \BibitemOpen
  \href@noop {} {}\bibinfo {note} {See Supporting Information Material at [URL will be inserted by publisher] for computational details, additional results, convergence calculations, atomic geometries, and comparison with existing literature.}\BibitemShut {Stop}%
\bibitem [{\citenamefont {Wing}\ \emph {et~al.}(2019)\citenamefont {Wing}, \citenamefont {Haber}, \citenamefont {Noff}, \citenamefont {Barker}, \citenamefont {Egger}, \citenamefont {Ramasubramaniam}, \citenamefont {Louie}, \citenamefont {Neaton},\ and\ \citenamefont {Kronik}}]{Wing2019}%
  \BibitemOpen
  \bibfield  {author} {\bibinfo {author} {\bibfnamefont {D.}~\bibnamefont {Wing}}, \bibinfo {author} {\bibfnamefont {J.~B.}\ \bibnamefont {Haber}}, \bibinfo {author} {\bibfnamefont {R.}~\bibnamefont {Noff}}, \bibinfo {author} {\bibfnamefont {B.}~\bibnamefont {Barker}}, \bibinfo {author} {\bibfnamefont {D.~A.}\ \bibnamefont {Egger}}, \bibinfo {author} {\bibfnamefont {A.}~\bibnamefont {Ramasubramaniam}}, \bibinfo {author} {\bibfnamefont {S.~G.}\ \bibnamefont {Louie}}, \bibinfo {author} {\bibfnamefont {J.~B.}\ \bibnamefont {Neaton}}, \ and\ \bibinfo {author} {\bibfnamefont {L.}~\bibnamefont {Kronik}},\ }\bibfield  {title} {\enquote {\bibinfo {title} {Comparing time-dependent density functional theory with many-body perturbation theory for semiconductors: Screened range-separated hybrids and the gw plus bethe-salpeter approach},}\ }\href {\doibase 10.1103/PhysRevMaterials.3.064603} {\bibfield  {journal} {\bibinfo  {journal} {Phys. Rev. Materials}\ }\textbf {\bibinfo {volume} {3}},\ \bibinfo {pages} {064603}
  (\bibinfo {year} {2019})}\BibitemShut {NoStop}%
\bibitem [{\citenamefont {Wing}\ \emph {et~al.}(2020)\citenamefont {Wing}, \citenamefont {Neaton},\ and\ \citenamefont {Kronik}}]{Wing2020}%
  \BibitemOpen
  \bibfield  {author} {\bibinfo {author} {\bibfnamefont {D.}~\bibnamefont {Wing}}, \bibinfo {author} {\bibfnamefont {J.~B.}\ \bibnamefont {Neaton}}, \ and\ \bibinfo {author} {\bibfnamefont {L.}~\bibnamefont {Kronik}},\ }\bibfield  {title} {\enquote {\bibinfo {title} {Time-dependent density functional theory of narrow band gap semiconductors using a screened range-separated hybrid functional},}\ }\href {\doibase https://doi.org/10.1002/adts.202000220} {\bibfield  {journal} {\bibinfo  {journal} {Adv. Theory Simul.}\ }\textbf {\bibinfo {volume} {3}},\ \bibinfo {pages} {2000220} (\bibinfo {year} {2020})}\BibitemShut {NoStop}%
\bibitem [{\citenamefont {Perdew}\ \emph {et~al.}(1996{\natexlab{b}})\citenamefont {Perdew}, \citenamefont {Burke},\ and\ \citenamefont {Ernzerhof}}]{PBE96}%
  \BibitemOpen
  \bibfield  {author} {\bibinfo {author} {\bibfnamefont {J.~P.}\ \bibnamefont {Perdew}}, \bibinfo {author} {\bibfnamefont {K.}~\bibnamefont {Burke}}, \ and\ \bibinfo {author} {\bibfnamefont {M.}~\bibnamefont {Ernzerhof}},\ }\bibfield  {title} {\enquote {\bibinfo {title} {Generalized gradient approximation made simple},}\ }\href {\doibase 10.1103/PhysRevLett.77.3865} {\bibfield  {journal} {\bibinfo  {journal} {Phys. Rev. Lett.}\ }\textbf {\bibinfo {volume} {77}},\ \bibinfo {pages} {3865--3868} (\bibinfo {year} {1996}{\natexlab{b}})}\BibitemShut {NoStop}%
\bibitem [{\citenamefont {Krukau}\ \emph {et~al.}(2006)\citenamefont {Krukau}, \citenamefont {Vydrov}, \citenamefont {Izmaylov},\ and\ \citenamefont {Scuseria}}]{HSE06}%
  \BibitemOpen
  \bibfield  {author} {\bibinfo {author} {\bibfnamefont {A.~V.}\ \bibnamefont {Krukau}}, \bibinfo {author} {\bibfnamefont {O.~A.}\ \bibnamefont {Vydrov}}, \bibinfo {author} {\bibfnamefont {A.~F.}\ \bibnamefont {Izmaylov}}, \ and\ \bibinfo {author} {\bibfnamefont {G.~E.}\ \bibnamefont {Scuseria}},\ }\bibfield  {title} {\enquote {\bibinfo {title} {Influence of the exchange screening parameter on the performance of screened hybrid functionals},}\ }\href {\doibase 10.1063/1.2404663} {\bibfield  {journal} {\bibinfo  {journal} {J. Chem. Phys.}\ }\textbf {\bibinfo {volume} {125}},\ \bibinfo {pages} {224106} (\bibinfo {year} {2006})}\BibitemShut {NoStop}%
\bibitem [{\citenamefont {Armiento}\ and\ \citenamefont {Mattsson}(2005)}]{AM05}%
  \BibitemOpen
  \bibfield  {author} {\bibinfo {author} {\bibfnamefont {R.}~\bibnamefont {Armiento}}\ and\ \bibinfo {author} {\bibfnamefont {A.~E.}\ \bibnamefont {Mattsson}},\ }\bibfield  {title} {\enquote {\bibinfo {title} {Functional designed to include surface effects in self-consistent density functional theory},}\ }\href {\doibase 10.1103/PhysRevB.72.085108} {\bibfield  {journal} {\bibinfo  {journal} {Phys. Rev. B}\ }\textbf {\bibinfo {volume} {72}},\ \bibinfo {pages} {085108} (\bibinfo {year} {2005})}\BibitemShut {NoStop}%
\bibitem [{\citenamefont {DelloStritto}\ \emph {et~al.}(2023)\citenamefont {DelloStritto}, \citenamefont {Kaplan}, \citenamefont {Perdew},\ and\ \citenamefont {Klein}}]{DKPK23}%
  \BibitemOpen
  \bibfield  {author} {\bibinfo {author} {\bibfnamefont {M.~J.}\ \bibnamefont {DelloStritto}}, \bibinfo {author} {\bibfnamefont {A.~D.}\ \bibnamefont {Kaplan}}, \bibinfo {author} {\bibfnamefont {J.~P.}\ \bibnamefont {Perdew}}, \ and\ \bibinfo {author} {\bibfnamefont {M.~L.}\ \bibnamefont {Klein}},\ }\bibfield  {title} {\enquote {\bibinfo {title} {Predicting the properties of nio with density functional theory: Impact of exchange and correlation approximations and validation of the r2scan functional},}\ }\href {\doibase 10.1063/5.0146967} {\bibfield  {journal} {\bibinfo  {journal} {APL Materials}\ }\textbf {\bibinfo {volume} {11}},\ \bibinfo {pages} {060702} (\bibinfo {year} {2023})}\BibitemShut {NoStop}%
\bibitem [{\citenamefont {Furness}\ \emph {et~al.}(2020)\citenamefont {Furness}, \citenamefont {Kaplan}, \citenamefont {Ning}, \citenamefont {Perdew},\ and\ \citenamefont {Sun}}]{r2SCAN}%
  \BibitemOpen
  \bibfield  {author} {\bibinfo {author} {\bibfnamefont {J.~W.}\ \bibnamefont {Furness}}, \bibinfo {author} {\bibfnamefont {A.~D.}\ \bibnamefont {Kaplan}}, \bibinfo {author} {\bibfnamefont {J.}~\bibnamefont {Ning}}, \bibinfo {author} {\bibfnamefont {J.~P.}\ \bibnamefont {Perdew}}, \ and\ \bibinfo {author} {\bibfnamefont {J.}~\bibnamefont {Sun}},\ }\bibfield  {title} {\enquote {\bibinfo {title} {Accurate and numerically efficient r2scan meta-generalized gradient approximation},}\ }\href {\doibase 10.1021/acs.jpclett.0c02405} {\bibfield  {journal} {\bibinfo  {journal} {J. Phys. Chem. Lett.}\ }\textbf {\bibinfo {volume} {11}},\ \bibinfo {pages} {8208--8215} (\bibinfo {year} {2020})}\BibitemShut {NoStop}%
\bibitem [{\citenamefont {Casida}(1996)}]{Casida1996}%
  \BibitemOpen
  \bibfield  {author} {\bibinfo {author} {\bibfnamefont {M.~E.}\ \bibnamefont {Casida}},\ }\enquote {\bibinfo {title} {Time-dependent density functional response theory of molecular systems: Theory, computational methods, and functionals},}\ in\ \href {https://www.sciencedirect.com/science/article/pii/S1380732396800938} {\emph {\bibinfo {booktitle} {Theoretical and Computational Chemistry}}},\ \bibinfo {series} {Recent Developments and Applications of Modern Density Functional Theory}, Vol.~\bibinfo {volume} {4}\ (\bibinfo  {publisher} {Elsevier},\ \bibinfo {year} {1996})\ pp.\ \bibinfo {pages} {391--439}\BibitemShut {NoStop}%
\bibitem [{\citenamefont {Tretiak}\ and\ \citenamefont {Chernyak}(2003)}]{TC2003}%
  \BibitemOpen
  \bibfield  {author} {\bibinfo {author} {\bibfnamefont {S.}~\bibnamefont {Tretiak}}\ and\ \bibinfo {author} {\bibfnamefont {V.}~\bibnamefont {Chernyak}},\ }\bibfield  {title} {\enquote {\bibinfo {title} {Resonant nonlinear polarizabilities in the time-dependent density functional theory},}\ }\href {\doibase 10.1063/1.1614240} {\bibfield  {journal} {\bibinfo  {journal} {The Journal of Chemical Physics}\ }\textbf {\bibinfo {volume} {119}},\ \bibinfo {pages} {8809--8823} (\bibinfo {year} {2003})}\BibitemShut {NoStop}%
\bibitem [{\citenamefont {Sander}\ \emph {et~al.}(2015)\citenamefont {Sander}, \citenamefont {Maggio},\ and\ \citenamefont {Kresse}}]{TD-approx}%
  \BibitemOpen
  \bibfield  {author} {\bibinfo {author} {\bibfnamefont {T.}~\bibnamefont {Sander}}, \bibinfo {author} {\bibfnamefont {E.}~\bibnamefont {Maggio}}, \ and\ \bibinfo {author} {\bibfnamefont {G.}~\bibnamefont {Kresse}},\ }\bibfield  {title} {\enquote {\bibinfo {title} {Beyond the tamm-dancoff approximation for extended systems using exact diagonalization},}\ }\href {\doibase 10.1103/PhysRevB.92.045209} {\bibfield  {journal} {\bibinfo  {journal} {Phys. Rev. B}\ }\textbf {\bibinfo {volume} {92}},\ \bibinfo {pages} {045209} (\bibinfo {year} {2015})}\BibitemShut {NoStop}%
\bibitem [{\citenamefont {Byun}\ and\ \citenamefont {Ullrich}(2017)}]{Byun2016}%
  \BibitemOpen
  \bibfield  {author} {\bibinfo {author} {\bibfnamefont {Y.-M.}\ \bibnamefont {Byun}}\ and\ \bibinfo {author} {\bibfnamefont {C.~A.}\ \bibnamefont {Ullrich}},\ }\bibfield  {title} {\enquote {\bibinfo {title} {Excitons in solids from time-dependent density-functional theory: Assessing the tamm-dancoff approximation},}\ }\href {https://www.mdpi.com/2079-3197/5/1/9} {\bibfield  {journal} {\bibinfo  {journal} {Computation}\ }\textbf {\bibinfo {volume} {5}} (\bibinfo {year} {2017})}\BibitemShut {NoStop}%
\bibitem [{\citenamefont {Hernang{\'o}mez-P{\'e}rez}\ \emph {et~al.}(2023)\citenamefont {Hernang{\'o}mez-P{\'e}rez}, \citenamefont {Kleiner},\ and\ \citenamefont {Refaely-Abramson}}]{Hernangomez2023-02}%
  \BibitemOpen
  \bibfield  {author} {\bibinfo {author} {\bibfnamefont {Daniel}\ \bibnamefont {Hernang{\'o}mez-P{\'e}rez}}, \bibinfo {author} {\bibfnamefont {Amir}\ \bibnamefont {Kleiner}}, \ and\ \bibinfo {author} {\bibfnamefont {Sivan}\ \bibnamefont {Refaely-Abramson}},\ }\bibfield  {title} {\enquote {\bibinfo {title} {Reduced absorption due to defect-localized interlayer excitons in transition-metal dichalcogenide--graphene heterostructures},}\ }\href {\doibase 10.1021/acs.nanolett.3c01182} {\bibfield  {journal} {\bibinfo  {journal} {Nano Letters}\ }\textbf {\bibinfo {volume} {23}},\ \bibinfo {pages} {5995--6001} (\bibinfo {year} {2023})}\BibitemShut {NoStop}%
\bibitem [{\citenamefont {Chen}\ \emph {et~al.}(2018)\citenamefont {Chen}, \citenamefont {Miceli}, \citenamefont {Rignanese},\ and\ \citenamefont {Pasquarello}}]{Chen2018}%
  \BibitemOpen
  \bibfield  {author} {\bibinfo {author} {\bibfnamefont {W.}~\bibnamefont {Chen}}, \bibinfo {author} {\bibfnamefont {G.}~\bibnamefont {Miceli}}, \bibinfo {author} {\bibfnamefont {G.}~\bibnamefont {Rignanese}}, \ and\ \bibinfo {author} {\bibfnamefont {A.}~\bibnamefont {Pasquarello}},\ }\bibfield  {title} {\enquote {\bibinfo {title} {Nonempirical dielectric-dependent hybrid functional with range separation for semiconductors and insulators},}\ }\href {\doibase 10.1103/PhysRevMaterials.2.073803} {\bibfield  {journal} {\bibinfo  {journal} {Phys. Rev. Mater.}\ }\textbf {\bibinfo {volume} {2}},\ \bibinfo {pages} {073803} (\bibinfo {year} {2018})}\BibitemShut {NoStop}%
\bibitem [{\citenamefont {Song}\ \emph {et~al.}(2013)\citenamefont {Song}, \citenamefont {Giorgi}, \citenamefont {Yamashita},\ and\ \citenamefont {Hirao}}]{Song2013}%
  \BibitemOpen
  \bibfield  {author} {\bibinfo {author} {\bibfnamefont {J.-W.}\ \bibnamefont {Song}}, \bibinfo {author} {\bibfnamefont {G.}~\bibnamefont {Giorgi}}, \bibinfo {author} {\bibfnamefont {K.}~\bibnamefont {Yamashita}}, \ and\ \bibinfo {author} {\bibfnamefont {K.}~\bibnamefont {Hirao}},\ }\bibfield  {title} {\enquote {\bibinfo {title} {Communication: Singularity-free hybrid functional with a gaussian-attenuating exact exchange in a plane-wave basis},}\ }\href {\doibase 10.1063/1.4811775} {\bibfield  {journal} {\bibinfo  {journal} {J. of Chem. Phys.}\ }\textbf {\bibinfo {volume} {138}},\ \bibinfo {pages} {241101} (\bibinfo {year} {2013})}\BibitemShut {NoStop}%
\end{thebibliography}%

%%%%%%%%%%%%%%%%%%%%%%%%%%%%%%%%%%%%%%%%%
%old figure formatting %
%\begin{figure*}[!htbp]
%    \centering
%    \subfigure (a)
%\includegraphics[width=0.46\textwidth]{bands_Si_RSH.png}
%    \hfill
%    \centering
%    \subfigure (b)
%\includegraphics[width=0.46\textwidth]{bands_Ge_RSH.png}
%    \hfill
%    \subfigure (c) \includegraphics[width=0.46\textwidth]{optical_absorption_combined_si2x1.pdf}
%    \hfill
%    \subfigure (d)
%    \includegraphics[width=0.46\textwidth]{ge_optical_spectra_placeholder.pdf}
%    \caption{Projected band structure for the reconstructed (a) Si (111) and (b) Ge (111) 2$\times$1 surfaces. The projected bands are shown in pink for Si and in green for Ge. The blue bands are the bulk states, using the SRSH functional with tuned bulk parameters. (c) Linear absorption spectrum, Im$[(\epsilon_{XX}+\epsilon_{YY}+\epsilon_{ZZ})/3]$, for Si(111) 2$\times$1 and (d) Ge (111) 2$\times$1. Experimental reported values is denoted by the vertical black bar \cite{Rohlfing2000}, the previous $GW$-BSE \cite{JCL11,Rohlfing2000} is the blue vertical line. The purple, magenta, and orange lines are the SRSH, HSE, and PBE0 calculations, respectively, for the reconstructed Si (and Ge)(111) 2$\times$1 surface.}
%\end{figure*}
%%%%%%%%%%%%%%%%%%%%%%%%%%%%%%%%%%%%%%%%%%%%%%%%
\label{page:end}
\end{document}

\maketitle

%%%%%%%%%%%%%%%%%%%%%%%%%%%%%%%%%%%%%%%%%%%%%%%%%%%%%%%%%%%%%%%%%%%%%%%%%
%%%%%%%%%%%%%%%%%%%%%%%%%%%%%%%%%%%%%%%%%%%%%%%
\sec{Static DFT calculations}

Ground state, static density functional theory calculations were performed with the Vienna \textit{Ab Initio} Simulation Package (VASP) \cite{vasp1, vasp2}, using a $4\times8\times1$ k point grid (in order to compare with previous calculations \cite{JCL11}) with an energy cutoff of $520$ eV. The VASP recommended pseudopotentials were used for all DFT calculations. All structures were relaxed with PBE.
The general k-point path for this system is $\Gamma$ ($0.0,0.0,0.0$) - J ($0.0,0.5,0.0$) - K ($0.5,0.5,0.0$) - J$'$ ($0.5,0.0,0.0$) - $\Gamma$ ($0.0,0.0,0.0$), as seen in the literature. The surface state fundamental gaps occur at $J$. The bulk Si fundamental gap occurs at $\Gamma-X$, and for bulk Ge at $\Gamma-L$. The bulk-tuned ($\alpha, \gamma$) parameters used for the Si \cite{WOHF21}, are ($\alpha =0.25, \gamma = 0.24$ Bohr\textsuperscript{$-1$}) and for Ge slab ($\alpha =0.25, \gamma= 0.19$ Bohr\textsuperscript{$-1$}). Note that VASP uses units of \AA\textsuperscript{-1}. \\

For comparison, we also performed ground state calculations using the QUANTUM ESPRESSO (QE) package~\cite{giannozzi2009,giannozzi2017}. We use norm conserving scalar-relativistic ONCVPSP pseudopotentials from the PseudoDojo~\cite{hamann2013,vansetten2018}, and tested various exchange-correlation functionals. A plane wave cut-off energy of 120\,Ry, a uniform $4 \times 8 \times 1$ $\mathbf{k}$-grid, and at least 13\,\AA\ of vacuum between the repeated slabs (to avoid spurious interactions were used).
The structures were relaxed in three Cartesian directions, until the force on each atom was lower than $10^{-5}$ Ha/bohr. Following previous works~\cite{northrup1991,zhu1991,rohlfing1999,rohlfing2000,nie2004,pulci2006}, we perform the relaxation starting from structures with either a positive or negative buckling. We find a ground-state structure with a positive (negative) buckling of 0.56\,\AA\ (0.85\,\AA), for the Si-surface (Ge-surface).\\

In Table SI~\ref{table1}, we report fundamental gaps for Si obtained with both VASP and QE (in parenthesis). QE gaps are in agreement with the VASP computed gaps to within 0.05 eV, with the exception of bulk Ge for the HSE and SRSH functionals. These gaps provided the largest deviation of almost 0.2 eV. We note that this deviation occurs when using both the DFT and $GW$ recommended pseudopotentials on VASP, and is in agreement with previous reports. We also note that using QE with the reported ($\alpha, \gamma$) values from \cite{WOHF21}, we can reproduce the bulk fundamental gap of Ge and Si.

\begin{table}[h]
\caption{Fundamental gaps of the Si and Ge (111)-($2 \times1$) surfaces, in eV. Black color is used for VASP calculated values while QE calculated values are given in blue color. 
Note that for the Ge surface, the surface gap could only be extracted through the projected band structures due to the band mixing at the $\Gamma$ point. For this study, the projected band structures were only calculated on VASP.}
%Those values are indicated in the main text.
\centering
\begin{tabular}{ c c c c |c c  }
\hline
\hline
VASP (\textcolor{blue}{QE}) & {Si(111)-($2 \times1$)} &  & {}  & Ge(111)-($2 \times1$)&\\ \hline
{} & {Bulk gap} & & {Surface gap} & Bulk gap &\\ \hline
PBE & 0.61 \textcolor{blue}{($0.59$)} & & 0.37 \textcolor{blue}{($0.36$)} & 0.07 \textcolor{blue}{($0.18$)}   \\
PBE0 & 1.78 \textcolor{blue}{($1.80$)} & & 1.20 \textcolor{blue}{($1.16$)} &  1.47 \textcolor{blue}{($1.35$)} \\
HSE & 1.16 \textcolor{blue}{($1.17$)} & & 0.64 \textcolor{blue}{($0.62$)} & 0.83 \textcolor{blue}{($0.70$)}\\
SRSH(Bulk Si($\alpha,\gamma$)\cite{WOHF21}) & 1.15
 \textcolor{blue}{($1.20$)} & & 0.64 \textcolor{blue}{($0.69$)}  & 0.86 \textcolor{blue}{($0.68$)} \\
%AM05 & 0.50 & & {0.37 } \\ 
%SCAN & 0.84 & & 0.45 \\ \hline
 \hline
 \hline
\bottomrule
\label{table1}
\end{tabular}
\end{table}
%Then, we compute the electronic band structure along the following high-symmetry path: 

 % using the \verb|projwfc.x| functionality of Quantum Espresso to project the bands on the individual atoms in the cell.
 
\subsection{Convergence with respect to the number of layers and vacuum size}

Here we check the convergence of the fundamental gap of the Si surface with respect to both the vacuum size and the number of layers. In the referenced text by Jain \textit{et al.} a 24 atom Si Pandey reconstructed surface was used. We note that the code PARATEC is not readily  available anymore, and all values for all functionals are systematically off by $0.1$ eV for the 24 atom slabs. To fix this issue we converge the gap of the Si surface with respect to number of layers in it, and find that with a converged surface of 48 atoms, it is possible to reproduce the results of Jain \textit{et al}. Ref.~\cite{JCL11} also mentions that the surface gap is invariant to vacuum after a minimum of $12$ \AA ~and we find this trend to hold as well in Table SI III.

The surface state gap of 24-atom relaxed Si(111)-(2$\times$1) slabs with three vacuum sizes, 12.125 $\text{\AA}$ \cite{JCL11}, 16 $\text{\AA}$, and 20 $\text{\AA}$ vacuum, were calculated using PBE~\cite{PBE1996} functional using VASP.  We find the surface state gap to be invariant within 0.001 eV. Next we check the convergence with respect to the Si slab and we compute the surface state gaps of a 24-atom, 36-atom, and 48-atom Si(111)-(2$\times$1) slabs with 12.125 $\text{\AA}$ vacuum sizes, using PBE functional in VASP. The 36-atom and 48-atom slabs are constructed by inserting perfect Si unit cells in the central part of 24-atom relaxed Si(111)-(2$\times$1) slabs. For 48-atom slab, the gap of the slab is converged to within 0.01 eV, with respect to the number of atoms in the unit cell. The 48 atom structures, with 12.125 $\text{\AA}$ vacuum, relaxed with PBE, were used in all following gap and band structure calculations for Si and Ge. 

%\begin{table}[htbp!]
%\caption{Si (111)-(2$\times$1) 24-atom Si surface surface fundamental gap convergence checked with respect to vacuum and slab size}
%\begin{tabular}{ c c c }
%\hline
% & 24-atom Si surface &       \\ \hline
% \hline
%vacuum size ( $\text{\AA}$)  & & surface state fundamental gap (eV)   \\ \hline
%12.125  & & 0.292    \\ \hline
%16  & & 0.292     \\ \hline
%20  & & 0.292    \\ \hline
%\hline
%\end{tabular}
%\end{table}
%
%\begin{table}[htbp!]
%\caption{Si (111)-(2$\times$1) convergence of atoms per unitcell checked using with 12 \text{\AA} vacuum, using the PBE XC functional.}
%\begin{tabular}{ c c c }
%\hline
%atoms in slab & & surface state fundamental gap (eV)  \\ \hline
%\hline
%24  & & 0.292  \\ \hline
%36  & & 0.367 \\ \hline
%48  & & 0.376 \\ \hline
%\end{tabular}
%\end{table}

\begin{table}[htbp!]
\centering
\begin{minipage}[b]{0.45\textwidth}
\centering
\caption{Convergence of surface state gap with vacuum size for a 24-atom Si surface.}
\begin{tabular}{ c c c }
\hline
Vacuum size ($\text{\AA}$)  & & Surface state gap (eV) \\ \hline \hline
12.125  & & 0.292 \\ \hline
16  & & 0.292 \\ \hline
20  & & 0.292 \\ \hline \hline
\end{tabular}
\end{minipage}
\hfill
\begin{minipage}[b]{0.45\textwidth}
\centering
\caption{Convergence of surface state gap with slab atoms.}
\begin{tabular}{ c c c }
\hline
Atoms in slab & & Surface state gap (eV) \\ \hline
\hline
24  & & 0.292 \\ \hline
36  & & 0.367 \\ \hline
48  & & 0.376 \\ \hline \hline
\end{tabular}
\end{minipage}
\end{table}

\sec{Projected band structure and optoelectronic calculations}

\begin{figure*}[htb]
    \centering
    \subfigure (a)\includegraphics[width=0.45\textwidth]{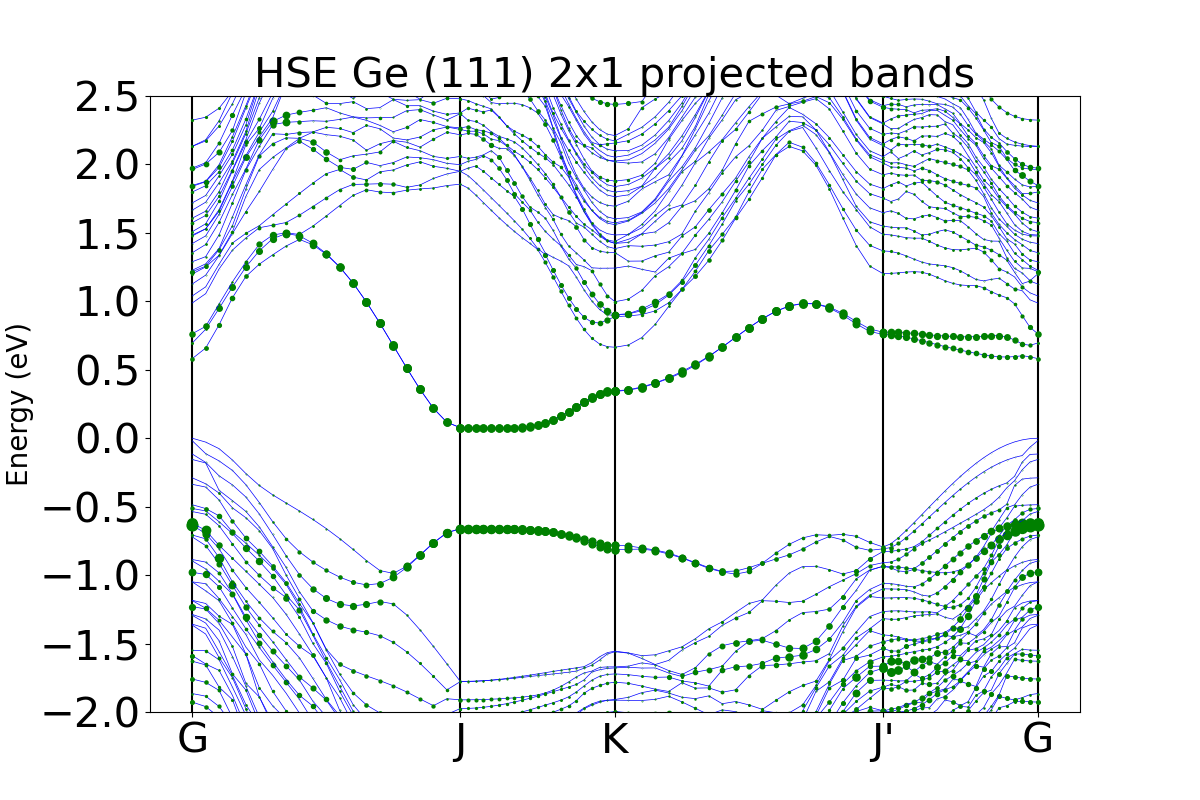}
    \centering
    \subfigure (b)\includegraphics[width=0.45\textwidth]{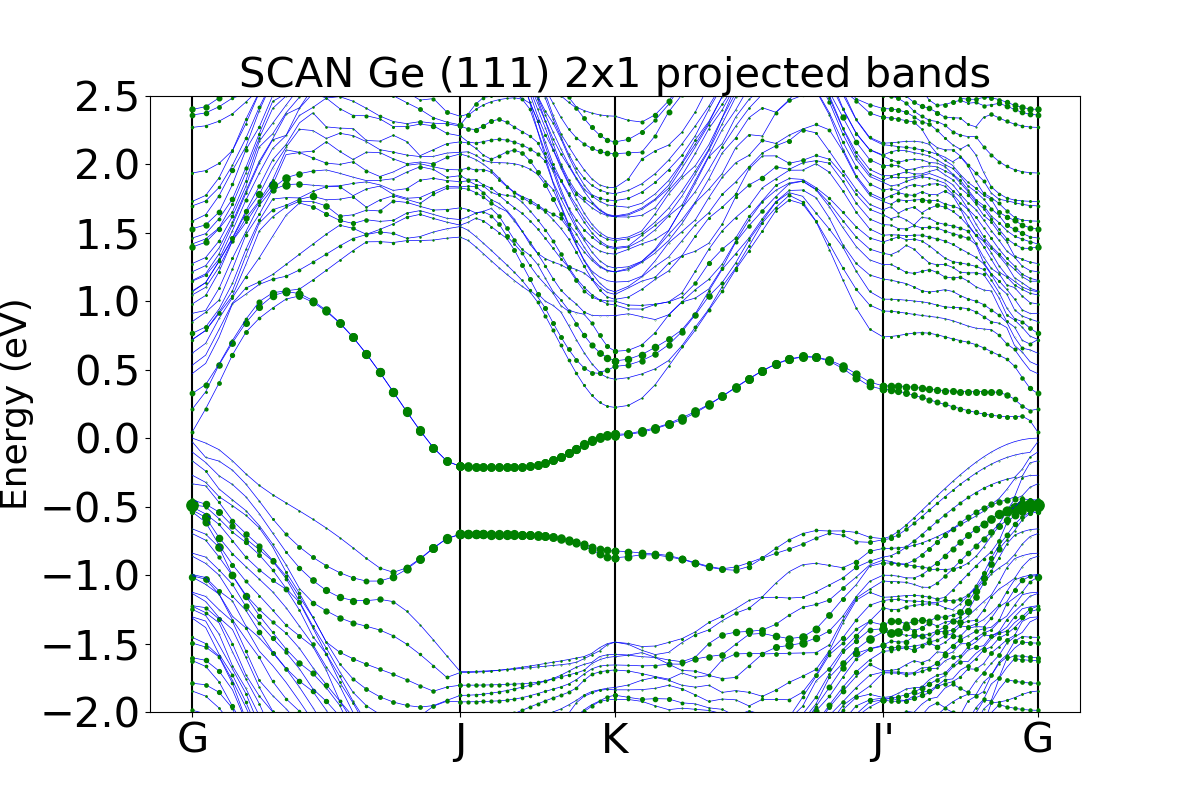}
    
    \subfigure (c)\includegraphics[width=0.45\textwidth]{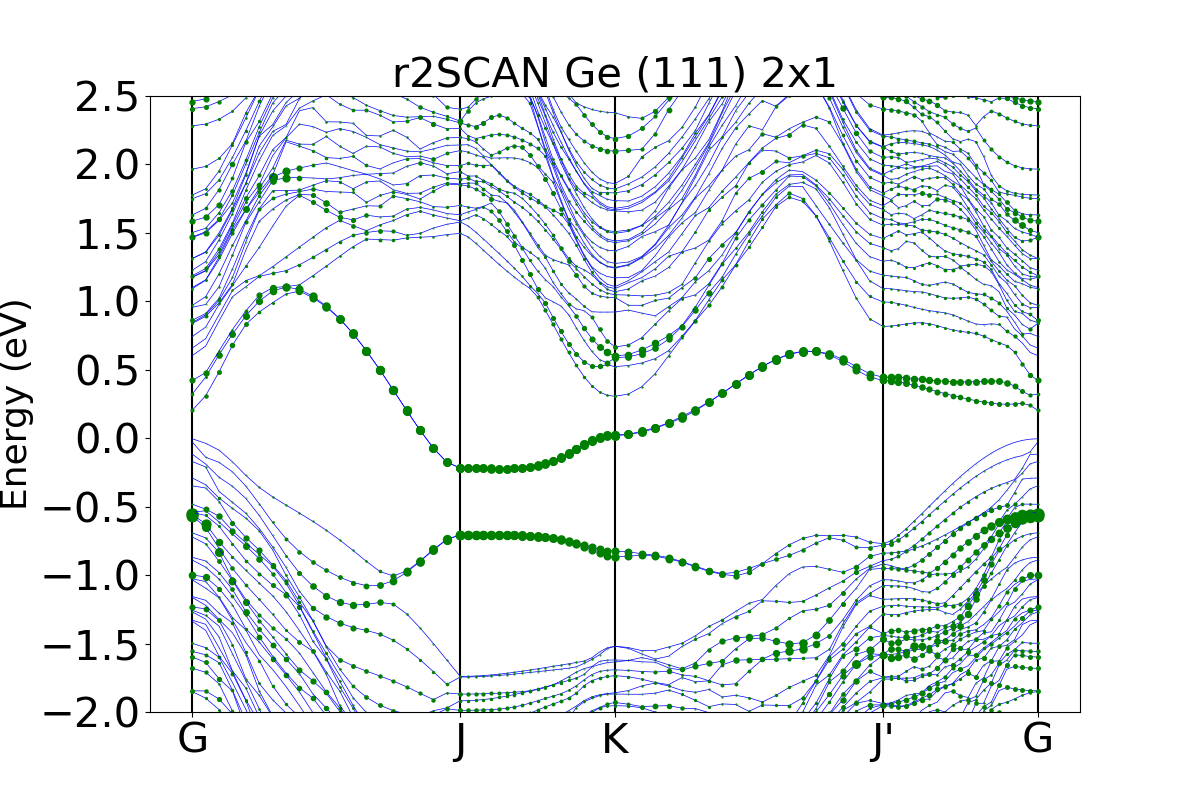}
    \subfigure (d)\includegraphics[width=0.45\textwidth]{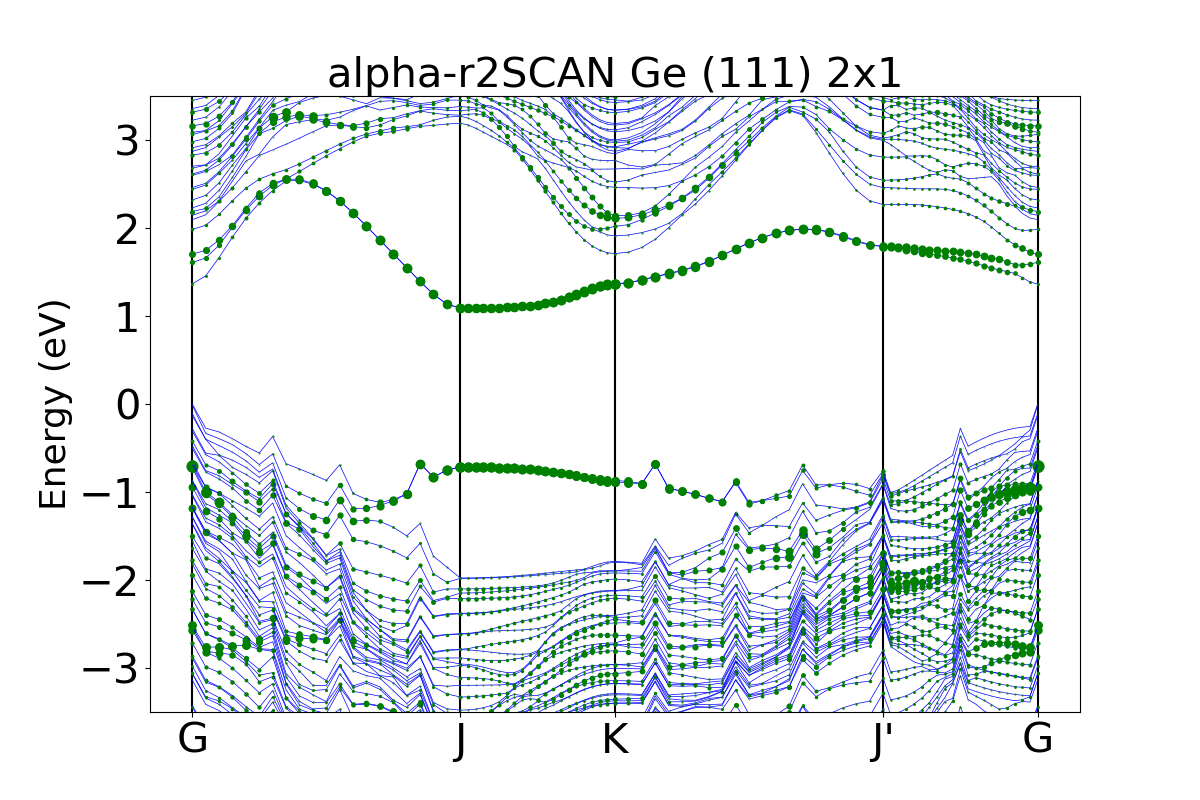}
    
    \subfigure{ (e)\includegraphics[width=0.45\textwidth]{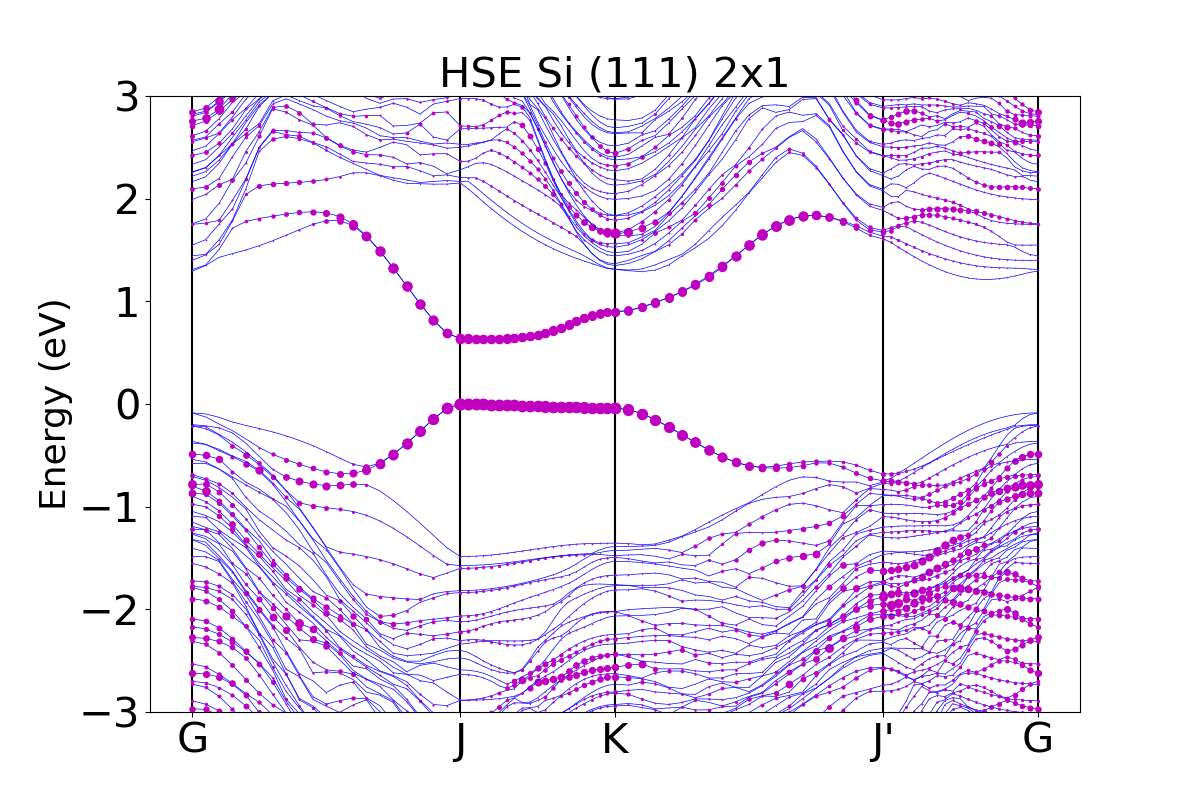}}
    \subfigure {(f)\includegraphics[width=0.45\textwidth]{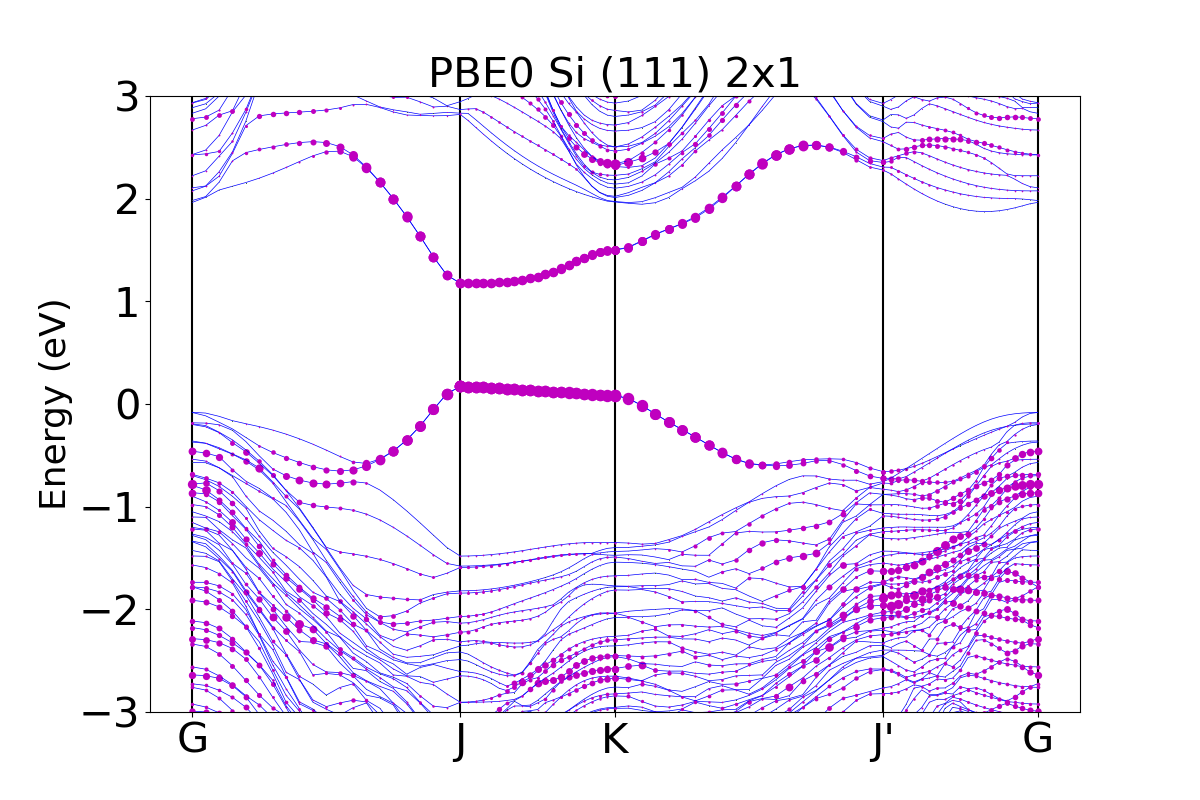}}
    \subfigure{ (g)\includegraphics[width=0.45\textwidth]{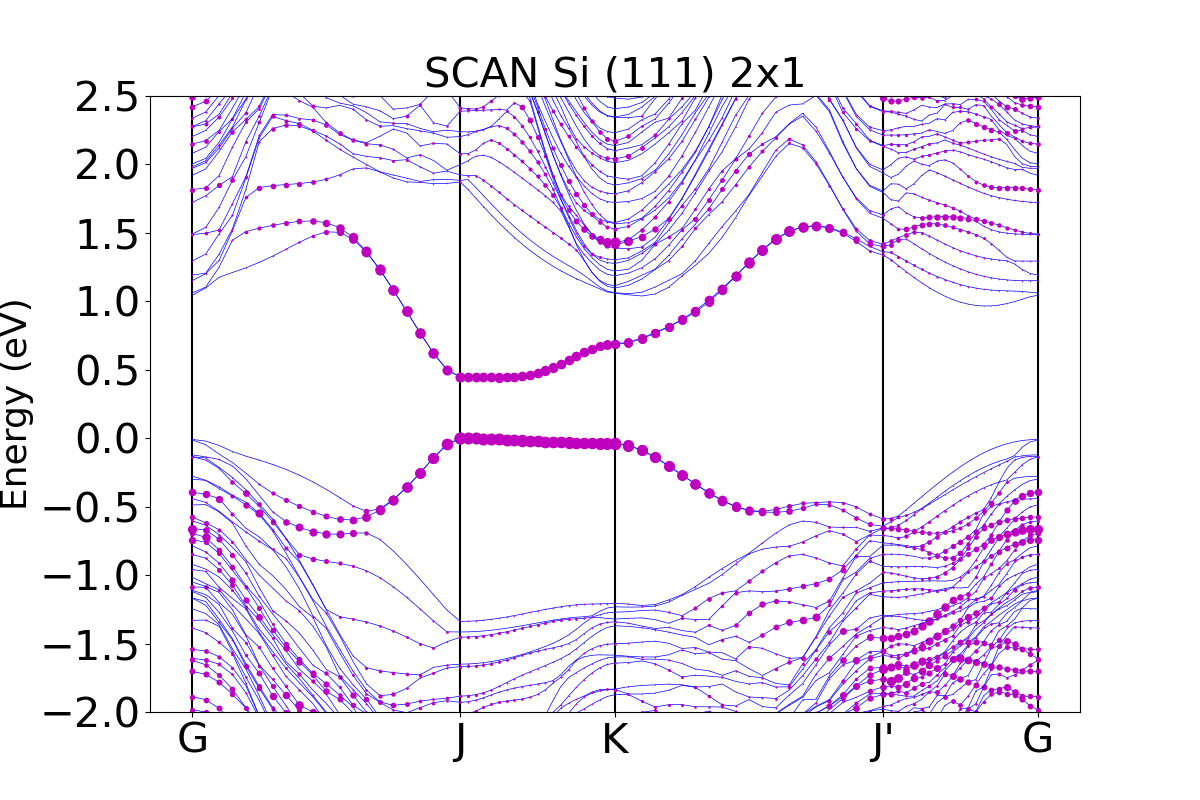}}
    \caption{Projected band structures for Ge(111)-(2$\times$1) using (a) HSE, (b) SCAN, (c) r\textsuperscript{2}SCAN,  (d) $\alpha$-r\textsuperscript{2}SCAN and for Si(111)-(2$\times$1) for (e) HSE, (f) PBE0, and (g) SCAN.}
    \label{fig proj bands}
\end{figure*}

Additional projected band structures obtained with various functionals are shown in Fig. SI~\ref{fig proj bands}. In particular (a-d) represent the band structures for the Ge (111) 2$\times$1 slab, with HSE, SCAN, r\textsuperscript{2}SCAN, and $\alpha$-r\textsuperscript{2}SCAN functionals, respectively; while, (e-g) represent the band structures for the Si (111) 2$\times$1 slab, with HSE, PBE0, and SCAN functionals, respectively. Overall, we observe that in all the band structures, the contribution (represented as overlaid colored dots) of the 8 sites at the surfaces is located mostly on the isolated bands close to the band gap, which therefore are usually named as surface states. Also, we note that the band shapes do not change much with various functionals, but rather the gap size varies. 

Projected band structures were obtained by combining the LORBIT=11 flag in VASP and post-precessing its output with pymatgen package\cite{pymatgen}. The LORBIT flag provides a decomposition of the KS orbitals into local quantum numbers (l and m) and obtain local properties, by using the projector functions of the PAW method. Then, to understand 
the contribution to the band structure of the 8 (Ge or Si) sites located at the surfaces, we used the function \verb|get_projected_plots_dots_patom_pmorb()| of the class \verb|BSPlotterProjected| in pymatgen package. This function takes care of summing both \textit{s} and \textit{p} orbitals of these sites and represent their total contribution as colored dots over the lines, with their size being proportional to the contribution.
The band structures with hybridfunctionals computed in VASP were obtained using k-points sampling on a uniform grid ($4\times8\times1$), with weights that depends on the symmetries, with a list of k-points on a specific path (20 points per segment), with zero-weight. The k-path $\Gamma$-J-K-J'-$\Gamma$ is representative of the surface cell and includes k-points with in-plane reduced coordinates exclusively ($z=0$). Finally, an energy cutoff of 520 eV and an energy difference of 1e-6 eV for electronic convergence were used.

\subsection{TDDFT}

\begin{figure*}[!htbp]{
    \centering
    \subfigure(a)\includegraphics[width=0.45\textwidth]{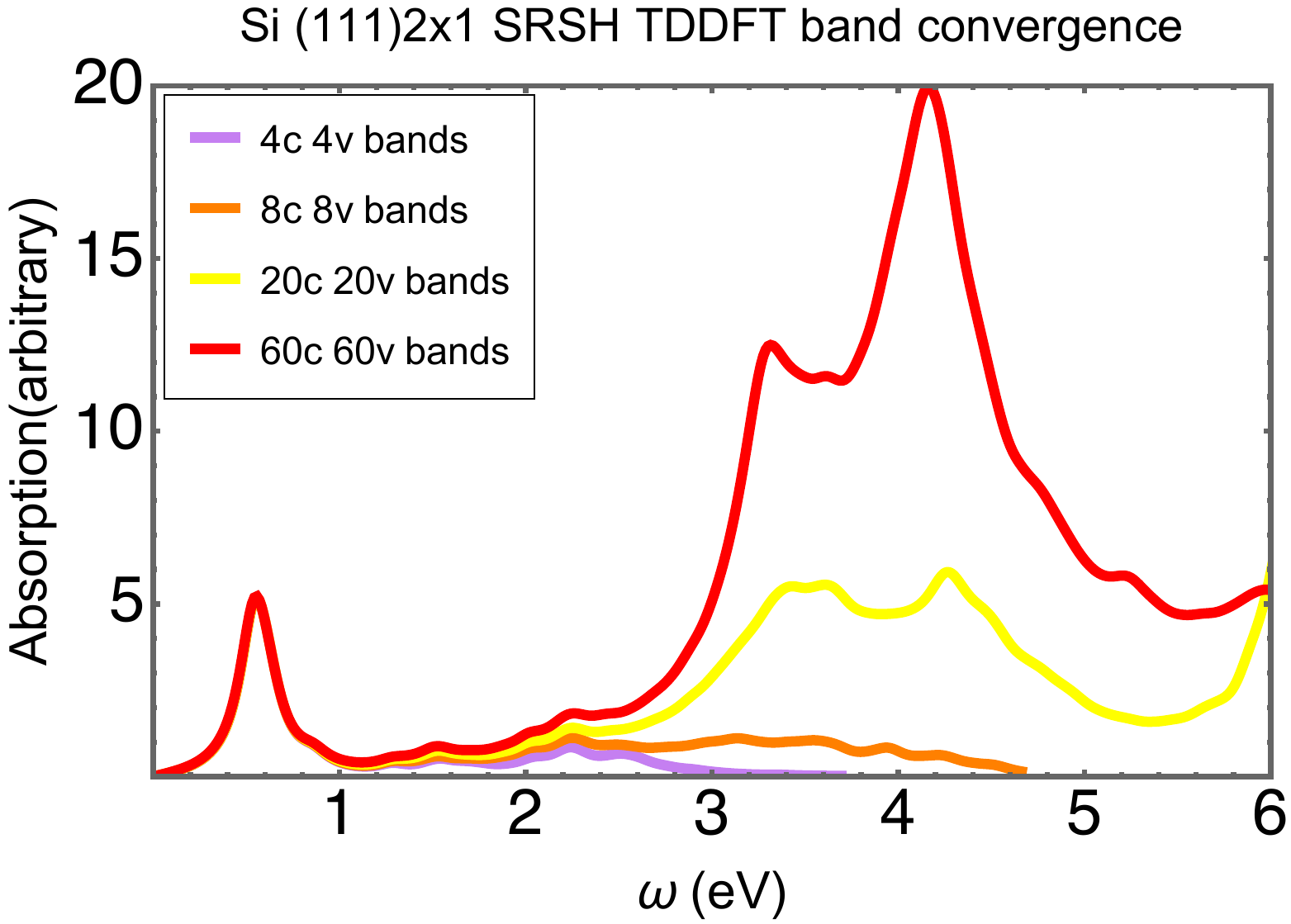}
    \hfill
    \centering
    \subfigure (b)
\includegraphics[width=0.45\textwidth]{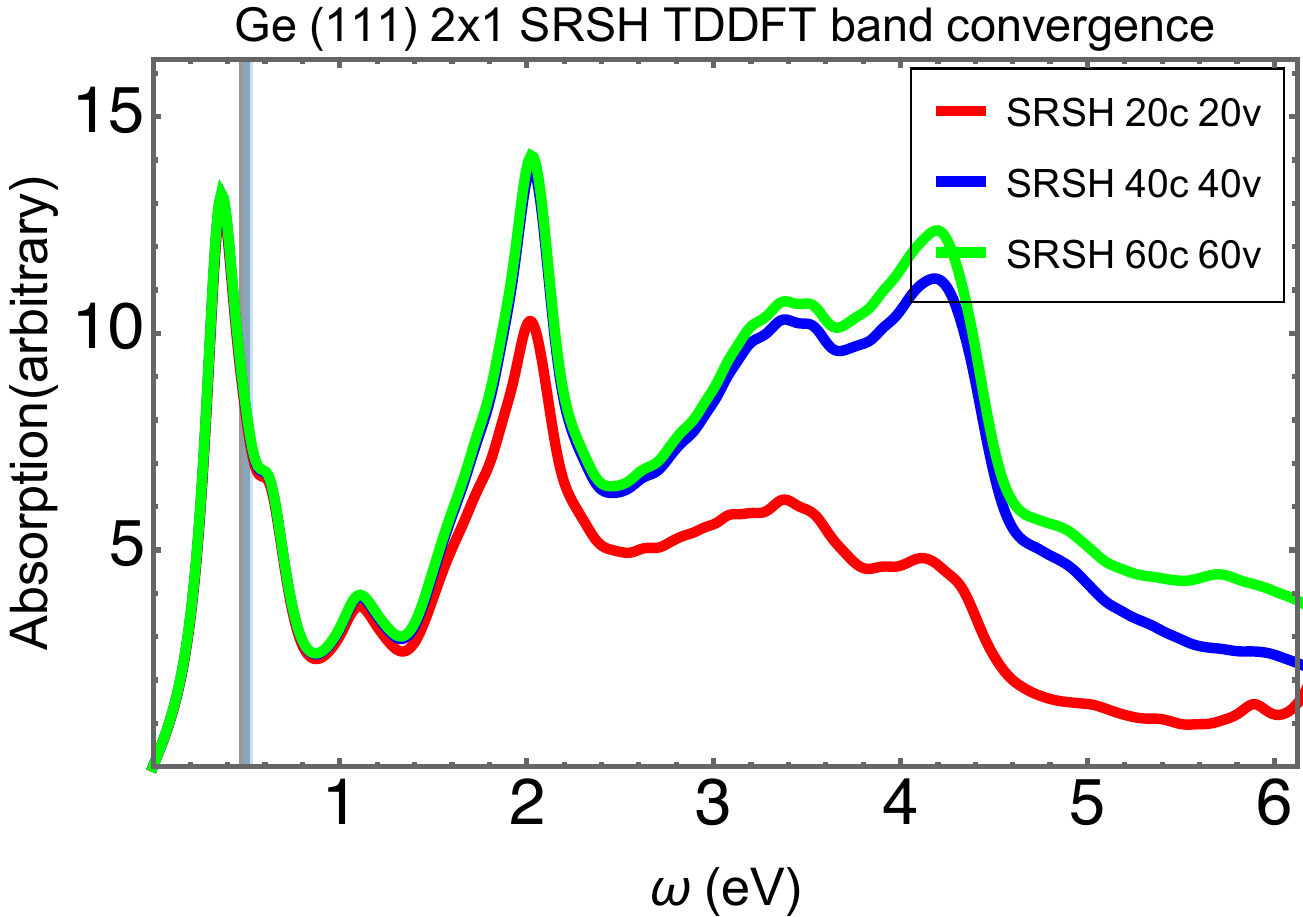}
   
    \hfill
    \centering
    \subfigure(c)
\includegraphics[width=0.45\textwidth]{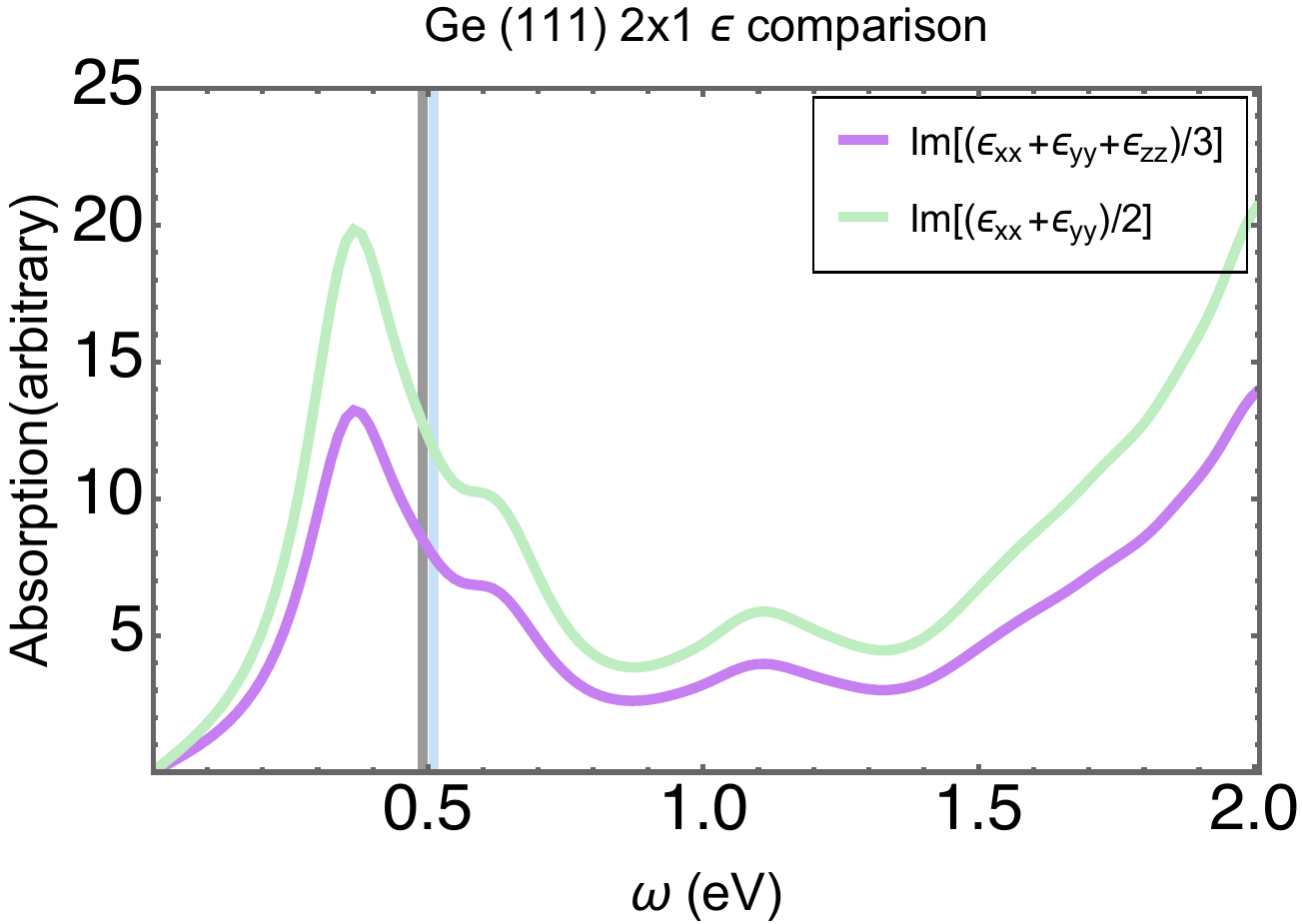}
  \hfill
    \centering
    \subfigure(d)
\includegraphics[width=0.45\textwidth]{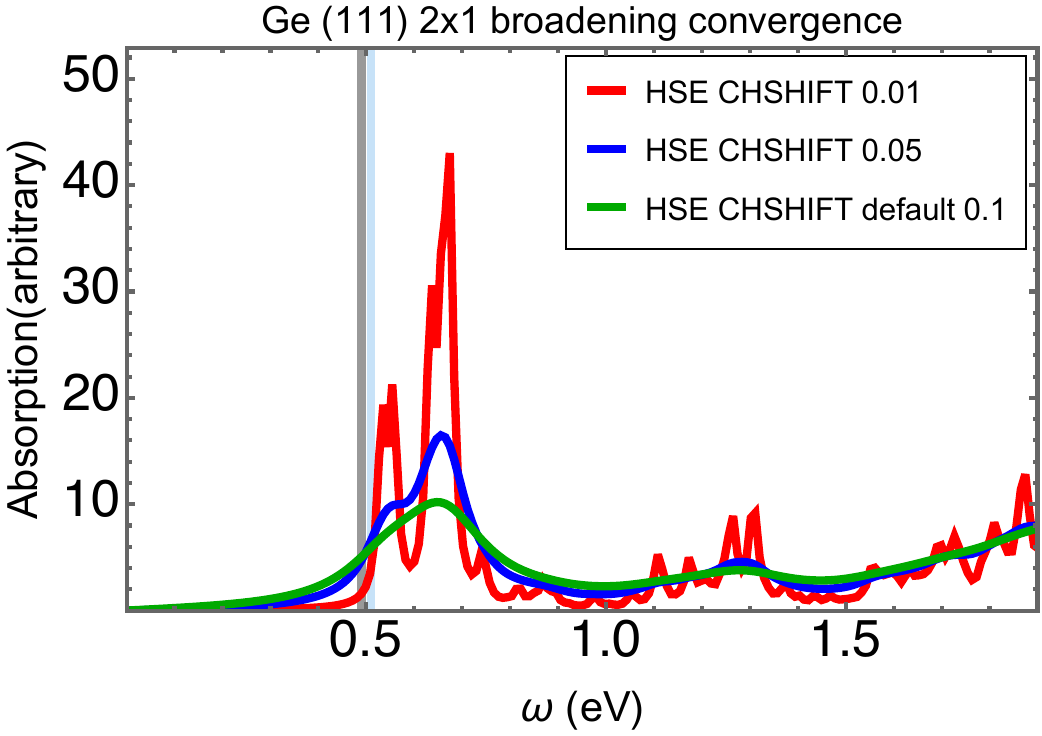}
 
  \hfill
    \centering
    \subfigure(e)
\includegraphics[width=0.45\textwidth]{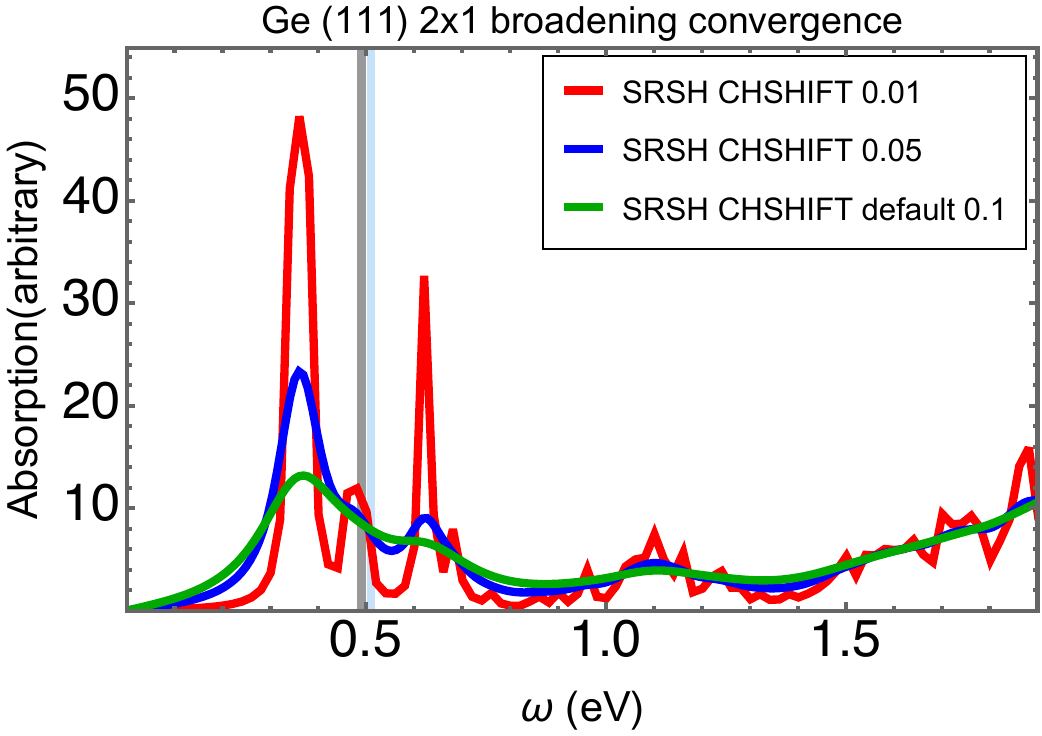}
\caption{Convergence of the SRSH-TDDFT computed optical spectra for (a) Si(111)$2\times1$ and (b) Ge(111)$2\times1$. For Ge(111)$2\times1$ slab: (c) Comparison between Im[$(\epsilon_{xx}+\epsilon_{yy}+\epsilon_{zz})/3$] and Im[$(\epsilon_{xx}+\epsilon_{yy})/2$, and effect of the Lorentzian broadening (CSHIFT flag in VASP) on the (d) HSE and (e) SRSH TDDFT computed optical spectra.}
    \label{fig converge}}
\end{figure*}

To compute the excited state spectra we use
linear response TDDFT and solve the so-called Casida equations using the Tamm Dancoff approximation, within the adiabatic approximation. All TDDFT calculations are done with VASP using a $4\times8\times1$ k-point grid and a 520 eV plane wave cut-off energy. A convergence study of the TDDFT calculations with respect to the number of valence (v) and conduction (c) bands included in the calculations is shown in  Fig. SI~\ref{fig converge}a and \ref{fig converge}b for the Si and Ge slabs, respectively. Here, we focus on the convergence of the lowest energy peaks, which correspond to the surface state optical gaps. For the Si Slab, we use 8 conduction, and 8 valence bands, for the Ge slab we use 20 conduction, and 20 valence bands for the final spectra in the main text. Indeed, previous literature \cite{Wing2019} reported the bulk direct optical gaps of Si and Ge at around $\sim2-3$ eV.

In Fig. SI~\ref{fig converge}c, we compare the different contributions of the Im[$\epsilon$]. We find the only difference between Im[$(\epsilon_{xx}+\epsilon_{yy}+\epsilon_{zz})/3$] vs. Im[$(\epsilon_{xx}+\epsilon_{yy})/2$] to be the magnitude of oscillator strengths. In the main text, we chose to plot Im[$(\epsilon_{xx}+\epsilon_{yy}+\epsilon_{zz})/3$] for easy comparison as has been done in literature \cite{Camarasa-Gomez2023,Camarasa-Gomez2024,Ohad2023,Sagredo2024}, and because of the choice in thickness of the slab itself.\  %In Fig. \ref{fig proj bands} d, we use a finer broadening (CSHIFT flag in VASP) to decouple the HSE and SRSH optical surface state gap. 

For all functionals we find a small number of close-in-energy low-lying excited states with large oscillator strength. These states are at energies far below the direct bulk gap, and are associated with the surface bands. Thus, we report the optical gap as the average of the lowest set of excitations with finite oscillator strength – that is, the lowest bright excitations – over a window of $\eta= 0.1$ eV (consistent with the Lorenzian broadening in the usual Kramers-Kronig transformation $\epsilon_{1}(\omega) = 1+ \frac{2}{\pi} \mathcal{P} \int_{0}^{\infty} \frac{\epsilon_{2}(\omega')\omega'}{\omega^{2'} - \omega^{2}+i\eta} , d\omega' $, where Im[$\epsilon_{2}$], is plotted in the spectrum). Complete set of exciton energies and oscillator strengths are reported in Table SI~\ref{table_excitons}. For the Si (111) slab, for SRSH we average over 5 low-lying excitations with nonzero and large oscillator strength close in energy, appearing at energies of $0.547, 0.619$ and $0.623$ eV, respectively, coming from our TDDFT calculations. For the Ge(111) slab, for PBE0 we average over six low-lying excitations with nonzero and large oscillator strength close in wavelength, appearing at the energies of 0.79, 0.8, 0.8, 0.81, 0.86, and 0.89 eV, respectively coming from our TDDFT calculations; for SRSH, we average over two excitations with energies of 0.35 and 0.37 eV, respectively; and for HSE06, we average over seven excitations, with energies of 0.53, 0.56, 0.63, 0.65, 0.66, 0.67, and 0.67 eV, respectively.
The optical gaps we obtain from these averages are reported in the main text and are tabulated below. In all absorption spectra presented in the main text, we use a Lorentzian broadening value of 0.1 eV, the default value in VASP (labeled as the flag CSHIFT=0.1). In Fig. SI~\ref{fig converge}d and Fig. SI~\ref{fig converge}e, we vary the Lorentzian broadening between 0.01 and 0.1 eV for SRSH and HSE for the Ge slab. In both cases, the individual excited states discussed above are increasingly resolved as the broadening is reduced. In particular, in  Fig. SI~\ref{fig converge}d, one can resolve several of the seven excitations mentioned above at a broadening of 0.01 eV in the range 0.53-0.67 eV. In Fig. SI~\ref{fig converge} e, we see that the two excited states at a wavelength of 0.35 and 0.37 eV are unresolved even with a small 0.01 eV broadening. As discussed above, for the purposes of this study we elect to report an average of the states with non-zero oscillator strength in a 0.1 eV energy range as the optical gap.

\begin{table}[h]
\caption{Peaks that appear in vasprun.xml file that show the wavelength ($\omega$ in eV) and their corresponding oscillator strength (arb.units). These optical transitions were used to make the optical spectra that appears in the main text.}
\label{table_excitons}
\centering
\begin{tabular}{ c c c c c }
\hline
\hline
 & &{Si(111)-($2 \times1$)} & & {}\\ \hline
SRSH& &{$\omega$ (eV)} & {optical tran. (arb. units)} & \\ \hline
labeled peak 0.55 eV & & 0.547 &  146410.932  \\
& & 0.547  &   426.304\\
& & 0.619 & 33777.981\\
& & 0.619 &  995.436 \\
& & 0.623 &  1686.491 \\
 \hline \hline
& & {Ge(111)-($2 \times1$)} & & {}\\ \hline
SRSH & &{$\omega$ (eV)} & {optical tran. (arb. units)} & \\ \hline
labeled peak 0.36 eV& & 0.349 & 237493.963 \\
& & 0.373    & 218997.240\\
& & 0.458 &29.607\\
& & 0.469   & 63776.400 \\ & & 0.494 & 39308.934\\
 \hline
 HSE & &{$\omega$ (eV)} & {optical tran. (arb. units)} & \\ \hline
labeled peak 0.65 eV & & 0.533 & 64188.776 \\
& & 0.556    & 70217.525\\
& & 0.634 & 97838.808\\
& & 0.646   & 2501.769 \\ 
& & 0.657 & 77646.238 \\
& & 0.673 & 71427.675 \\
& & 0.673 & 67820.036 \\
 \hline
  PBE0 &  &{$\omega$ (eV)} & {optical tran. (arb. units)} & \\ \hline
labeled peak 0.81 eV& & 0.788 & 44343.101 \\
& & 0.800    & 40047.793\\
& & 0.801 & 40140.240\\
& & 0.814   & 48567.651 \\ 
& & 0.860 & 29724.070 \\
& & 0.887 & 19814.673 \\
 \hline \hline
\bottomrule
\end{tabular}
\end{table}

\clearpage

\sec{Atomic structure}

For convenience we provide the atomic structures for both Si(111)$2\times1$ and Ge(111)$2\times1$ slabs, used in our DFT and TDDFT calculations.

\begin{minipage}{0.44\linewidth}
\noindent \textbf{POSCAR for VASP calculations of Pandey reconstructed surface Si(111)-(2$\times$1)}\\
\tiny
\noindent Si\\
1.0\\
        6.6884665489         0.0000000000         0.0000000000\\
        0.0000000000         3.8615860939         0.0000000000\\
        0.0000000000         0.0000000000        48.4192314148\\
   Si\\
   48\\
Cartesian\\
     4.433917834         1.930793047         8.132477660\\
     3.310091319         0.000000000         8.688234651\\
     6.641571342         1.930793047         7.407239668\\
     1.341486814         0.000000000         7.340852033\\
     5.648178412         1.930793047         5.188376940\\
     2.140914021         0.000000000         5.090118724\\
     4.455170993         0.000000000         4.475193456\\
     1.118030144         1.930793047         4.211863342\\
     4.454841298         0.000000000         2.040276001\\
     1.116338614         1.930793047         1.899929101\\
     3.344233274         1.930793047         1.182364832\\
     0.000000000         0.000000000         1.182364832\\
     0.000000000         0.000000000        28.319015555\\
     3.344233274         1.930793047        28.319015555\\
     5.572127735         1.930793047        27.601452188\\
     2.233625650         0.000000000        27.461105468\\
     5.570435807         1.930793047        25.289517767\\
     2.233295357         0.000000000        25.026186571\\
     4.547552528         0.000000000        24.411261302\\
     1.040287738         1.930793047        24.313004169\\
     5.346979635         0.000000000        22.160527632\\
     0.046894945         1.930793047        22.094140719\\
     3.378375230         0.000000000        20.813147179\\
     2.254548715         1.930793047        21.368902006\\
     3.344234869         1.930794888        37.777939986\\
     3.344234869         1.930794888        47.236864418\\
     1.114745156         1.930794888        31.471991327\\
     1.114745156         1.930794888        40.930915759\\
     4.458980224         0.000000000        31.471991327\\
     4.458980224         0.000000000        40.930915759\\
     4.458980224         0.000000000        33.836722435\\
     4.458980224         0.000000000        43.295646867\\
     3.344234869         1.930794888        30.683746663\\
     3.344234869         1.930794888        40.142673980\\
     1.114745156         1.930794888        33.836722435\\
     1.114745156         1.930794888        43.295646867\\
     0.000000000         0.000000000        37.777939986\\
     0.000000000         0.000000000        47.236864418\\
     2.229489713         0.000000000        34.624964214\\
     2.229489713         0.000000000        44.083891531\\
     5.573725181         1.930794888        34.624964214\\
     5.573725181         1.930794888        44.083891531\\
     2.229489713         0.000000000        36.989695322\\
     2.229489713         0.000000000        46.448622639\\
     5.573725181         1.930794888        36.989695322\\
     5.573725181         1.930794888        46.448622639\\
     0.000000000         0.000000000        30.683746663\\
     0.000000000         0.000000000        40.142673980\\
     \end{minipage}
\begin{minipage}{0.45\linewidth}
\includegraphics[width=.8\linewidth]{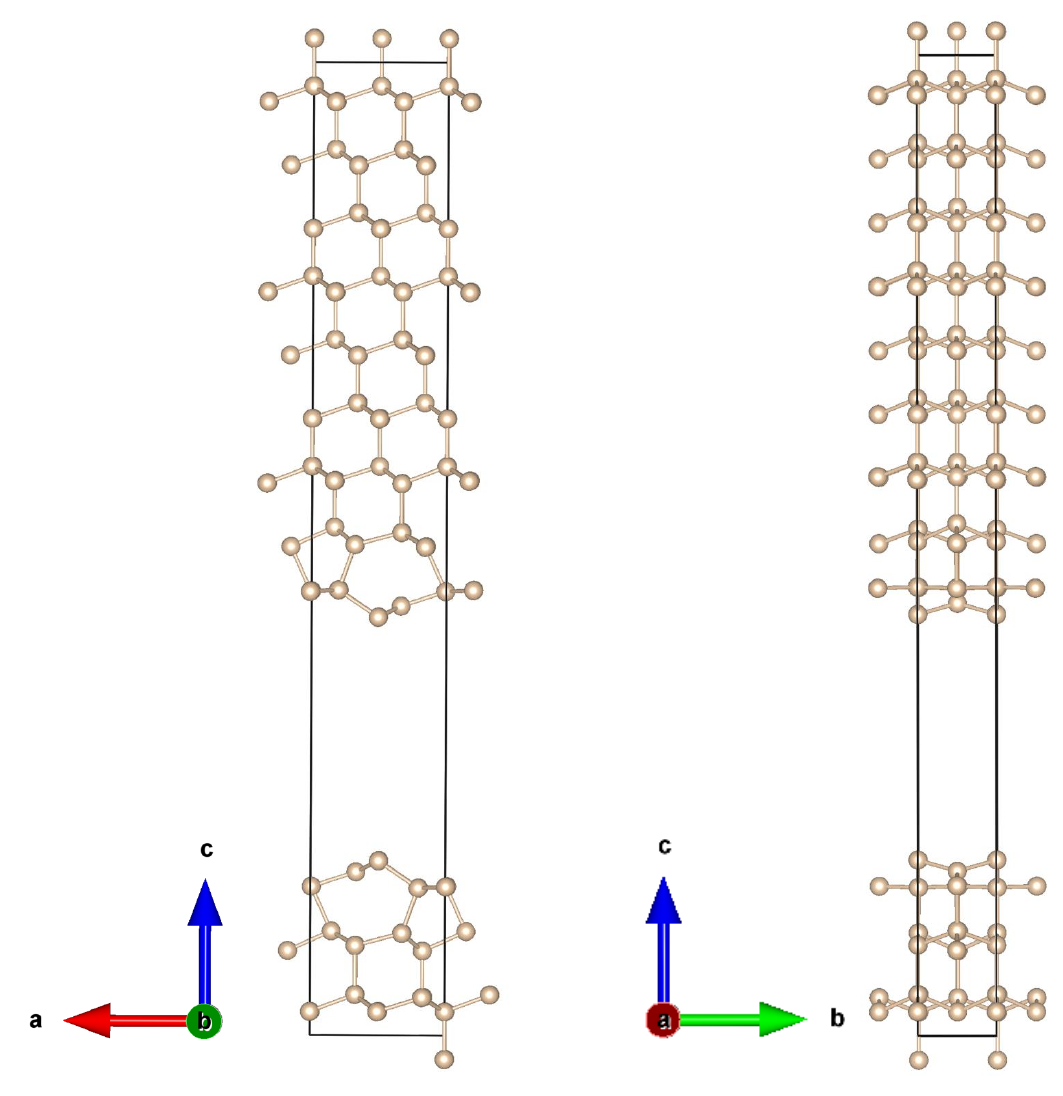}
\caption{Geometric structure of the Pandey reconstructed surface of Si(111)-(2x1) of the 48-atom slab. \label{fsi3}}
\end{minipage}
     
\begin{minipage}{0.45\linewidth}
\noindent \textbf{POSCAR for VASP calculations of Pandey reconstructed surface Ge(111)-(2$\times$1)}\\
\tiny
\noindent Ge\\
1.0\\
        6.9259290695         0.0000000000         0.0000000000\\
        0.0000000000         3.9986848831         0.0000000000\\
        0.0000000000         0.0000000000        52.6382713318\\
   Ge\\
   48\\
Cartesian\\
     4.655013992         1.999342442        18.259479862\\
     3.503571332         0.000000000        17.410160978\\
     6.616963211         1.999342442        16.659749754\\
     1.169706327         0.000000000        16.670698010\\
     5.856690354         1.999342442        14.283342964\\
     2.198241439         0.000000000        14.293133487\\
     4.600451485         0.000000000        13.573041355\\
     1.167413848         1.999342442        13.276161499\\
     4.604419900         0.000000000        11.032560547\\
     1.170752097         1.999342442        10.826113219\\
     3.503744303         1.999342442        10.081282013\\
     6.889609382         0.000000000        10.066016578\\
     0.036312647         0.000000000        40.123732860\\
     3.422046060         1.999342442        40.108468994\\
     5.755052198         1.999342442        39.363689557\\
     2.321418969         0.000000000        39.157029664\\
     5.758154935         1.999342442        36.913746382\\
     2.325283767         0.000000000        36.616601409\\
     4.727555942         0.000000000        35.896510845\\
     1.068871721         1.999342442        35.906670023\\
     5.756492932         0.000000000        33.519049859\\
     0.308910283         1.999342442        33.530366769\\
     3.422184767         0.000000000        32.782063936\\
     2.271559279         1.999342442        31.931902637\\
     3.463393865         1.999342442        50.170905700\\
     3.488424231         1.999342442         7.588596090\\
     1.150244457         1.999342442        43.496846621\\
     1.158472426         1.999342442         0.883007046\\
     4.617385260         0.000000000        43.438574103\\
     4.615009908         0.000000000         0.879322363\\
     4.617115277         0.000000000        45.938472321\\
     4.610667067         0.000000000         3.368007110\\
     3.437518668         1.999342442        42.601047458\\
     3.462597540         1.999342442         0.018739225\\
     1.152862442         1.999342442        45.977530878\\
     1.159885294         1.999342442         3.374902911\\
     0.000907297         0.000000000        50.166798732\\
     6.906612511         0.000000000         7.574489171\\
     2.315379655         0.000000000        46.821533293\\
     2.308896975         0.000000000         4.251013960\\
     5.766106214         1.999342442        46.814583762\\
     5.773142689         1.999342442         4.212061686\\
     2.310974685         0.000000000        49.310218138\\
     2.308626786         0.000000000         6.750911002\\
     5.767484199         1.999342442        49.306531592\\
     5.775753760         1.999342442         6.692798496\\
     0.019399527         0.000000000        42.615206930\\
     6.925056373         0.000000000         0.022792372\\
     \end{minipage}
\begin{minipage}{0.5\linewidth}  \includegraphics[width=.7\linewidth]{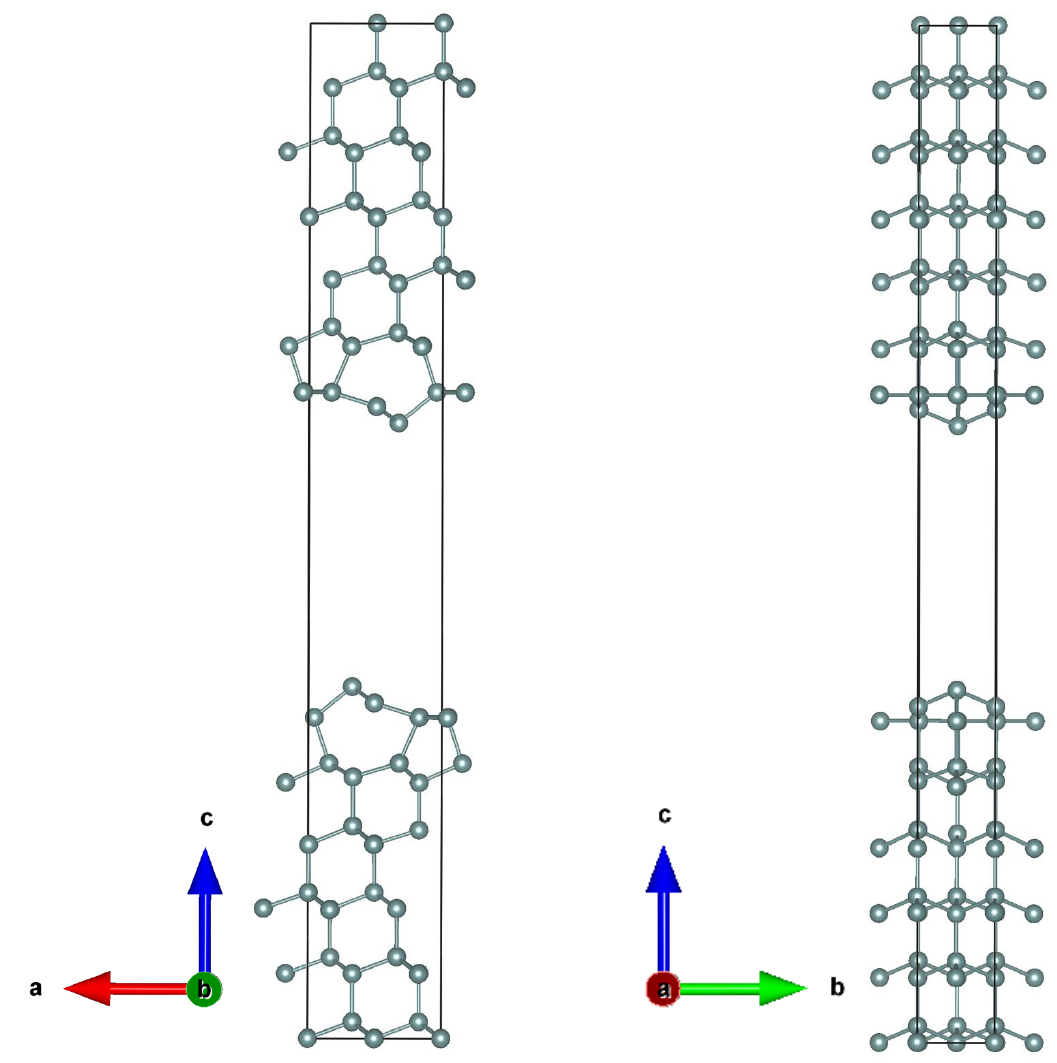} \caption{Geometric structure in the POSCAR format of VASP for the Pandey reconstructed surface of Ge(111)-(2x1) of the 48-atom slab. \label{fsi4}}
\end{minipage}
\clearpage

\bibliography{wot-srsh-surfaces}
\label{page:end}